\documentclass[12pt]{book}
\usepackage{sussp} 
\usepackage{graphicx}     
\usepackage{makeidx}
\makeindex
\def \As{\not \! \! A}
\def \ba{\bar \alpha}
\def \bam{{\bf A}_\mu}
\def \ban{{\bf A}_\nu}
\def \bAs{\not \! \! {\bf A}}
\def \bea{\begin{eqnarray}}
\def \Bee{{\cal B}_{e^+ e^-}}
\def \bD{{\bf D}}
\def \bDm{{\bf D}_\mu}
\def \bDM{{\bf D}^\mu}
\def \bDn{{\bf D}_\nu}
\def \bDs{\not \! \! {\bf D}}
\def \beq{\begin{equation}}
\def \bFmn{{\bf F}_{\mu \nu}}
\def \bg{\bar{g}}

\def \bT{{\bf T}}
\def \cB{{\cal B}}
\def \cft{\cos^4 \theta}
\def \cH{{\cal H}}
\def \cL{{\cal L}}
\def \cM{{\cal M}}
\def \cst{\cos^2 \theta}
\def \dm{\partial_\mu}
\def \dM{\partial^\mu}

\def \dn{\partial_\nu}
\def \ds{\not \! \partial}
\def \dst{\not \! \partial}
\def \Ds{\not \! \! D}

\def \eea{\end{eqnarray}}
\def \eeq{\end{equation}}

\def \ew{SU(2) $\otimes$ U(1)}
\def \g{{\rm~GeV}}
\def \gf{\gamma_5}
\def \gm{\gamma_\mu}
\def \gM{\gamma^\mu}
\def \gmn{g_{\mu \nu}}

\def \gz{(g^2 + {g'}^2)^{1/2}}
\def \hc{{\rm h.c.}}

\def \im{{\rm Im}}
\def \ite{{\it et al.}}

\def \ks{\not \! k}
\def \m{{\rm MeV}}

\def \MSb{\overline{\rm MS}}
\def \nb{\bar \nu}
\def \ob{\overline{B}^0}

\def \ok{\overline{K}^0}

\def \pb{\overline{\psi}}
\def \Pmn{\Pi_{\mu \nu}}
\def \pr{\parallel}
\def \ps{\not \! p}
\def \qs{\not \! q}
\def \re{{\rm Re}}
\def \s{\sqrt{2}}
\def \sef{\sin^2 \theta^{\rm eff}}
\def \sft{\sin^4 \theta}
\def \sst{\sin^2 \theta}
\def \st{\sqrt{3}}
\def \SUL{SU(2)$_L$}
\def \sx{\sqrt{6}}
\def \tcm{\theta_{\rm c.m.}}

\def \U1Y{U(1)$_Y$}
\def \vev#1{\langle #1 \rangle}
\def \Zs{\not \! Z}
\begin{document}

\headings{The Standard Model in 2001} 
{Standard Model}
{Jonathan L. Rosner}
{Enrico Fermi Institute and Department of Physics \\
University of Chicago \\
5640 South Ellis Avenue, Chicago IL 60637, USA} 

\section{Introduction}

The ``Standard Model'' of elementary particle physics encompasses the
progress that has been made in the past half-century in understanding the
weak, electromagnetic, and strong interactions.  The name was apparently
bestowed by my Ph.\ D. thesis advisor, Sam B. Treiman, whose dedication
to particle physics kindled the light for so many of his students during
those times of experimental and theoretical discoveries.  These lectures
are dedicated to his memory.

As graduate students at Princeton in the 1960s, my colleagues and I had no
idea of the tremendous strides that would be made in bringing quantum field
theory to bear upon such a wide variety of phenomena.  At the time, its only
domain of useful application seemed to be in the quantum electrodynamics (QED)
of photons, electrons, and muons.

Our arsenal of techniques for understanding the strong interactions included
analyticity, unitarity, and crossing symmetry (principles still of great use),
and the emerging SU(3) and SU(6) symmetries.  The quark model (Gell-Mann 1964,
Zweig 1964)
was just beginning to emerge, and its successes at times seemed mysterious.
The ensuing decade gave us a theory of the strong interactions, quantum
chromodynamics (QCD), based on the exchange of self-interacting vector quanta.
QCD has permitted quantitative calculations of a wide range of hitherto
intractable properties of the {\it hadrons} (Lev Okun's name for the
strongly interacting particles),
and has been validated by the discovery of its force-carrier, the {\it gluon}.

In the 1960s the weak interactions were represented by a phenomenological (and
unrenormalizable) four-fermion theory which was of no use for higher-order
calculations.  Attempts to describe weak interactions in terms of heavy boson
exchange eventually bore fruit when they were unified with electromagnetism and
a suitable mechanism for generation of heavy boson mass was found.  This
{\it electroweak theory} has been spectacularly successful, leading to
the prediction and observation of the $W$ and $Z$ bosons and to precision
tests which have confirmed the applicability of the theory to higher-order
calculations.

In this introductory section we shall assemble the ingredients of the
standard model --- the quarks and leptons and their interactions.
We shall discuss both the theory of the strong interactions, quantum
chromodynamics (QCD), and the unified theory of weak and electromagnetic
interactions based on the gauge group \ew.  Since QCD is
an unbroken gauge theory, we shall discuss it first, in the general
context of gauge theories in Section 2.  We then discuss the theory of
charge-changing weak interactions (Section 3) and its unification with
electromagnetism (Section 4).  The unsolved part of the puzzle, the
Higgs boson, is treated in Section 5, while Section 6 concludes.

These lectures are based in part on courses that I have taught at
the University of Minnesota and the University of Chicago, as well as at
summer schools (e.g., Rosner 1988, 1997).  They owe a significant
debt to the fine book by Quigg (1983).
  
\subsection{Quarks and leptons}

\begin{figure}
\centerline{\includegraphics[height=3in]{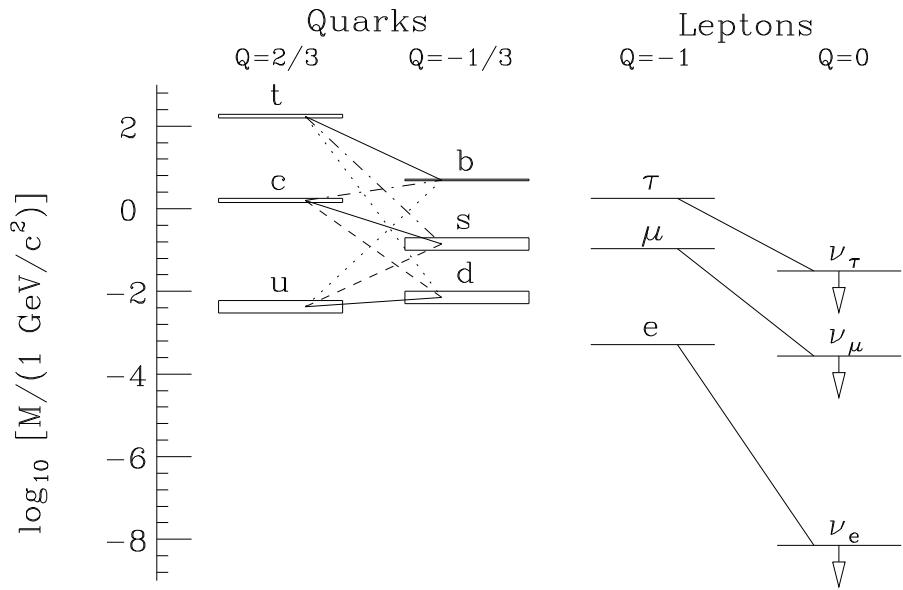}}
\caption{Patterns of charge-changing weak transitions among quarks and leptons.
The strongest inter-quark
transitions correspond to the solid lines, with dashed, dot-dashed, and dotted
lines corresponding to successively weaker transitions. \label{fig:qls}}
\end{figure}

The fundamental building blocks of strongly interacting particles, the
{\it quarks}, and the fundamental fermions lacking strong interactions,
the {\it leptons}, are summarized in Table \ref{tab:ql}.  Masses are as quoted
by the Particle Data Group (2000).  These are illustrated, along with their interactions, in
Figure \ref{fig:qls}.  The relative strengths of the charge-current weak
transitions between the quarks are summarized in Table \ref{tab:tr}.

\begin{table}
\caption{The known quarks and leptons.  Masses in GeV except where
indicated otherwise.  Here and elsewhere we take $c=1$.
\label{tab:ql}}
\begin{center}
\begin{tabular}{|c c|c c|c c|c c|} \hline
\multicolumn{4}{|c}{Quarks} & \multicolumn{4}{|c|}{Leptons} \\ \hline
\multicolumn{2}{|c}{Charge $2/3$} & \multicolumn{2}{|c}{Charge $-1/3$} &
\multicolumn{2}{|c}{Charge $-1$} & \multicolumn{2}{|c|}{Charge 0} \\ \hline
    & Mass &      & Mass &      & Mass &     & Mass \\ \hline
$u$ & 0.001--0.005 & $d$ & 0.003--0.009 & $e$ & 0.000511 & $\nu_e$ & 
 $< 3$ eV \\
$c$ & 1.15--1.35 & $s$ & 0.075--0.175 & $\mu$ & 0.106 & $\nu_\mu$ & 
 $<190$ keV \\
$t$ & $174.3 \pm 5.1$ & $b$ & 4.0--4.4 & $\tau$ & 1.777 & $\nu_\tau$ &
 $<18.2$ MeV \\ \hline
\end{tabular}
\end{center}
\end{table}

\begin{table}
\caption{Relative strengths of charge-changing weak transitions.
\label{tab:tr}}
\bigskip
\begin{center}
\begin{tabular}{|c c l|} \hline
Relative & Transition & Source of information \\
amplitude & & ~~~~~~~~(example) \\ \hline
$\sim$ 1 & $u \leftrightarrow d$ & Nuclear $\beta$-decay \\
$\sim$ 1 & $c \leftrightarrow s$ & Charmed particle decays \\
$\sim 0.22$ & $u \leftrightarrow s$ & Strange particle decays \\
$\sim 0.22$ & $c \leftrightarrow d$ & Neutrino prod.\ of charm \\
$\sim 0.04$ & $c \leftrightarrow b$ & $b$ decays \\
$\sim 0.003$--0.004 & $u \leftrightarrow b$ & Charmless $b$ decays \\
$\sim$ 1 & $t \leftrightarrow b$ & Dominance of $t \to W b$ \\
$\sim 0.04$ & $t \leftrightarrow s$ & Only indirect evidence \\
$\sim 0.01$ & $t \leftrightarrow d$ & Only indirect evidence \\ \hline
\end{tabular}
\end{center}
\end{table}

The quark masses quoted in Table \ref{tab:ql} are those which emerge when
quarks are probed at distances short compared with 1 fm, the characteristic
size of strongly interacting particles and the scale at which QCD becomes too
strong to utilize perturbation theory.  When regarded as constituents of
strongly interacting particles, however, the $u$ and $d$ quarks act as
quasi-particles with masses of about 0.3 GeV.  The corresponding
``constituent-quark'' masses of $s$, $c$, and $b$ are about 0.5, 1.5, and
4.9 GeV, respectively.

\subsection{Color and quantum chromodynamics}

The quarks are distinguished from the leptons by possessing a three-fold
charge known as ``color'' which enables them to interact strongly with one
another.  (A gauged color symmetry was first proposed by Nambu 1966.)
We shall also speak of quark and lepton ``flavor'' when
distinguishing the particles in Table \ref{tab:ql} from one another.  The
experimental evidence for color comes from several quarters.

{\em 1.  Quark statistics.}  One of the lowest-lying hadrons is a particle
known as the $\Delta^{++}$, an excited state of the nucleon first produced in
$\pi^+ p$ collisions in the mid-1950s at the University of Chicago cyclotron.
It can be represented in the quark model as $uuu$, so it is totally symmetric
in flavor.  It has spin $J= 3/2$, which is a totally symmetric combination of
the three quark spins (each taken to be 1/2).  Moreover, as a ground state, it
is expected to contain no relative orbital angular momenta among the quarks.

This leads to a paradox if there are no additional degrees of freedom.  A
state composed of fermions should be totally antisymmetric under the
interchange of any two fermions, but what we have described so far is totally
symmetric under flavor, spin, and space interchanges, hence totally symmetric
under their product.  Color introduces an additional degree of freedom
under which the interchange of two quarks can produce a minus sign, through
the representation $\Delta^{++} \sim \epsilon_{abc} u^a u^b u^c$.  The
totally antisymmetric product of three color triplets is a color singlet.

\begin{figure}
\centerline{\includegraphics[height=4in]{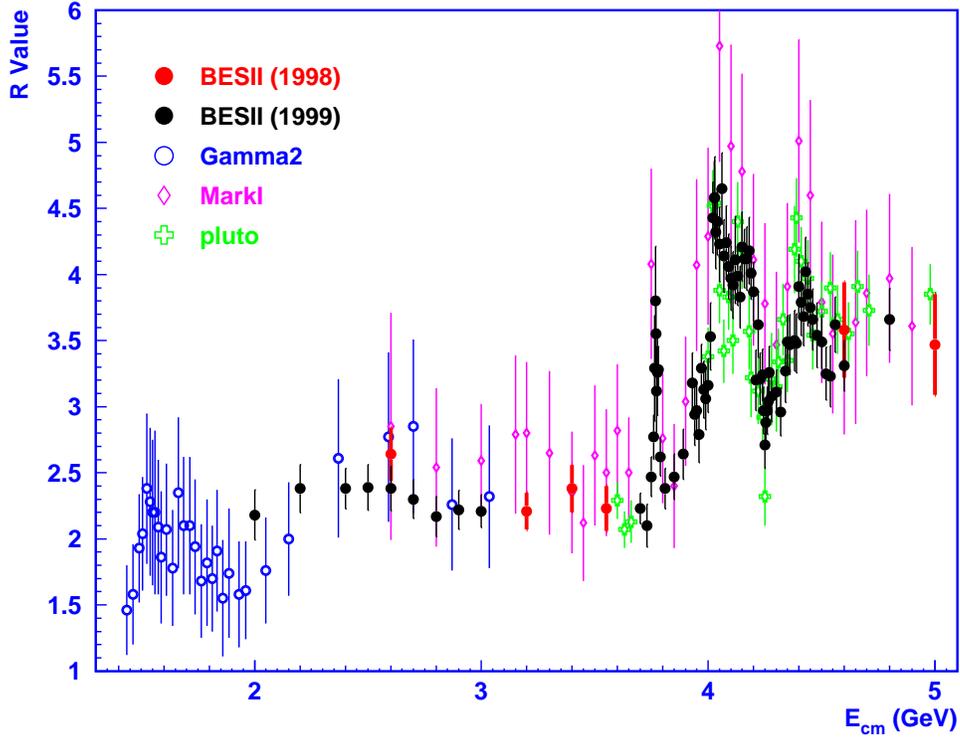}}
\caption{Values of $R$ measured by the BES Collaboration.
\label{fig:bes}}
\end{figure}

{\em 2.  Electron-positron annihilation to hadrons.}  The charges of all
quarks which can be produced in pairs below a given center-of-mass energy
is measured by the ratio
\beq
R \equiv \frac{\sigma(e^+ e^- \to {\rm hadrons})}{\sigma(e^+ e^- \to
\mu^+ \mu^-)} = \sum_i Q_i^2~~~.
\eeq
For energies at which only $u \bar u$, $d \bar d$, and $s \bar s$ can be
produced, i.e., below the charmed-pair threshold of about 3.7 GeV, one expects
\beq
R = N_c \left[ \left( \frac{2}{3} \right)^2 + \left( \frac{-1}{3} \right)^2
+ \left( \frac{-1}{3} \right)^2 \right] = \frac{2}{3} N_c
\eeq
for $N_c$ ``colors'' of quarks.  Measurements first performed at the Frascati
laboratory in Italy and most recently at the Beijing Electron-Positron
Collider (Bai \ite~2001; see Fig.\ \ref{fig:bes}) indicate $R = 2$ in
this energy range (with a small positive correction associated with
the strong interactions of the quarks), indicating $N_c = 3$.

{\em 3.  Neutral pion decay.}  The $\pi^0$ decay rate is governed by a quark
loop diagram in which two photons are radiated by the quarks in $\pi^0 =
(u \bar u - d \bar d)/\sqrt{2}$.  The predicted rate is
\beq \label{eqn:piz}
\Gamma(\pi^0 \to \gamma \gamma) = \frac{S^2 m_\pi^3}{8 \pi f_\pi^2}
\left( \frac{\alpha}{2 \pi} \right)^2~~~,
\eeq
where $f_\pi = 131$ MeV and $S = N_c(Q_u^2 - Q_d^2) = N_c/3$.  The
experimental rate is $7.8 \pm 0.6$ eV, while Eq.~(\ref{eqn:piz})
gives $7.6 S^2$ eV, in accord with experiment if $S=1$ and $N_c = 3$.

{\em 4.  Triality.}  Quark composites appear only in multiples of three.
Baryons are composed of $qqq$, while mesons are $q \bar q$ (with total
quark number zero).  This is compatible with our current understanding
of QCD, in which only color-singlet states can appear in the spectrum.  Thus,
mesons $M$ and baryons $B$ are represented by
\beq
M = \frac{1}{\st}(q^a \bar q'_a)~~,~~~
B = \frac{1}{\sx}(\epsilon_{abc} q^a {q'}^b {q''}^c)~~~.
\eeq

Direct evidence for the quanta of QCD, the gluons, was first presented in
1979 on the basis of extra ``jets'' of particles produced in electron-positron
annihilations to hadrons.  Normally one sees two clusters of energy associated
with the fragmentation of each quark in $e^+ e^- \to q \bar q$ into hadrons.
However, in some fraction of events an extra jet was seen, corresponding to
the radiation of a gluon by one of the quarks.

The transformations which take one color of quark into another are those of
the group SU(3).  We shall often refer to this group as SU(3)$_{\rm color}$
to distinguish it from the SU(3)$_{\rm flavor}$ associated with the
quarks $u$, $d$, and $s$.

\subsection{Electroweak unification}

The electromagnetic interaction is described in terms of photon exchange,
for which the Born approximation leads to a matrix element behaving as
$1/q^2$.  Here $q$ is the four-momentum transfer, and $q^2$ is its invariant
square.  The quantum electrodynamics of photons and charged pointlike
particles (such as electrons) initially encountered calculational problems in
the form of divergent quantities, but these had been tamed by the late 1940s
through the procedure known as {\it renormalization},
leading to successful estimates of such quantities as the anomalous magnetic
moment of the electron and the Lamb shift in hydrogen.

By contrast, the weak interactions as formulated up to the mid-1960s involved
the pointlike interactions of two currents, with an interaction Hamiltonian
${\cal H}_W = G_F J_\mu J^{\mu \dag}/\s$, with $G_F = 1.16637(1) \times
10^{-5}$ GeV$^{-2}$ the current value for the {\it Fermi coupling constant}.
This interaction is very singular and cannot be renormalized.  The weak
currents $J_\mu$ in this theory were purely charge-changing.  As a result of
work by Lee and Yang, Feynman and Gell-Mann, and Marshak and Sudarshan in
1956--7 they were identified as having (vector)--(axial) or ``$V-A$'' form.

Hideki Yukawa (1935) and Oskar Klein (1938) proposed a boson-exchange
model for the charge-changing weak interactions.  Klein's model attempted a
unification with electromagnetism and was based on a local isotopic gauge
symmetry, thus anticipating the theory of Yang and Mills (1954).  Julian
Schwinger and others studied such models in the 1950s, but Glashow (1961) was
the first to realize that a new {\it neutral} heavy boson had to be introduced
as well in order to successfully unify the weak and electromagnetic 
interactions.  The breaking of the electroweak symmetry (Weinberg 1967, Salam
1968) via the Higgs (1964) mechanism converted this
phenomenological theory into one which could be used for higher-order
calculations, as was shown by 't Hooft and Veltman in the early 1970s.

The boson-exchange model for charge-changing interactions replaces the
Fermi interaction constant with a coupling constant $g$ at each vertex and
the low-$q^2$ limit of a propagator, $1/(M_W^2 - q^2) \to 1/M_W^2$, with
factors of 2 chosen so that $G_F/\s = g^2/8 M_W^2$.  The $q^2$ term in
the propagator helps the theory to be more convergent, but it is not the
only ingredient needed, as we shall see.

The normalization of the charge-changing weak currents $J_\mu$ suggested
well in advance of electroweak unification that one regard the corresponding
integrals of their time components (the so-called {\it weak charges}) as
members of an SU(2) algebra (Gell-Mann and L\'evy 1960, Cabibbo 1963).
However, the identification of
the neutral member of this multiplet as the electric charge was problematic.
In the $V-A$ theory the $W$'s couple only to left-handed fermions $\psi_L
\equiv (1 - \gamma_5)\psi/2$, while the photon couples to $\psi_L + \psi_R$,
where $\psi_R \equiv (1 + \gamma_5) \psi/2$.  Furthermore, the high-energy
behavior of the $\nu \bar \nu \to W^+ W^-$ scattering amplitude based on
charged lepton exchange leads to unacceptable divergences if we
incorporate it into the one-loop contribution to $\nu \bar \nu \to \nu \bar
\nu$ (Quigg 1983).

A simple solution was to add a neutral boson $Z$ coupling to $W^+
W^-$ and $\nu \bar \nu$ in such a way as to cancel the leading high-energy
behavior of the charged-lepton-exchange diagram.  This relation between
couplings occurs naturally in a theory based on the gauge group \ew.
The $Z$ leads to {\it neutral current interactions}, in which (for
example) an incident neutrino scatters inelastically on a hadronic target
without changing its charge.  The discovery of neutral-current interactions
of neutrinos and many other manifestations of the $Z$ proved to be striking
confirmations of the new theory. 

If one identifies the $W^+$ and $W^-$ with raising and lowering operations in
an SU(2), so that $W^\pm = (W^1 \mp i W^2)\s$, then left-handed fermions
may be assigned to doublets of this ``weak isospin,'' with $I_{3L} (u,c,t)
= I_{3L} (\nu_e,\nu_\mu,\nu_\tau) = +1/2$, $I_{3L} (d,s,b) = I_{3L} (e^-,
\mu^-,\tau^-) = -1/2$.  All the right-handed fermions have $I_L = I_{3L} = 0$.
As mentioned, one cannot simply identify the photon with $W^3$, which also
couples only to left-handed fermions.  Instead, one must introduce another
boson $B$ associated with a U(1) gauge group.  It will mix with the $W^3$
to form physical states consisting of the massless photon $A$ and the massive
neutral boson $Z$:

\beq
A = B \cos \theta + W^3 \sin \theta~~,~~~
Z = - B \sin \theta + W^3 \cos \theta~~~.
\eeq

The mixing angle $\theta$ appears in many electroweak processes.  It has been
measured to sufficiently great precision that one must specify the
renormalization scheme in which it is quoted.  For present purposes we shall
merely note that $\sin^2 \theta \simeq 0.23$.  The corresponding SU(2) and U(1)
coupling constants $g$ and $g'$ are related to the electric charge $e$ by
$e = g \sin \theta = g' \cos \theta$, so that 
\beq
\frac{1}{e^2} = \frac{1}{g^2} + \frac{1}{{g'}^2}~~~.
\eeq
The electroweak theory successfully predicted the masses of the $W^\pm$ and
$Z$:
\beq \label{eqn:MWZ}
M_W \simeq 38.6 \g/\sin \theta \simeq 80.5 \g~~,~~~
M_Z \simeq M_W/\cos \theta \simeq 91.2 \g~~~,
\eeq
where we show the approximate {\it experimental} values.  The detailed check
of these predictions has reached the precision that one can begin to look into
the deeper structure of the theory.  A key ingredient in this structure is
the {\it Higgs boson}, the price that had to be paid for the breaking of the
electroweak symmetry.

\subsection{Higgs boson}

An unbroken \ew~theory involving the photon would require
{\it all} fields to have zero mass, whereas the $W^\pm$ and $Z$ are massive.
The symmetry-breaking which generates $W$ and $Z$ masses must not destroy the
renormalizability of the theory.  However, a massive vector boson propagator is
of the form
\beq
D_{\mu \nu}(q) = \frac{-g_{\mu \nu} + q_\mu q_\nu/M^2}{q^2 - M^2}
\eeq
where $M$ is the boson mass.  The terms $q_\mu q_\nu$, when appearing in loop
diagrams, will destroy the renormalizability of the theory.  They are
associated with longitudinal vector boson polarizations, which are only
present for massive bosons.  For massless bosons like the photon,
there are only transverse polarization states $J_z = \pm J$.

The {\it Higgs mechanism}, to be discussed in detail later in these lectures,
provides the degrees of freedom needed to add a longitudinal polarization
state for each of $W^+$, $W^-$, and $W^0$.  In the simplest model, this is
achieved by introducing a doublet of complex Higgs fields:
\beq
\phi = \left[ \begin{array}{c} \phi^+ \\ \phi^0 \end{array} \right]~~,~~~
\phi^* = \left[ \begin{array}{c} \bar \phi^0 \\ \phi^- \end{array} \right]~~~.
\eeq
Here the charged Higgs fields $\phi^\pm$ provide the longitudinal component
of $W^\pm$ and the linear combination $(\phi^0 - \bar \phi^0)/i \s$ provides
the longitudinal component of the $Z$.  The additional degree of freedom
$(\phi^0 + \bar \phi^0)/\s$ corresponds to a physical particle, the
{\it Higgs particle}, which is the subject of intense searches.

Discovering the nature of the Higgs boson is a key to further progress in
understanding what may lie beyond the Standard Model.  There may exist one
Higgs boson or more than one.  There may exist other particles in the spectrum
related to it.  The Higgs boson may be elementary or composite.  If composite,
it points to a new level of substructure of the elementary particles.
Much of our discussion will lead up to strategies for the next few years
designed to address these questions.  First, we introduce the necessary topic
of {\it gauge theories}, which have been the platform for all the developments
of the past thirty years.

\section{Gauge theories}

\subsection{Abelian gauge theories}

The Lagrangian describing a free fermion of mass $m$ is $\cL_{\rm free} =
\pb(i \dst - m) \psi$.  It is invariant under the global phase
change $\psi \to \exp(i \alpha) \psi$.  (We shall always consider the fermion
fields to depend on $x$.) Now consider independent phase changes at each point:
\beq \label{eqn:lph}
\psi \to \psi' \equiv \exp[i \alpha(x)] \psi.
\eeq
Because of the derivative, the Lagrangian then acquires an additional phase
change at each point:  $\delta \cL_{\rm free} = \pb i \gamma^\mu [i
\dm \alpha(x)] \psi$.  The free Lagrangian is not invariant
under such changes of phase, known as {\it local gauge transformations}.

Local gauge invariance can be restored if we make the replacement
$\dm \to D_\mu \equiv \dm + i e A_\mu$ in the free-fermion
Lagrangian, which now is
\beq
\cL = \pb (i \Ds - m) \psi = \pb (i \ds - m) \psi - e \pb \As (x) \psi~~~.
\eeq
The effect of a local phase in $\psi$ can be compensated if we allow the
{\it vector potential} $A_\mu$ to change by a total divergence, which does not
change the electromagnetic field strength (defined as in Peskin and Schroeder
1995; Quigg 1983 uses the opposite sign)
\beq
F_{\mu \nu} \equiv \dm A_\nu - \dn A_\nu~~~.
\eeq
Indeed, under the transformation $\psi \to \psi'$ and with $A \to A'$ with
$A'$ yet to be determined, we have
\beq
\cL ' = \pb ' (i \ds - m) \psi' - e \pb ' {\As}' \psi'
= \pb (i \ds - m) \psi - \pb [\ds \alpha(x)] \psi - e \pb {\As}' \psi~~~.
\eeq
This will be the same as $\cL$ if
\beq \label{eqn:gta}
A'_\mu(x) = A_\mu(x) - \frac{1}{e} \dm \alpha(x)~~~.
\eeq
The derivative $D_\mu$ is known as the {\it covariant derivative}.  One can 
check that under a local gauge transformation, $D_\mu \psi \to e^{i \alpha(x)}
D_\mu \psi$. 

Another way to see the consequences of local gauge invariance suggested by
Yang (1974) and discussed by Peskin and Schroeder (1995, pp 482--486)
is to define $-e A_\mu(x)$ as the local change in phase undergone by a particle
of charge $e$ as it passes along an infinitesimal space-time increment between
$x^\mu$ and $x^\mu + dx^\mu$.  For a space-time trip from point $A$ to point
$B$, the phase change is then
\beq
\Phi_{AB} = \exp \left( - i e \int_A ^B A_\mu(x) dx^\mu \right)~~~.
\eeq
The phase in general will depend on the path in space-time taken from point $A$
to point $B$.  As a consequence, the phase $\Phi_{AB}$ is not uniquely defined.
However, one can compare the result of a space-time trip along one path,
leading to a phase $\Phi_{AB}^{(1)}$, with that along another, leading to
a phase $\Phi_{AB}^{(2)}$.  The two-slit experiment in quantum mechanics
involves such a comparison; so does the Bohm-Aharonov effect in which a
particle beam traveling past a solenoid on one side interferes with a
beam traveling on the other side.  Thus, phase differences
\beq \label{eqn:cp}
\Phi_{AB}^{(1)} \Phi_{AB}^{(2)*} = \Phi_C = \exp \left(- i e \oint A_\mu(x)
dx^\mu \right)~~~,
\eeq
associated with {\it closed paths} in space-time (represented by the
circle around the integral sign), are the ones which correspond to physical
experiments.
The phase $\Phi_C$ for a closed path $C$ is independent of the phase convention
for a charged particle at any space-time point $x_0$, since any change in the
contribution to $\Phi_C$ from the integral up to $x_0$ will be compensated by
an equal and opposite contribution from the integral departing from $x_0$.

The closed path integral (\ref{eqn:cp}) can be expressed as a surface integral
using Stokes' theorem:
\beq
\oint A_\mu(x) dx^\mu = \int F_{\mu \nu}(x) d \sigma^{\mu \nu}~~~,
\eeq
where the electromagnetic field strength $F_{\mu \nu}$ was defined previously
and $d \sigma^{\mu \nu}$ is an element of surface area.  It is also clear
that the closed path integral is invariant under changes (\ref{eqn:gta})
of $A_\mu(x)$ by a total divergence.  Thus $F_{\mu \nu}$ suffices to
describe all physical experiments as long as one integrates over a suitable
domain.  In the Bohm-Aharonov effect, in which
a charged particle passes on either side of a solenoid, the surface
integral will include the solenoid (in which the magnetic field is non-zero).

If one wishes to describe the energy and momentum of free electromagnetic
fields, one must include a kinetic term ${\cL}_K = -(1/4)F_{\mu \nu}
F^{\mu \nu}$ in the Lagrangian, which now reads

\beq
\cL = - \frac{1}{4} F_{\mu \nu} F^{\mu \nu} + \pb(i \ds - m) \psi
- e \pb \As \psi~~~.
\eeq
If the electromagnetic current is defined as $J^{\rm em}_\mu \equiv \pb
\gamma_\mu \psi$, this Lagrangian leads to Maxwell's equations.

The local phase changes (\ref{eqn:lph}) form a U(1) group of transformations.
Since such transformations commute with one another, the group is said to
be {\it Abelian}.  Electrodynamics, just constructed here, is an example of
an {\it Abelian gauge theory.}
 
\subsection{Non-Abelian gauge theories}

One can imagine that a particle traveling in space-time undergoes not only
phase changes, but also changes of identity.  Such transformations were first
considered by Yang and Mills (1954).  For example, a quark can change in color
(red to blue) or flavor ($u$ to $d$).  In that case we replace the
coefficient $e A_\mu$ of the infinitesimal displacement $dx_\mu$ by an
$n \times n$ matrix $-g \bam(x) \equiv - g A^i_\mu(x) \bT_i$ acting in
the $n$-dimensional space of the particle's degrees of freedom.  (The sign
change follows the convention of Peskin and Schroeder 1995.)  For colors,
$n=3$.  The $\bT_i$ form a linearly independent basis set of matrices for
such transformations, while the $A^i_\mu$ are their coefficients.  The phase
transformation then must take account of the fact that the matrices
$\bam(x)$ in general do not commute with one another for different
space-time points, so that a {\it path-ordering} is needed:
\beq \label{eqn:nona}
\Phi_{AB} = {\cal P} \left[ \exp \left( i g \int_A^B \bam(x) dx^\mu
\right) \right]~~~.
\eeq
When the basis matrices $\bT_i$ do not commute with one another, the
theory is {\it non-Abelian}.

We demand that changes in phase or identity conserve probability, i.e., that
$\Phi_{AB}$ be {\it unitary}: $\Phi^\dag_{AB} \Phi_{AB} = 1$.  When $\Phi_{AB}$
is a matrix, the corresponding matrices $\bam(x)$ in (\ref{eqn:nona}) 
must be Hermitian.  If we wish to separate out pure phase changes, in which
$\bam(x)$ is a multiple of the unit matrix, from the remaining
transformations, one may consider only transformations such that
det$(\Phi_{AB}) =1$, corresponding to {\it traceless} $\bam(x)$.

The $n \times n$ basis matrices $\bT_i$ must then be Hermitian and
traceless.  There will be $n^2 - 1$ of them, corresponding to the number of
independent SU(N) generators.  (One can generalize this approach to other
invariance groups.)  The matrices will satisfy the commutation relations
\beq \label{eqn:str}
[\bT_i, \bT_j] = i c_{ijk} \bT_k~~~,
\eeq
where the $c_{ijk}$ are {\it structure constants} characterizing the group.
For SU(2), $c_{ijk} = \epsilon_{ijk}$ (the Kronecker symbol), while for
SU(3), $c_{ijk} = f_{ijk}$, where the $f_{ijk}$ are defined in Gell-Mann
and Ne'eman (1964).
A $3 \times 3$ representation in SU(3) is $\bT_i = \lambda_i/2$, where
$\lambda_i/2$ are the Gell-Mann matrices normalized such that Tr $\lambda_i
\lambda_j = 2 \delta_{ij}$.  For this representation, then, Tr $\bT_i
\bT_j = \delta_{ij}/2$.

In order to define the field-strength tensor $\bFmn = F^i_{\mu \nu}
\bT_i$ for a non-Abelian transformation, we may consider an infinitesimal
closed-path transformation analogous to Eq.\ (\ref{eqn:cp}) for the case in
which the matrices $\bam(x)$ do not commute with one another.  The
result (see, e.g., Peskin and Schroeder 1995, pp 486--491) is
\beq
\bFmn = \dm \ban - \dn \ban - i g [\bam,\ban]~~,~~~
F^i_{\mu \nu} = \dm A^i_\nu - \dn A^i_\mu + g c_{ijk}A^j_\mu A^k_\nu~~~.
\eeq

An alternative way to introduce non-Abelian gauge fields is to demand that,
by analogy with Eq.\ (\ref{eqn:lph}), a theory involving fermions $\psi$ be
invariant under local transformations
\beq \label{eqn:lgt}
\psi(x) \to \psi'(x) = U(x) \psi(x)~~,~~~U^\dag U = 1~~~,
\eeq
where for simplicity we consider unitary transformations.  Under this
replacement, $\cL \to \cL '$, where
$$
\cL ' \equiv \pb ' (i \ds - m) \psi' = \pb U^{-1}(i \ds - m) U \psi
$$
\beq
= \pb (i \ds - m) \psi + i \psi U^{-1} \gamma^\mu (\dm U) \psi~~~.
\eeq
As in the Abelian case, an extra term is generated by the local transformation.
It can be compensated by replacing $\dm$ by
\beq
\dm \to \bD_\mu \equiv \dm - i g \bam(x)~~~.
\eeq
In this case $\cL = \pb (i \bDs - m) \psi$ and under the change (\ref{eqn:lgt})
we find
$$
\cL ' \equiv \pb ' (i \bDs ' - m) \psi' = \pb U^{-1}(i \ds + g \bAs ' - m) U
 \psi
$$
\beq
=  \cL + \pb [ g(U^{-1} \bAs ' U - \bAs) + i U^{-1} (\ds U) ] \psi ~~~.
\eeq
This is equal to $\cL$ if we take
\beq \label{eqn:gtn}
\bam ' = U \bam U^{-1} - \frac{i}{g}(\dm U) U^{-1}~~~.
\eeq
This reduces to our previous expressions if $g = -e$ and $U =
e^{i \alpha(x)}$.

The covariant derivative acting on $\psi$ transforms in the same way as
$\psi$ itself under a gauge transformation: $\bDm \psi \to \bDm ' \psi' = U
\bDm \psi$.  The field strength $\bFmn$ transforms as
$\bFmn \to \bFmn ' = U \bFmn U^{-1}$.
It may be computed via $[\bDm,\bDn] = - i g \bFmn$; both sides
transform as $U(~~)U^{-1}$ under a local gauge transformation.

In order to obtain propagating gauge fields, as in electrodynamics, one must
add a kinetic term ${\cL}_K = -(1/4) F^i_{\mu \nu} F^{i \mu \nu}$ to the
Lagrangian.  Recalling the representation $\bFmn = F^i_{\mu \nu}$ in terms of
gauge group generators normalized such that Tr$(\bT_i \bT_j) = \delta_{ij}/2$,
we can write the full Yang-Mills Lagrangian for gauge fields interacting with
matter fields as
\beq
\cL = - \frac{1}{2} {\rm Tr}(\bFmn {\bf F}^{\mu \nu}) + \pb (i \bDs - m)
\psi~~~.
\eeq
We shall use Lagrangians of this type to derive the strong, weak, and
electromagnetic interactions of the ``Standard Model.''

The interaction of a gauge field with fermions then corresponds to a term in
the interaction Lagrangian $\Delta \cL = g \pb(x) \gamma^\mu \bam(x) \psi(x)$.
The $[\bam,\ban]$ term in $\bFmn$ leads to self-interactions of
non-Abelian gauge fields, arising solely from the kinetic term.  Thus, one has
three- and four-field vertices arising from
\beq
\Delta {\cL}_K^{(3)} = (\dm A^i_\nu) g c_{ijk} A^{\mu j} A^{\nu k}~~,~~~
\Delta {\cL}_K^{(4)} = -\frac{g^2}{4}c_{ijk} c_{imn} A^{\mu j} A^{\nu k}
A^m_\mu A^n_\nu~~~.
\eeq
These self-interactions are an important aspect of non-Abelian gauge theories
and are responsible in particular for the remarkable {\it asymptotic freedom}
of QCD which leads to its becoming weaker at short distances, permitting the
application of perturbation theory.

\subsection{Elementary divergent quantities}

In most quantum field theories, including quantum electrodynamics, divergences
occurring in higher orders of perturbation theory must be removed using
charge, mass, and wave function renormalization.  This is conventionally done
at intermediate calculational stages by introducing a cutoff momentum scale
$\Lambda$ or analytically continuing the number of space-time dimensions
away from four.  Thus, a vacuum polarization graph in QED associated with
external photon momentum $k$ and a fermion loop will involve an integral
\beq \label{eqn:vp}
\Pmn(k) \sim \int \frac{d^4 p}{(2 \pi)^4} {\rm Tr} \left( \frac{1}{\ps -m}
\gamma_\mu \frac{1}{\ps + \ks - m} \gamma_\nu \right)~~~;
\eeq
a self-energy of a fermion with external momentum $p$ will involve
\beq \label{eqn:se}
\Sigma(p) \sim \int \frac{d^4 q}{(2 \pi)^4} \frac{1}{q^2} \gamma_\mu
\frac{1}{\ps + \qs - m} \gamma^\mu~~~,
\eeq
and a fermion-photon vertex function with external fermion momenta $p,p'$
will involve
\beq \label{eqn:vtx}
\Lambda_\mu(p',p) \sim \int \frac{d^4 k}{(2 \pi)^4} \frac{1}{k^2} \gamma_\nu
\frac{1}{\ps ' + \ks - m} \gamma_\mu \frac{1}{\ps + \ks - m} \gamma^\nu~~~.
\eeq
The integral (\ref{eqn:vp}) appears to be quadratically divergent.  However,
the gauge invariance of the theory translates into the requirement $k^\mu
\Pmn = 0$, which requires $\Pmn$ to have the form
\beq
\Pmn(k) = (k^2 \gmn - k_\mu k_\nu) \Pi(k^2)~~~.
\eeq
The corresponding integral for $\Pi(k^2)$ then will be only logarithmically
divergent.  The integral in (\ref{eqn:se}) is superficially linearly divergent
but in fact its divergence is only logarithmic, as is the integral in
(\ref{eqn:vtx}).

Unrenormalized functions describing vertices and self-energies involving $n_B$
external boson lines and $n_F$ external fermion lines may be defined in terms
of a momentum cutoff $\Lambda$ and a bare coupling constant $g_0$
(Coleman 1971, Ellis 1977, Ross 1978):
\beq
\Gamma^{U}_{n_B,n_F} \equiv \Gamma^{U}_{n_B,n_F}(p_i,g_0,\Lambda)~~,
\eeq
where $p_i$ denote external momenta.  {\it Renormalized} functions $\Gamma^R$
may be defined in terms of a scale parameter $\mu$, a renormalized coupling
constant $g = g(g_0,\Lambda/\mu)$, and renormalization constants
$Z_B(\Lambda)$ and $Z_F(\Lambda)$ for the external boson and fermion wave
functions:
\beq \label{eqn:GamR}
\Gamma^R(p_i,g,\mu) \equiv \lim_{\Lambda \to \infty} [Z_B(\Lambda)]^{n_B}
[Z_F(\Lambda)]^{n_F} \Gamma^{U}_{n_B,n_F}(p_i,g_0,\Lambda)~~~.
\eeq
The scale $\mu$ is typically utilized by demanding that $\Gamma^R$ be equal
to some predetermined function at a Euclidean momentum $p^2 = - \mu^2$.
Thus, for the one-boson, two-fermion vertex, we take
\beq \label{eqn:coup}
\Gamma^R_{1,2}(0,p,-p)|_{p^2 = - \mu^2} = \lim_{\Lambda \to \infty}
Z_F^2 Z_B \Gamma^U_{1,2}(0,p,-p)|_{p^2 = - \mu^2} \equiv g~~~.
\eeq

The unrenormalized function $\Gamma^U$ is {\it independent} of $\mu$,
while $\Gamma^R$ and the renormalization constants $Z_B(\Lambda),~
Z_F(\Lambda)$ will depend on $\mu$.  For example, in QED, the photon
wave function renormalization constant (known as $Z_3$) behaves as
\beq \label{eqn:Z3}
Z_3 = 1 - \frac{\alpha_0}{3 \pi} \ln \frac{\Lambda^2}{\mu^2}~~~.
\eeq
The bare charge $e_0$ and renormalized charge $e$ are related by $e = e_0
Z_3^{1/2}$.  To lowest order in perturbation theory, $e < e_0$.  The vacuum
behaves as a normal dielectric; charge is screened.  It is the exception rather
than the rule that in QED one can define the renormalized charge for $q^2 = 0$;
in QCD we shall see that this is not possible.

\subsection{Scale changes and the beta function}

We differentiate both sides of (\ref{eqn:GamR}) with respect to $\mu$
and multiply by $\mu$. Since the functions $\Gamma^U$ are independent of $\mu$, 
we find
$$
\left( \mu \frac{\partial}{\partial \mu} + \mu \frac{\partial g}{\partial \mu}
\frac{\partial}{\partial g} \right) \Gamma^R(p_i,g,\mu)
$$
\beq
= \lim_{\Lambda \to \infty} \left( \frac{n_B}{Z_B} \mu \frac{\partial Z_B}
{\partial \mu} + \frac{n_F}{Z_F} \mu \frac{\partial Z_F}{\partial \mu}
\right) Z_B^{n_B} Z_F^{n_F} \Gamma^U~~~,
\eeq
or
\beq
\left[ \mu \frac{\partial}{\partial \mu} + \beta(g) \frac{\partial}{\partial g}
+ n_B \gamma_B(g) + n_F \gamma_F(g) \right] \Gamma^R(p_i,g,\mu) = 0~~~,
\eeq
where
\beq \label{eqn:uni}
\beta(g) \equiv \mu \frac{\partial g}{\partial \mu}~~,~~~
\gamma_B(g) \equiv - \frac{\mu}{Z_B} \frac{\partial Z_B}{\partial \mu}~~,~~~
\gamma_F(g) \equiv - \frac{\mu}{Z_F} \frac{\partial Z_F}{\partial \mu}~~~.
\eeq
The behavior of any generalized vertex function $\Gamma^R$ under a change of
scale $\mu$ is then governed by the universal functions (\ref{eqn:uni}).

Here we shall be particularly concerned with the function $\beta(g)$.  Let us
imagine $\mu \to \lambda \mu$ and introduce the variables
$t \equiv \ln \lambda$, $\bar g(g,t) \equiv g(g_0, \Lambda/\lambda \mu)$,
Then the relation for the beta-function may be written
\beq
\frac{d \bar g(g,t)}{d t} = \beta(\bg)~~,~~~\bar g(g,0) = g(g_0,\Lambda/\mu)
= g~~~.  
\eeq
Let us compare the behavior of $\bar g$ with increasing $t$ (larger momentum
scales or shorter distance scales) depending on the sign of $\beta(\bg)$.
In general we will find $\beta(0) = 0$.  We take $\beta(\bg)$ to have
zeroes at $\bg = 0,~g_1,~g_2,~\ldots$.  Then:

\begin{enumerate}

\item Suppose $\beta(\bg) > 0$.  Then $\bg$ {\it increases} from its $t = 0$
value $\bg = g$ until a zero $g_i$ of $\beta(\bg)$ is encountered.  Then
$\bg \to g_i$ as $t \to \infty$.

\item Suppose $\beta(\bg) < 0$.  Then $\bg$ {\it decreases} from its $t = 0$
value $\bg = g$ until a zero $g_i$ of $\beta(\bg)$ is encountered.

\end{enumerate}

In either case $\bg$ approaches a point at which $\beta(\bg) = 0$, $\beta'(\bg)
< 0$ as $t \to \infty$.  Such points are called {\it ultraviolet fixed
points}.  Similarly, points for which $\beta(\bg) = 0$, $\beta'(\bg) > 0$ are
{\it infrared fixed points}, and $\bg$ will tend to them for $t \to
- \infty$ (small momenta or large distances).  The point $e = 0$ is an
infrared fixed point for quantum electrodynamics, since $\beta'(e) > 0$
at $e=0$.

It may happen that $\beta'(0) < 0$ for specific theories.  In that case
$\bg = 0$ is an ultraviolet fixed point, and the theory is said to be
{\it asymptotically free}.  We shall see that this property is particular to
non-Abelian gauge theories (Gross and Wilczek 1973, Politzer 1974).

\subsection{Beta function calculation}

In quantum electrodynamics a loop diagram involving a fermion of unit charge 
contributes the following expression to the relation between the bare charge
$e_0$ and the renormalized charge $e$:
\beq
e = e_0 \left( 1 - \frac{\alpha_0}{3 \pi} \ln \frac{\Lambda}{\mu} \right)~~~,
\eeq
as implied by (\ref{eqn:coup}) and (\ref{eqn:Z3}), where $\alpha_0 \equiv
e_0^2/4 \pi$.  We find
\beq \label{eqn:betaf}
\beta(e) = \frac{e_0^3}{12 \pi^2} \simeq \frac{e^3}{12 \pi^2}~~~,
\eeq
where differences between $e_0$ and $e$ correspond to higher-order terms in
$e$.  (Here $\alpha \equiv e^2/4 \pi$.)  Thus $\beta(e) > 0$ for small
$e$ and the coupling constant becomes stronger at larger momentum scales
(shorter distances).

We shall show an extremely simple way to calculate (\ref{eqn:betaf}) and the
corresponding result for a charged scalar particle in a loop.  From this we
shall be able to first calculate the effect of a charged vector particle in a
loop (a calculation first performed by Khriplovich 1969) and then generalize
the result to Yang-Mills fields.  The method follows that of Hughes (1980).

When one takes account of vacuum polarization, the electromagnetic interaction
in momentum space may be written
\beq \label{eqn:chg}
\frac{e^2}{q^2} \to \frac{e^2}{q^2[1+\Pi(q^2)]}~~~
\eeq
Here the long-distance ($q^2 \to 0$) behavior has been defined such that $e$
is the charge measured at macroscopic distances, so $\Pi(0) = 0$.  Following
Sakurai (1967), we shall reconstruct $\Pi_i(q^2)$ for a loop involving the
fermion species $i$ from its imaginary part, which is measurable through the
cross section for $e^+ e^- \to i \bar{i}$:
\beq
{\rm Im}~\Pi_i(s) = \frac{s}{4 \pi \alpha} \sigma(e^+ e^- \to i \bar{i})~~~,
\eeq
where $s$ is the square of the center-of-mass energy.  For fermions $f$ of
charge $e_f$ and mass $m_f$,
\beq
{\rm Im}~\Pi_f(s) = \frac{\alpha e_f^2}{3} \left( 1 + \frac{2 m_f^2}{s} \right)
\left( 1 - \frac{4 m_f^2}{s} \right)^{1/2} \theta(s - 4 m_f^2)~~~,
\eeq
while for scalar particles of charge $e_s$ and mass $m_s$,
\beq
{\rm Im}~\Pi_s(s) = \frac{\alpha e_s^2}{12}\left( 1 - \frac{4 m_s^2}{s} \right)
^{3/2} \theta(s - 4 m_s^2)~~~.
\eeq

The corresponding cross section for $e^+ e^- \to \mu^+ \mu^-$, neglecting the
muon mass, is $\sigma(e^+ e^- \to \mu^+ \mu^-) = 4 \pi \alpha^2/3s$,
so one can define
\beq
R_i \equiv \sigma(e^+ e^- \to i \bar{i})/\sigma(e^+ e^- \to \mu^+ \mu^-)~~~,
\eeq
in terms of which Im $\Pi_i(s) = \alpha R_i(s)/3$.  For $s \to \infty$ one has
$R_f(s) \to e_f^2$ for a fermion and $R_s(s) \to e_s^2/4$ for a scalar.
 
The full vacuum polarization function $\Pi_i(s)$ cannot directly be
reconstructed in terms of its imaginary part via the dispersion relation
\beq
\Pi_i(s) = \frac{1}{\pi}\int_{4m^2}^\infty \frac{ds'}{s'-s}{\rm Im}~
\Pi_i(s')~~~,
\eeq
since the integral is logarithmically divergent.  This divergence is exactly
that encountered earlier in the discussion of renormalization.  For
quantum electrodynamics we could deal with it by defining the charge at
$q^2 = 0$ and hence taking $\Pi_i(0) = 0$.  The {\it once-subtracted}
dispersion relation for $\Pi_i(s) - \Pi_i(0)$ would then converge:
\beq
\Pi_i(s) = \frac{1}{\pi}\int_{4m^2}^\infty \frac{ds'}{s'(s'-s)}{\rm Im}~
\Pi_i(s')~~~.
\eeq
However, in order to be able to consider cases such as Yang-Mills fields
in which the theory is not well-behaved at $q^2 = 0$, let us instead
define $\Pi_i(- \mu^2) = 0$ at some spacelike scale $q^2 = - \mu^2$.  The
dispersion relation is then
\beq
\Pi_i(s) = \frac{1}{\pi}\int_{4m^2}^\infty ds' \left[ \frac{1}{s'-s} -
\frac{1}{s' + \mu^2} \right] {\rm Im}~\Pi_i(s')~~~.
\eeq
For $|q^2| \gg \mu^2 \gg m^2$, we find
\beq
\Pi_i(q^2) \to - \frac{\alpha}{3 \pi} R_i(\infty) \left[ \ln \frac{-q^2}
{\mu^2} + {\rm const.} \right]~~~,
\eeq
and so, from (\ref{eqn:chg}), the ``charge at scale $q$'' may be written as
\beq
e_q^2 \equiv \frac{e^2}{1 + \Pi_i(q^2)} \simeq e^2 \left[ 1 + \frac{\alpha}
{3 \pi} R_i(\infty) \ln \frac{-q^2}{\mu^2} \right]~~~.
\eeq
The beta-function here is defined by $\beta(e) = \mu (\partial e/\partial \mu)
|_{{\rm fixed}~e_q}$.  Thus, expressing $\beta(e) = - \beta_0 e^3/(16 \pi^2)
+ {\cal O}(e^5)$, one finds $\beta_0 = -(4/3)e_f^2$ for spin-1/2 fermions and
$\beta_0 = -(1/3) e_s^2$ for scalars.

These results will now be used to find the value of $\beta_0$ for a single
charged massless vector field.
We generalize the results for spin 0 and 1/2 to higher spins by splitting
contributions to vacuum polarization into ``convective'' and ``magnetic''
ones.  Furthermore, we take into account the fact that a closed
fermion loop corresponds to an extra minus sign in $\Pi_f(s)$ (which is
already included in our result for spin 1/2).  The ``magnetic'' contribution
of a particle with spin projection $S_z$ must be proportional to $S_z^2$.
For a massless spin-$S$ particle, $S_z^2 = S^2$.  We may then write
\beq
\beta_0 = \left\{ \begin{array}{c} (-1)^{n_F}(aS^2 + b) (S = 0)~~~, \\
(-1)^{n_F}(aS^2 + 2b) (S \ne 0)~~~, \end{array} \right.
\eeq
where $n_F = 1$ for a fermion, 0 for a boson.  The factor of $2b$ for
$S \ne 0$ comes from the contribution of each polarization state $(S_z =
\pm S)$ to the convective term.  Matching the results for spins 0 and 1/2, 
\beq
- \frac{1}{3} = b~~~,~~- \frac{4}{3} = - \left( \frac{a}{4} + 2 b \right)~~~,
\eeq
we find $a = 8$ and hence for $S = 1$
\beq \label{eqn:khr}
\beta_0 = 8 - \frac{2}{3} = \frac{22}{3}~~~.
\eeq
The magnetic contribution is by far the dominant one (by a factor of 12), and
is of opposite sign to the convective one.  A similar separation of
contributions, though with different interpretations, was obtained in the
original calculation of Khriplovich (1969).  The reversal of sign with respect
to the scalar and spin-1/2 results is notable.

\subsection{Group-theoretic techniques}

The result (\ref{eqn:khr}) for a charged, massless vector field interacting
with the photon is also the value of $\beta_0$ for the Yang-Mills group
SO(3) $\sim$ SU(2) if we identify the photon with $A^3_\mu$ and the charged
vector particles with $A^{\pm}_\mu \equiv (A^1_\mu \mp i A^2_\mu)/\s$.
We now generalize it to the contribution of gauge fields in an arbitrary
group $G$.

The value of $\beta_{0~{\rm gauge~fields}}$ depends on a sum over all
possible self-interacting gauge fields that can contribute to the loop
with external gauge field labels $i$ and $m$:
\beq \label{eqn:rat}
\frac{\beta_0[G]}{\beta_0[SU(2)]} = \frac{c^G_{ijk} c^G_{mjk}}
{c^{\rm SU(2)}_{ijk} c^{\rm SU(2)}_{mjk}}~~~,
\eeq
where $c^G_{ijk}$ is the structure constant for $G$, introduced in Eq.\
(\ref{eqn:str}).  The sums in (\ref{eqn:rat}) are proportional to
$\delta_{im}$:
\beq \label{eqn:Cas}
c_{ijk} c_{mjk} = \delta_{im} C_2(A)~~~.
\eeq
The quantity $C_2(A)$ is the {\it quadratic Casimir operator} for the adjoint
representation of the group $G$.

Since the structure constants for SO(3) $\sim$ SU(2) are just
$c^{\rm SU(2)}_{ijk} = \epsilon_{ijk}$, one finds $C_2(A) = 2$ for SU(2),
so the generalization of (\ref{eqn:khr}) is that $\beta_{0~{\rm gauge~fields}}
= (11/3) C_2(A)$.

The contributions of arbitrary scalars and spin-1/2 fermions in
representations $R$ are proportional to $T(R)$, where
\beq \label{eqn:Tdef}
{\rm Tr}~ (T_i T_j) \equiv \delta_{ij} T(R)
\eeq
for matrices $T_i$ in the representation $R$.  For a
single charged scalar particle (e.g., a pion) or fermion (e.g., an electron),
$T(R) = 1$.  Thus $\beta_{0~{\rm spin~0}} = -(1/3)T_0(R)$, while
$\beta_{0~{\rm spin~1/2}} = -(4/3)T_{1/2}(R)$, where the subscript on $T(R)$
denotes the spin.  Summarizing the contributions of gauge bosons, spin 1/2
fermions, and scalars, we find
\beq
\beta_0 = \frac{11}{3} C_2(A) - \frac{4}{3} \sum_f T_{1/2}(R_f) - \sum_s
\frac{1}{3}T_0(R_s)~~~.
\eeq

One often needs the beta-function to higher orders, notably in QCD where
the perturbative expansion coefficient is not particularly small.  It is
\beq \label{eqn:NLO}
\beta(\bar g) = - \beta_0 \frac{\bar g^3}{16 \pi^2} - \beta_1 \frac{\bar g^5}
{(16 \pi^2)^2} + \ldots~~~,
\eeq
where the result for gauge bosons and spin 1/2 fermions (Caswell 1974) is
\beq
\beta_1 = \frac{2}{3} \left\{ 17[C_2(A)]^2 - 10 T(R) C_2(A) - 6 T(R) C_2(R)
\right\}~~~.
\eeq
The first term involves loops exclusively of gauge bosons.  The second
involves single-gauge-boson loops with a fermion loop on one
of the gauge boson lines.  The third involves fermion loops with a fermion
self-energy due to a gauge boson.  The quantity $C_2(R)$ is defined such that
\beq
[T^i(R) T^i(R)]_{\alpha \beta} = C_2(R) \delta_{\alpha \beta}~~~,
\eeq
where $\alpha$ and $\beta$ are indices in the fermion representation.
 
We now illustrate the calculation of $C_2(A)$, $T(R)$, and $C_2(R)$ for
SU(N).  More general techniques are given by Slansky (1981).

Any SU(N) group contains an SU(2) subgroup, which we may take to be generated
by $T_1$, $T_2$, and $T_3$.  The isospin projection $I_3$ may be identified
with $T_3$.  Then the $I_3$ value carried by each generator $T_i$ (written
for convenience in the fundamental N-dimensional representation) may be
identified as shown below:
\medskip

\begin{center}
\begin{tabular}{|c c|c c c|} \hline
\multicolumn{2}{|c}{$\leftarrow 2 \rightarrow$} &
\multicolumn{3}{|c|}{$\leftarrow N-2 \rightarrow$} \\ \hline
0 & 1 & 1/2 & $\cdots$ & 1/2 \\
$-1$ & 0 & $-1/2$ & $\cdots$ & $-1/2$ \\ \hline
$-1/2$ & 1/2 & 0 & $\cdots$ & 0 \\
$\cdots$ & $\cdots$ & $\cdots$ & $\cdots$ & $\cdots$ \\
$-1/2$ & 1/2 & 0 & $\cdots$ & 0 \\ \hline
\end{tabular}
\end{center}
\medskip

Since $C_2(A)$ may be calculated for any convenient value of the index
$i = m$ in (\ref{eqn:Cas}), we chose $i = m = 3$.  Then
\beq \label{eqn:C2A}
C_{2}(A) = \sum_{\rm adjoint}(I_3)^2 = 1 + 1 + 4(N-2) \left(
\frac{1}{2} \right)^2 = N~~~.
\eeq
As an example, the octet (adjoint) representation of SU(3) has two
members with $|I_3| = 1$ (e.g., the charged pions) and four with $|I_3|
= 1/2$ (e.g., the kaons).

For members of the fundamental representation of SU(N), there will be
one member with $I_3 = + 1/2$, another with $I_3 = -1/2$, and all the rest
with $I_3 = 0$.  Then again choosing $i = m = 3$ in Eq.\ (\ref{eqn:Tdef}),
we find $T(R)|_{\rm fundamental} = 1/2$.  The SU(N) result for $\beta_0$ in
the presence of $n_f$ spin 1/2 fermions and $n_s$ scalars in the fundamental
representation then may be written
\beq
\beta_0 = \frac{11}{3} N - \frac{2}{3} n_f - \frac{1}{6} n_s~~~.
\eeq

The quantity $C_2(A)$ in (\ref{eqn:C2A}) is most easily calculated by
averaging over all indices $\alpha = \beta$.  If all generators $T^i$ are
normalized in the same way, one may calculate the result for an {\it
individual} generator (say, $T_3$) and then multiply by the number of
generators [$N^2 - 1$ for SU(N)].  For the fundamental representation one
then finds
\beq
C_2(R) = \frac{1}{N}(N^2 - 1) \left[ \left( \frac{1}{2} \right)^2
+ \left(- \frac{1}{2} \right)^2 \right] = \frac{N^2 - 1}{2N}~~~.
\eeq

\subsection{The running coupling constant}

One may integrate Eq.\ (\ref{eqn:NLO}) to obtain the coupling constant as a
function of momentum scale $M$ and a scale-setting parameter $\Lambda$.
In terms of $\bar \alpha \equiv \bar g^2/4 \pi$, one has
\beq
\frac{d \ba}{d t'} = - \beta_0 \frac{\ba^2}{4 \pi} - \beta_1 \frac{\ba^3}
{(4 \pi)^2}~~~,~~ t' \equiv 2 t = \ln \left( \frac{M^2}{\Lambda^2}
\right)~~~.
\eeq
For large $t'$ the result can be written as
\beq
\ba(M^2) = \frac{ 4 \pi}{\beta_0 t'} \left[ 1 - \frac{\beta_1}{\beta_0^2}
\frac{\ln t'}{t'} \right] + {\cal O}(t'^{-2})~~~.
\eeq

Suppose a process involves $p$ powers of $\ba$ to leading order and a
correction of order $\ba^{p+1}$:
\beq
\Gamma = A \ba^p[1 + B \ba + {\cal O}(\ba^2)]~~~.
\eeq
If $\Lambda$ is rescaled to $\lambda \Lambda$, then $t' \to t' - 2 \ln \lambda
= t'(1 - 2 \ln \lambda /t')$, so
\beq
\ba^p \to \ba^p \left( 1 + \frac{p \beta_0}{2 \pi} \ba \ln \lambda \right)~~~.
\eeq
The coefficient $B$ thus depends on the scale parameter used to
define $\ba$.

Many prescriptions have been adopted for defining $\Lambda$.  In one ('t Hooft
1973), the ``minimal subtraction'' or MS scheme, ultraviolet logarithmic
divergences are parametrized by continuing the space-time dimension $d=4$
to $d = 4 - \epsilon$ and subtracting pole terms $\int d^{4-\epsilon}/p^4
\sim 1/\epsilon$.  In another (Bardeen \ite~1978) (the ``modified
minimal subtraction or $\MSb$ scheme) a term
\beq
\frac{1}{\hat \epsilon} = \frac{1}{\epsilon} + \frac{\ln 4 \pi - \gamma_E}{2}
\eeq
containing additional finite pieces is subtracted.  Here $\gamma_E = 0.5772$
is Euler's constant, and one can show that $\Lambda_{\MSb}
= \Lambda_{{\rm MS}} \exp[(\ln 4 \pi - \gamma_E)/2]$.  Many ${\cal O}(\ba)$
corrections are quoted in the $\MSb$ scheme.  Specification of $\Lambda$
in any scheme is equivalent to specification of $\ba(M^2)$.

\subsection{Applications to quantum chromodynamics}

\begin{figure}
\centerline{\includegraphics[height=3in]{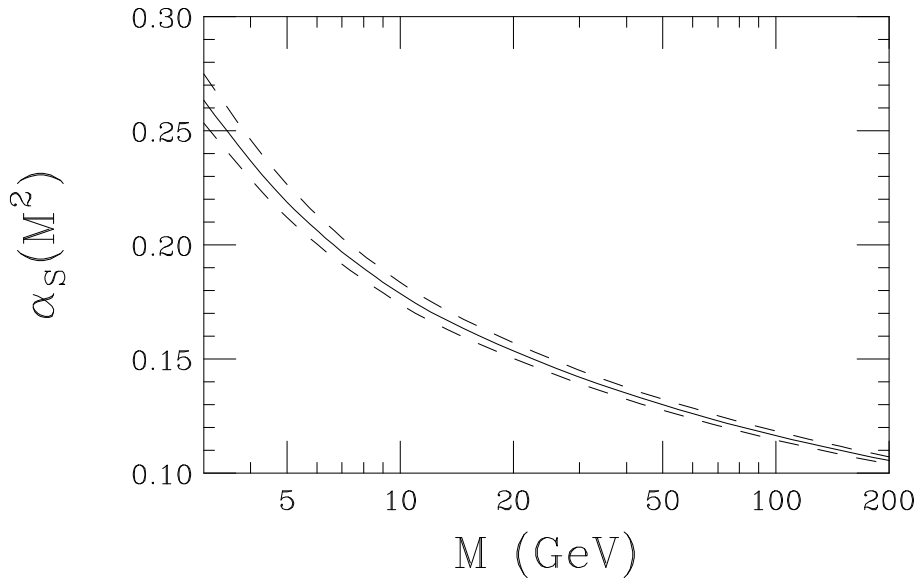}}
\caption{Scale-dependence of the strong-coupling constant $\ba_S(M^2)$
subject to the constraint $\ba_S(M_Z^2) = 0.118 \pm 0.002$.  The solid line
shows the central value; dashed lines indicate $\pm 1 \sigma$ limits.
\label{fig:als}}
\end{figure}

A ``golden application'' of the running coupling constant to QCD is the
effect of gluon radiation on the value of $R$ in $e^+ e^-$ annihilations.
Since $R$ is related to the imaginary part of the photon vacuum polarization
function $\Pi(s)$ which we have calculated for fermions and scalar particles,
one calculates the effects of gluon radiation by calculating the correction
to $\Pi(s)$ due to internal gluon lines.  The leading-order result for
color-triplet quarks is $R(s) \to R(s)[1 + \ba(s)/\pi]$.  There are many
values of $s$ at which one can measure such effects.  For example, at
the mass of the $Z$, the partial decay rate of the $Z$ to hadrons involves
the same correction, and leads to the estimate $\ba_S(M_Z^2) =
0.118 \pm 0.002$.  The dependence of $\ba_S(M^2)$ satisfying this constraint
on $M^2$ is shown in Figure \ref{fig:als}.  As we shall see in Section 5.1,
the {\it electromagnetic} coupling constant also runs, but much more
slowly, with $\alpha^{-1}$ changing from 137.036 at $q^2 = 0$ to about 129
at $q^2 = M_Z^2$.

\begin{figure}
\centerline{\includegraphics[height=3.5in]{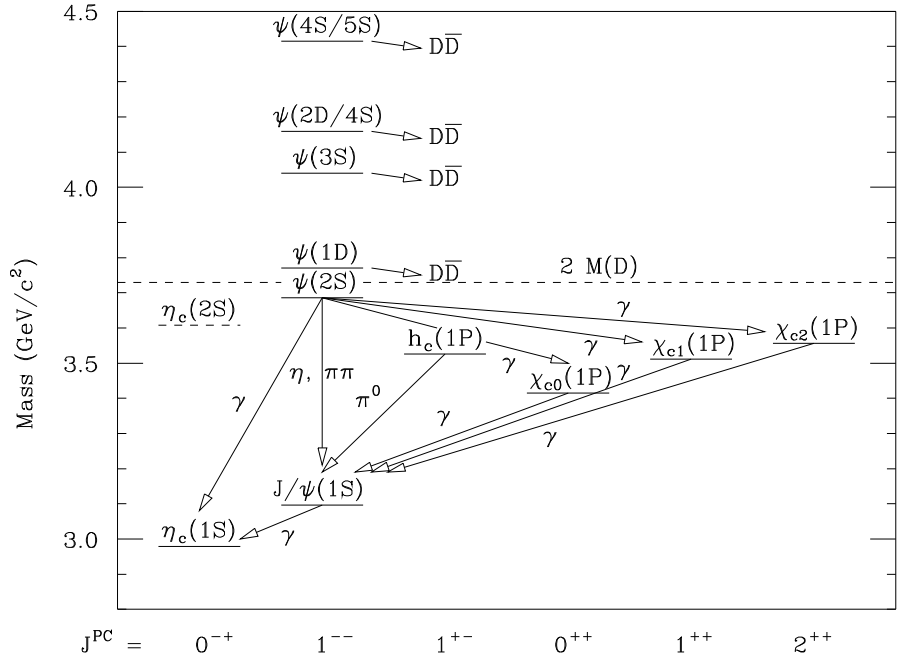}}
\caption{Charmonium ($c \bar c$) spectrum.  Observed and predicted levels are
denoted by solid and dashed horizontal lines, respectively.  Arrows denote
electromagnetic transitions (labeled by $\gamma$) and hadronic transitions
(labeled by emitted hadrons). \label{fig:chm}}
\end{figure}

\begin{figure}
\centerline{\includegraphics[height=3.5in]{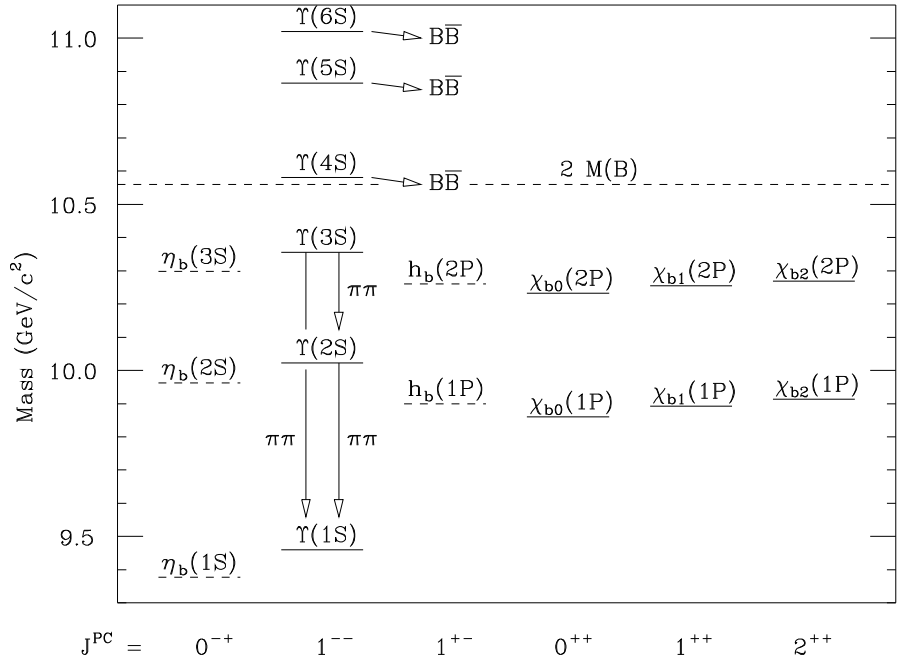}}
\caption{Spectrum of $b \bar b$ states.  Observed and predicted levels are
denoted by solid and dashed horizontal lines, respectively.  In addition to the
transitions labeled by arrows, numerous electric dipole transitions and decays
of states below $B \bar B$ threshold to hadrons containing light quarks have
been seen. \label{fig:ups}}
\end{figure}

A system which illustrates both perturbative and non-perturbative aspects
of QCD is the bound state of a heavy quark and a heavy antiquark, known as
{\it quarkonium} (in analogy with positronium, the bound state of a positron
and an electron).  We show in Figures \ref{fig:chm} and \ref{fig:ups} the
spectrum of the $c \bar c$ and $b \bar b$ bound states (Rosner 1997).  The
charmonium ($c \bar c$) system was an early laboratory of QCD (Appelquist and
Politzer 1975).

The S-wave ($L=0$) levels have total angular momentum $J$, parity $P$,
and charge-conjugation eigenvalue $C$ equal to $J^{PC} = 0^{+-}$ and $1^{--}$
as one would expect for $^1S_0$ and $^3S_1$ states, respectively, of a quark
and antiquark.  The P-wave ($L=1$) levels have $J^{PC} = 1^{+-}$ for the
$^1P_1$, $0^{++}$ for the $^3P_0$, $1^{++}$ for the $^3P_1$, and $2^{++}$ for
the $^3P_2$. The $J^{PC} = 1^{--}$ levels are identified as such by their
copious production through single virtual photons in $e^+ e^-$ annihilations. 
The $0^{-+}$ level $\eta_c$ is produced via single-photon emission from the
$J/\psi$ (so its $C$ is positive) and has been directly measured to have
$J^{P}$ compatible with $0^-$.  Numerous studies have been made
of the electromagnetic (electric dipole) transitions between the S-wave and
$P$-wave levels and they, too, support the assignments shown.

The $b \bar b$ and $c \bar c$ levels have a very similar structure,
aside from an overall shift.  The similarity of the $c \bar c$ and $b \bar b$
spectra is in fact an accident of the fact that for the interquark distances
in question (roughly 0.2 to 1 fm), the interquark potential interpolates
between short-distance Coulomb-like and long-distance linear behavior.  The
Coulomb-like behavior is what one would expect from single-gluon exchange,
while the linear behavior is a particular feature of non-perturbative QCD
which follows from Gauss' law if chromoelectric flux lines are confined to a
fixed area between two widely separated sources (Nambu 1974).  It has
been explicitly demonstrated by putting QCD on a space-time lattice, which
permits it to be solved numerically in the non-perturbative regime.

\begin{figure}
\centerline{\includegraphics[height=3.6in]{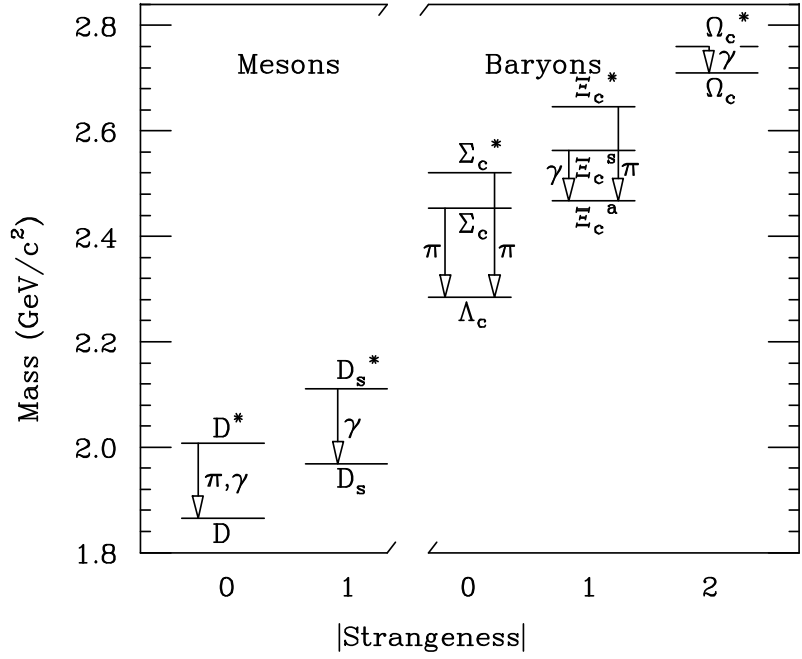}}
\caption{Spectrum of lowest-lying states containing one charmed and one light
quark. Observed and predicted levels are denoted by solid and broken horizontal
lines, respectively. \label{fig:Ds}}
\end{figure}

\begin{figure}
\centerline{\includegraphics[height=3.6in]{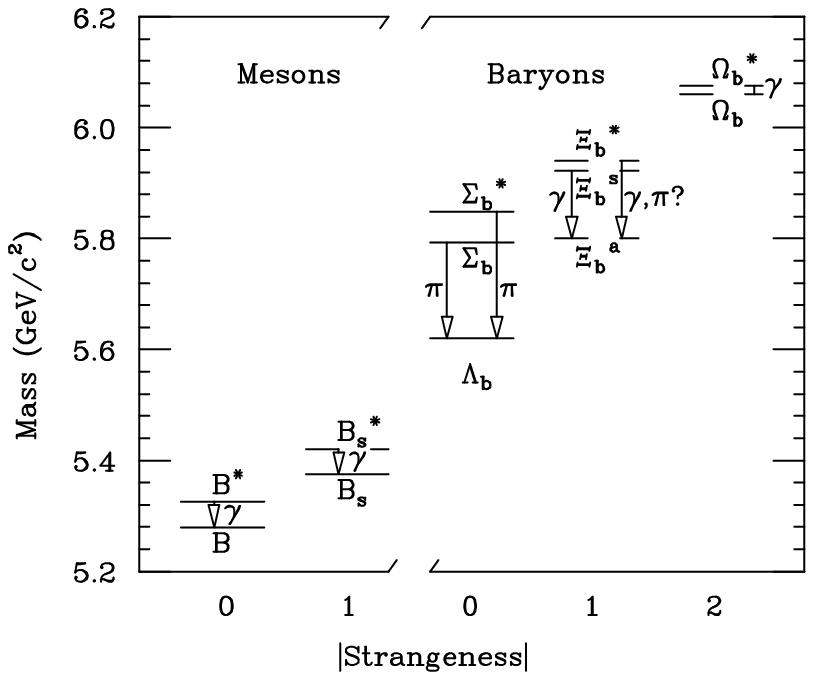}}
\caption{Spectrum of lowest-lying states containing one bottom and one light
quark. Observed and predicted levels are denoted by solid and broken horizontal
lines, respectively. \label{fig:Bs}}
\end{figure}

States consisting of a single charmed quark and light ($u,~d$, or $s$) quarks
or antiquarks are shown in Figure \ref{fig:Ds}.  Finally, the pattern of states
containing a single $b$ quark (Figure \ref{fig:Bs}) is very
similar to that for singly-charmed states, though not as well fleshed-out.
In many cases the splittings between states containing a single $b$ quark
is less than that between the corresponding charmed states by roughly a
factor of $m_c/m_b \simeq 1/3$ as a result of the smaller chromomagnetic
moment of the $b$ quark.  Pioneering work in understanding the spectra
of such states using QCD was done by De R\'ujula \ite~(1975), building on
earlier observations on light-quark systems by Zel'dovich and Sakharov (1966),
Dalitz (1967), and Lipkin (1973).

\section{$W$ bosons}

\subsection{Fermi theory of weak interactions}

The effective four-fermion Hamiltonian for the $V-A$ theory of the weak
interactions is
\beq \label{eqn:4f}
\cH_W = \frac{G_F}{\s}[\pb_1 \gm (1 - \gf) \psi_2][\pb_3 \gM(1-\gf)\psi_4]
= 4 \frac{G_F}{\s}(\pb_{1L} \gm \psi_{2L})(\pb_{3L} \gM \psi_{4L})~~~,
\eeq
where $G_F$ and $\psi_L$ were defined in Section 1.3.  We wish to write
instead a Lagrangian for interaction of particles with charged $W$ bosons
which reproduces (\ref{eqn:4f}) when taken to second order at low momentum
transfer.  We shall anticipate a result of Section 4 by introducing the
$W$ through an SU(2) symmetry, in the form of a gauge coupling.

In the kinetic term in the Lagrangian for fermions,
\beq \label{eqn:Kf}
\cL_{Kf} = \pb(i \ds - m) \psi = \pb_L(i \ds) \psi_L + \pb_R(i \ds) \psi_R
- m \pb \psi~~~,
\eeq
the $\dst$ term does not mix $\psi_L$ and $\psi_R$, so in the absence of
the $\pb \psi$ term one would have the freedom to introduce different
covariant derivatives $\bDs$ acting on left-handed and right-handed fermions.
We shall find that the same mechanism which allows us to give masses to the
$W$ and $Z$ while keeping the photon massless will permit the generation of
fermion masses even though $\psi_L$ and $\psi_R$ will transform differently
under our gauge group.  We follow the conventions of Peskin and Schroeder
(1995, p 700 ff).

We now let the left-handed spinors be doublets of an SU(2), such as
\beq
\left[ \begin{array}{c} \nu_e \\ e^- \end{array} \right]_L~~,~~~
\left[ \begin{array}{c} \nu_\mu \\ \mu^- \end{array} \right]_L~~,~~~
\left[ \begin{array}{c} \nu_\tau \\ \tau^- \end{array} \right]_L~~~.
\eeq
(We will postpone the question of neutrino mixing until the last Section.)
The $W$ is introduced via the replacement
\beq
\dm \to \bDm \equiv \dm - i g \bT^i W^i_\mu~~,~~
\bT^i \equiv \tau^i/2~~~,
\eeq
where $\tau^i$ are the Pauli matrices and $W^i_\mu$ are a triplet of massive
vector mesons.  Here we will be concerned only with the $W^\pm$, defined by
$W_\mu^\pm \equiv (W^1_\mu \mp i W^2_\mu)/\s$.  The field $W^+_\mu$
annihilates a $W^+$ and creates a $W^-$, while $W^-_\mu$ annihilates a $W^-$
and creates a $W^+$.  Then $W^1_\mu = (W^+_\mu + W^-_\mu)/\s$ and $W^2_\mu
= i(W^+_\mu - W^-_\mu)/\s$,
so
\beq \label{eqn:Wdef}
\bT^i W^i_\mu = \frac{1}{2} \left[ \begin{array}{c c} W^3_\mu & \s W^+_\mu \\
\s W^-_\mu & - W^3_\mu \end{array} \right]~~~.
\eeq
The interaction arising from (\ref{eqn:Kf}) for a lepton $l = e,\mu,\tau$ is
then
\beq \label{eqn:Lintl}
{\cL}^{(W^\pm)}_{{\rm int},~l} = \frac{g}{\s} \left[ \bar \nu_{lL} \gM
W^+_\mu l_L + \bar l_L \gM W^-_\mu \nu_{lL} \right]~~~,
\eeq
where we temporarily neglect the $W^3_\mu$ terms.  Taking this interaction
to second order and replacing the $W$ propagator $(M_W^2 - q^2)^{-1}$ by
its $q^2 = 0$ value, we find an effective interaction of the form
(\ref{eqn:4f}), with
\beq
\frac{G_F}{\s} = \frac{g^2}{8 M_W^2}~~~.
\eeq

\subsection{Charged-current quark interactions}

\begin{figure}
\centerline{\includegraphics[height=3.2in]{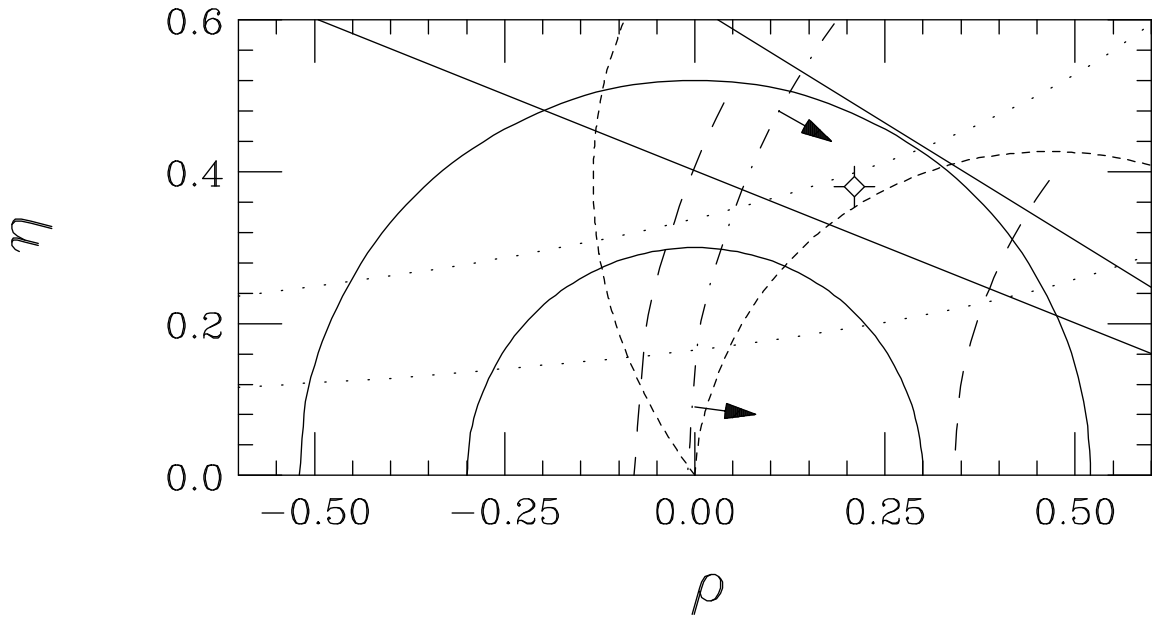}}
\caption{Constraints on parameters of the Cabibbo-Kobayashi-Maskawa (CKM)
matrix.  The plotted point at $\rho = 0.21, \eta=0.38$ lies in the middle of
the allowed region.  (See text.)
\label{fig:re}}
\end{figure}

The left-handed quark doublets may be written
\beq
\left[ \begin{array}{c} u \\ d' \end{array} \right]_L~~,~~~
\left[ \begin{array}{c} c \\ s' \end{array} \right]_L~~,~~~
\left[ \begin{array}{c} t \\ b' \end{array} \right]_L~~~,
\eeq
where $d'$, $s'$, and $b'$ are related to the mass eigenstates $d$, $s$,
$b$ by a unitary transformation
\beq
\left[ \begin{array}{c} d' \\ s' \\ b' \end{array} \right] = V
\left[ \begin{array}{c} d \\ s \\ b \end{array} \right]~~,~~~
V^\dag V = 1~~~.
\eeq
The rationale for the unitary matrix $V$ of Kobayashi and Maskawa (1973) will
be reviewed in the next Section when we discuss the origin of fermion masses in
the electroweak theory.  The interaction Lagrangian for $W$'s with quarks then
is
\beq \label{eqn:Lintq}
{\cL}^{(W^\pm)}_{\rm int,~quarks} = \frac{g}{\s}(\bar U_L \gM W^+_\mu V
D_L) + \hc~~,~~~
U \equiv \left[ \begin{array}{c} u \\ c \\ t \end{array} \right]~~,~~~
D \equiv \left[ \begin{array}{c} d \\ s \\ b \end{array} \right]~~~.
\eeq
A convenient parametrization of $V$
(conventionally known as the Cabibbo-Kobayashi-Maskawa matrix, or CKM matrix)
suggested by Wolfenstein (1983)
is
\beq \label{eqn:wp}
V \equiv \left[ \begin{array}{c c c} V_{ud} & V_{us} & V_{ub} \\
V_{cd} & V_{cs} & V_{cb} \\ V_{td} & V_{ts} & V_{tb} \end{array} \right]
= \left[ \begin{array}{c c c} 1 - \frac{\lambda^2}{2} & \lambda & 
A \lambda^3(\rho - i \eta) \\ - \lambda & 1 - \frac{\lambda^2}{2} &
A \lambda^2 \\ A \lambda^3 (1 - \rho - i \eta) & - A \lambda^2 & 1
\end{array} \right]~~~.
\eeq
Experimentally $\lambda \simeq 0.22$ and $A \simeq 0.85$.  Present constraints
on the parameters $\rho$ and $\eta$ are shown in Figure \ref{fig:re}.  The
solid circles denote limits on $|V_{ub}/|V_{cb}| = 0.090 \pm 0.025$ from
charmless $b$ decays.  The dashed arcs are associated with limits on $V_{td}$
from $B^0$--$\ob$ mixing.  The present lower limit on $B_s$--$\bar B_s$ mixing
leads to a lower bound on $|V_{ts}/V_{td}|$ and the dot-dashed arc.  The dotted
hyperbolae arise from limits on CP-violating $K^0$--$\ok$ mixing.  The phases
in the CKM matrix associated with $\eta \ne 0$ lead to CP violation in neutral
kaon decays (Christenson \ite~1964) and, as recently
discovered, in neutral $B$ meson decays (Aubert \ite~2001a, Abe \ite~2001).
These last results lead to a result shown by the two rays, $\sin(2
\beta)=0.79 \pm 0.10$, where $\beta={\rm Arg}(-V_{cd}V^*_{cb}/V_{td}V^*_{tb})$.
The small dashed lines represent $1 \sigma$ limits derived by Gronau and
Rosner (2002) (see also Luo and Rosner 2001) on the basis of CP asymmetry
data of Aubert \ite~(2001b) for $B^0 \to \pi^+ \pi^-$.  Our range of
parameters (confined by $1 \sigma$ limits)
is $0.10 \le \rho \le 0.32$, $0.33 \le \eta \le 0.43$.  Similar plots are
presented in several other lectures at this Summer School (see, e.g.,
Buchalla 2001, Nir 2001, Schubert 2001, Stone 2001), which may be consulted for
further details, and an ongoing analysis of CKM parameters by H\"ocker
\ite~(2001) is now incorporating several other pieces of data.

\subsection{Decays of the $\tau$ lepton}

The $\tau$ lepton (Perl \ite~1975) provides a good example of ``standard model''
charged-current physics.  The $\tau^-$ decays to a $\nu_\tau$ and a
virtual $W^-$ which can then materialize into any kinematically allowed final
state: $e^- \bar \nu_e$, $\mu^- \bar \nu_\mu$, or three colors of $\bar u d'$,
where, in accord with (\ref{eqn:wp}), $d' \simeq 0.975 d + 0.22 s$.

Neglecting strong interaction corrections and final fermion masses, the rate
for $\tau$ decay is expected to be
\beq
\Gamma(\tau^- \to {\rm all}) = 5 G_F^2 \frac{m_\tau^5}{192 \pi^3}
\simeq 2 \times 10^{-3}~{\rm eV}~~~,
\eeq
corresponding to a lifetime of $\tau_\tau \simeq 3 \times 10^{-13}$ s as
observed.  The factor of $5 = 1 + 1 + 3$ corresponds to equal rates into
$e^- \bar \nu_e$, $\mu^- \bar \nu_\mu$, and each of the three colors of $\bar u
d'$.  The
branching ratios are predicted to be
\beq
\cB(\tau^- \to \nu_\tau e^- \bar \nu_e) = \cB(\tau^- \to \nu_\tau \mu^-
\bar \nu_\mu) = \frac{1}{3}\cB(\tau^- \to \nu_\tau \bar u d') = 20\%~~~.
\eeq
Measured values for the purely leptonic branching ratios are slightly
under 18\%, as a result of the enhancement of the hadronic channels
by a QCD correction whose leading-order behavior is $1 + \alpha_S/\pi$,
the same as for $R$ in $e^+ e^-$ annihilation.  The $\tau$ decay is thus
further evidence for the existence of three colors of quarks.

\subsection{$W$ decays}

We shall calculate the rate for the process $W \to f \bar f'$ and then
generalize the result to obtain the total $W$ decay rate.
The interaction Lagrangian (\ref{eqn:Lintl}) implies that the covariant
matrix element for the process $W(k) \to f(p) \bar f'(p')$ is
\beq
\cM^{(\lambda)} = \frac{g}{2 \s} \bar u_f(p) \gM (1 - \gf) v_{f'}(p')
\epsilon^{(\lambda)}_\mu(k)~~~.
\eeq
Here $\lambda$ describes the polarization state of the $W$.  The partial
width is
\beq
\Gamma(W^- \to f \bar f') = \frac{1}{2 M_W} \frac{1}{3} \sum_{\rm pols}
|\cM^{(\lambda)}|^2 \frac{p^*}{4 \pi M_W}~~~,
\eeq
where $(2 M_W)^{-1}$ is the initial-state normalization, 1/3 corresponds to an
average of $W$ polarizations, the sum is over both $W$ and lepton
polarizations, and $p^*$ is the final center-of-mass (c.m.) 3-momentum.
We use the identity
\beq
\sum_\lambda \epsilon^{(\lambda)}_\mu(k) \epsilon^{(\lambda)*}_\nu(k)
= - \gmn + \frac{k_\mu k_\nu}{M_W^2}
\eeq
for sums over $W$ polarization states.  The result is that
\beq
\sum_{\rm pols}|\cM^{(\lambda)}|^2 = g^2 \left[ M_W^2 -
\frac{1}{2}(m^2 + {m'}^2) - \frac{(m^2 - {m'}^2)^2}{2 M_W^2} \right]
\eeq
for any process $W \to f \bar f'$, where $m$ is the mass of $f$ and $m'$ is
the mass of $f'$.  Recalling the relation between $G_F$ and $g^2$, this may
be written in the simpler form
\beq
\Gamma(W \to f \bar f') = \frac{G_F}{\s} \frac{M_W^3}{6 \pi} \Phi_{f f'}~~,~~~
\Phi_{f f'} \equiv \frac{2 p^*}{M_W} \frac{p^{*2} + 3 E E'}{M_W^2}~~~.
\eeq
Here $E = (p^{*2} + m^2)^{1/2}$ and $E' = (p^{*2} + {m'}^2)^{1/2}$ are the
c.m.\ energies of $f$ and $f'$.  The factor $\Phi_{f f'}$ reduces to 1 as $m,m'
\to 0$.

The present experimental average for the $W$ mass (Kim 2001) is $M_W =
80.451 \pm 0.033$ GeV.  Using this value, we predict $\Gamma(W \to e^-
\bar \nu_e) = 227.8 \pm 2.3$ MeV.  The widths to various channels are
expected to be in the ratios
\beq
e^- \bar \nu_e:~\mu^- \bar \nu_\mu:~\tau^- \bar \nu_\tau:~\bar u d':
~\bar c s' = 1:~1:~1:~3 \left[1 + \frac{\alpha_S(M_W^2)}{\pi} \right]:~
3 \left[1 + \frac{\alpha_S(M_W^2)}{\pi} \right]~~~,
\eeq
so $\alpha_S(M_W^2) = 0.120 \pm 0.002$ leads to the prediction $\Gamma_{\rm
tot}(W) = 2.10 \pm 0.02$ GeV.  This is to be compared with a value (Drees
2001) obtained at LEP II by direct reconstruction of $W$'s:
$\Gamma_{\rm tot}(W) = 2.150 \pm 0.091$ GeV.  Higher-order electroweak
corrections, to be discussed in Section 5, are not expected to play a major
role here.  This agreement means, among other things, that we are not missing
a significant channel to which the charged weak current can couple below the
mass of the $W$. 

\subsection{$W$ pair production}

We shall outline a calculation (Quigg 1983) which indicates that the weak
interactions cannot possibly be complete if described only by charged-current
interactions.  We consider the process $\nu_e(q) + \bar \nu_e(q') \to
W^+(k) + W^-(k')$ due to exchange of an electron $e^-$ with momentum $p$.  The
matrix element is
\beq
\cM^{(\lambda,\lambda')} = \frac{G_F M_W^2}{\s} \bar v(q') \not \!
\epsilon^{(\lambda')}(k')(1 - \gf) \frac{\ps}{p^2} \not \!
\epsilon^{(\lambda)}(k) u(q)~~~.
\eeq
For a longitudinally polarized $W^+$, this matrix element grows in an
unacceptable fashion for high energy.  In fact, an inelastic amplitude for any
given partial wave has to be bounded, whereas $\cM^{(\lambda,\lambda')}$ will
not be.

The polarization vector for a longitudinal $W^+$ traveling along the $z$
axis is
\beq
\epsilon^{(\lambda)}_\nu(k) = (|\vec{k}|,0,0,M_W) \simeq k_\nu/M_W~~~,
\eeq
with a correction which vanishes as $|\vec{k}| \to \infty$.  Replacing
$\epsilon^{(\lambda)}_\nu(k)$ by $k_\nu / M_W$, using $\ks = \qs - \ps$ and
$\qs u(q) = 0$, we find
\beq
\cM^{(\lambda,\lambda')} \simeq - \s G_F M_W \bar v(q') \not \! \epsilon^
{(\lambda')}(k')u(q)~~~,
\eeq
\beq
\sum_{\rm lepton~pol.} |\cM^{(\lambda,\lambda')}|^2 = 2 G_F^2 M_W^2
[ 8 q' \cdot \epsilon^{(\lambda')} q \cdot \epsilon^{(\lambda')}
- 4 q \cdot q' \epsilon^{(\lambda')} \cdot \epsilon^{(\lambda')}]~~~.
\eeq
This quantity contributes only to the lowest two partial waves, and grows
without bound as the energy increases.  Such behavior is not only
unacceptable on general grounds because of the boundedness of inelastic
amplitudes, but it leads to divergences in higher-order perturbation
contributions, e.g., to elastic $\bar \nu \nu$ scattering.

Two possible contenders for a solution of the problem in the early 1970s were
(1) a neutral gauge boson $Z^0$ coupling to $\nu \bar \nu$ and $W^+ W^-$
(Glashow 1961, Weinberg 1967, Salam 1968), or (2) a left-handed heavy lepton
$E^+$ (Georgi and Glashow 1972a)
coupling to $\nu_e W^+$.  Either can reduce the unacceptable high-energy
behavior to a constant.  The $Z^0$ alternative seems to be the one selected in
nature.  In what follows we will retrace the steps of the standard electroweak
theory, which led to the prediction of the $W$ and $Z$ and all the phenomena
associated with them.

\section{Electroweak unification}

\subsection{Guidelines for symmetry}

We now return to the question of what to do with the ``neutral $W$'' (the
particle we called $W^3$ in the previous Section), a puzzle since the time of
Oskar Klein in the 1930s.  The time component of the charged weak current
\beq \label{eqn:cwc}
J^{(+)}_\mu = \bar N_L \gm L_L + \bar U_L \gm V D_L~~~,
\eeq
where $N_L$ and $L_L$ are neutral and charged lepton column vectors defined
in analogy with $U_L$ and $D_L$, may be used to define operators
\beq
Q^{(+)} \equiv \int d^3x J_0^{(+)}~~,~~~Q^{(-)} \equiv Q^{(+)\dag}~~~
\eeq
which are charge-raising and -lowering members of an SU(2) triplet.  If we
define $Q_3 \equiv (1/2)[Q^{(+)},Q^{(-)}]$, the algebra closes:
$[Q_3, Q^{(\pm)}] = \pm Q^{(\pm)}$.  This serves to normalize the weak
currents, as mentioned in the Introduction.

The form (\ref{eqn:cwc}) (with unitary $V$) guarantees that the corresponding
neutral current will be
\beq
J_\mu^{(3)} = \frac{1}{2} \left[ \bar N_L \gm N_L - \bar L_L \gm L_L + \bar U_L
\gm U_L - \bar D_L \gm D_L \right]~~~,
\eeq
which is {\it diagonal} in neutral currents.
This can only succeed, of course, if there are equal numbers of charged and
neutral leptons, and equal numbers of charge 2/3 and charge $-1/3$ quarks.

It would have been possible to define an SU(2) algebra making use only of a
doublet (Gell-Mann and L\'evy 1960)
\beq
\left[ \begin{array}{c} u \\ d' \end{array} \right]_L = \left[
\begin{array}{c} u \\ V_{ud} d + V_{us} s \end{array} \right]_L
\eeq
which was the basis of the Cabibbo (1963) theory of the charge-changing weak
interactions of strange and nonstrange particles.  If one takes $V_{ud} =
\cos \theta_C$, $V_{us} = \sin \theta_C$, as is assumed in the Cabibbo
theory, the $u,~d,~s$ contribution to the neutral current $J^{(3)}_\mu$
is
$$
J^{(3)}_\mu|_{u,d,s} = \frac{1}{2} [ \bar u_L \gm u_L - \cos^2
\theta_C \bar d_L \gm d_L
$$
\beq \label{eqn:fcnc}
 - \sin^2 \theta_C \bar s_L \gm s_L -
\sin \theta_C \cos \theta_C (\bar d_L \gm s_L + \bar s_L \gm d_L)] ~~~.
\eeq
This expression contains {\it strangeness-changing neutral currents}, leading
to the expectation of many processes like $K^+ \to \pi^+ \nu \bar \nu$,
$K^0_L \to \mu^+ \mu^-$, $\ldots$, at levels far above those observed.
It was the desire to banish strangeness-changing neutral currents that led
Glashow \ite~(1970) to introduce the charmed quark $c$
(proposed earlier by several authors on the basis of a quark-lepton analogy)
and the doublet
\beq
\left[ \begin{array}{c} c \\ s'  \end{array} \right]_L = \left[
\begin{array}{c} c \\ V_{cd} d + V_{cs} s \end{array} \right]_L~~~.
\eeq
In this four-quark theory, one assumes the corresponding matrix $V$ is
unitary.  By suitable phase changes of the quarks, all elements can be made
real, making $V$ an orthogonal matrix with $V_{ud} = V_{cs} = \cos \theta_C$,
$V_{us} = -  V_{cd} = \sin \theta_C$.  Instead of (\ref{eqn:fcnc}) one then has
\beq \label{eqn:fpnc}
J^{(3)}_\mu|_{u,d,s,c} = \frac{1}{2} [ \bar u_L \gm u_L + \bar c_L \gm c_L
- \bar d_L \gm d_L - \bar s_L \gm s_L ]~~~,
\eeq
which contains no flavor-changing neutral currents. 

\begin{figure}
\centerline{\includegraphics[height=2in]{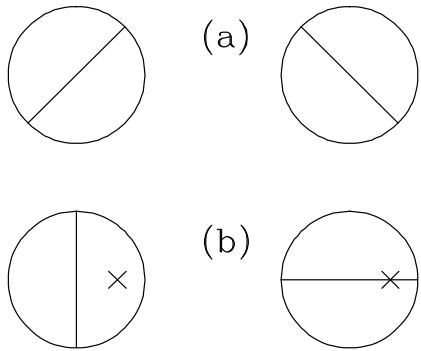}}
\caption{Basis states for first excitations of a drum head. (a) Nodal
lines at $\pm 45^\circ$ with respect to horizontal; (b) horizontal and
vertical nodal lines.
\label{fig:dh}}
\end{figure}

The charmed quark also plays a key role in higher-order charged-current
interactions.  Let us consider $K^0$--$\ok$ mixing.  The CP-conserving
limit in which the eigenstates are $K_1$ (even CP) and $K_2$ (odd CP) can be
illustrated using a degenerate two-state system such as
the first excitations of a drum head.  There is no way to
distinguish between the basis states illustrated in Fig.\ \ref{fig:dh}(a),
in which the nodal lines are at angles of $\pm 45^\circ$ with respect to
the horizontal, and those in Fig.\ \ref{fig:dh}(b), in which they are
horizontal and vertical.

If a fly lands on the drum-head at the point marked ``$\times$'', the
basis (b) corresponds to eigenstates.  One of the modes couples to the
fly; the other doesn't.  The basis in (a) is like that of $(K^0,\ok)$, while
that in (b) is like that of $(K_1,K_2)$.  Neutral kaons are produced as in
(a), while they decay as in (b), with the fly analogous to the $\pi \pi$
state.  The short-lived state ($K_1$, in this CP-conserving approximation)
has a lifetime of 0.089 ns, while the long-lived state ($\simeq K_2$) lives
$\sim 600$ times as long, for 52 ns.  Classical illustration of CP-violating
mixing is more subtle but can be achieved as well, for instance in a rotating
reference frame (Rosner and Slezak 2001, Kosteleck\'y and Roberts 2001).

\begin{figure}
\centerline{\includegraphics[height=1.5in]{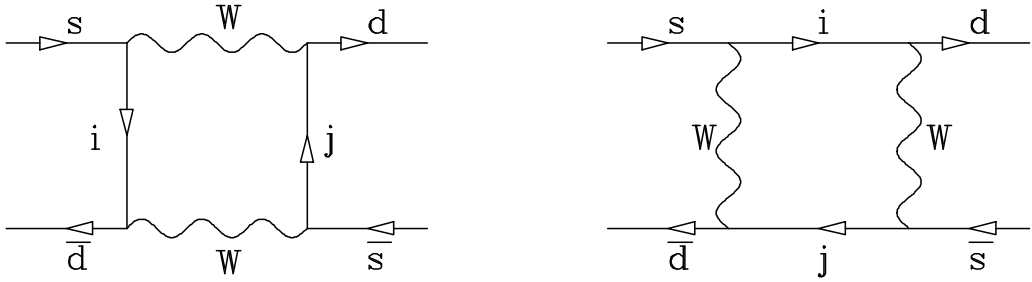}}
\caption{Higher-order weak contributions to $K^0$--$\ok$ mixing due to loops
with internal $u,c,t$ quarks.
\label{fig:kmix}}
\end{figure}

The shared $\pi \pi$ intermediate state and other low-energy states like
$\pi^0$, $\eta$, and $\eta'$ are chiefly responsible for CP-conserving
$K^0$--$\ok$ mixing.  However, one must ensure that large {\it short-distance}
contributions do not arise from diagrams such as those illustrated in
Figure \ref{fig:kmix}.

If the only charge 2/3 quark contributing to this process were the $u$ quark,
one would expect a contribution to $\Delta m_K$ of order
\beq
\Delta m_K |_u \sim g^4 f_K^2 m_K \sin^2 \theta_C \cos^2 \theta_C/ 16 \pi^2
M^2_W \sim G_F f_K^2 m_K (g^2/16 \pi^2)~~~,
\eeq
where $f_K$ is the amplitude for $d \bar s$ to be found in a $K^0$, and the
factor of $16 \pi^2$ is characteristic of loop diagrams.  This is far too
large, since $\Delta m_K \sim \Gamma_{K_S} \sim G_F^2 f_K^2 m_K^3$.  However,
the introduction of the charmed quark, coupling to $-d \sin \theta_C + s \cos
\theta_C$, cancels the leading contribution, leading to an additional factor
of $[(m_c^2 - m_u^2)/M_W^2] \ln(M_W^2/m_c^2)$ in the above expression.  Using
such arguments Glashow \ite~(1970) and Gaillard and Lee
(1974) estimated the mass of the charmed quark to be less than several GeV.
(Indeed, early candidates for charmed particles had been seen by Niu, Mikumo,
and Maeda 1971.) The discovery of the $J/\psi$ (Aubert \ite~1974,
Augustin \ite~1974) confirmed
this prediction; charmed hadrons produced in neutrino interactions (Cazzoli
\ite~1975) and in $e^+ e^-$ annihilations (Goldhaber \ite~1976, Peruzzi
\ite~1976) followed soon after.

\begin{figure}
\centerline{\includegraphics[height=1.5in]{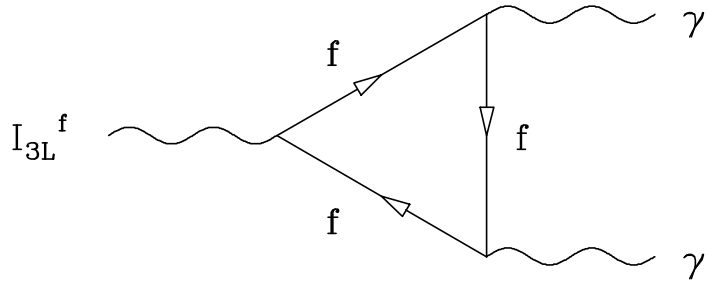}}
\caption{Example of triangle diagram for which leading behavior must cancel
in a renormalizable electroweak theory.
\label{fig:anom}}
\end{figure}

An early motivation for charm relied on an analogy between quarks and leptons.
Hara (1964), Maki and Ohnuki (1964), and Bjorken and Glashow (1964)
inferred the existence of a charmed quark coupling mainly to the
strange quark from the existence of the $\mu - \nu_\mu$ doublet:
\beq \label{eqn:secfam}
\left( \begin{array}{c} \nu_\mu \\ \mu^- \end{array} \right): {\rm leptons}
\Rightarrow \left( \begin{array}{c} c \\ s \end{array} \right): {\rm
quarks}~~~.
\eeq
Further motivation for the quark-lepton analogy was noted by Bouchiat
\ite~(1972), Georgi and Glashow (1972b), and Gross and
Jackiw (1972).  In a gauge theory of the electroweak interactions, triangle
anomalies associated with graphs of the type shown in Figure \ref{fig:anom}
have to be avoided.  This cancellation requires the fermions $f$
in the theory to contribute a total of zero to the sum over $f$ of $Q_f^2
I^f_{3L}$.  Such a cancellation can be achieved by requiring quarks and leptons
to occur in complete families so that the terms
\beq
{\rm Leptons}:~~(0)^2 \left( \frac{1}{2} \right) + (-1)^2 \left( -\frac{1}{2}
\right) = - \frac{1}{2}
\eeq
\beq
{\rm Quarks}:~~3 \left[ \left( \frac{2}{3} \right)^2 \left( \frac{1}{2} \right)
+ \left( -\frac{1}{3} \right)^2 \left( -\frac{1}{2}\right) \right] =
\frac{1}{2}
\eeq
sum to zero for each family.

We are then left with a flavor-preserving neutral current $J^{(3)}_\mu$,
given by (\ref{eqn:fpnc}), whose interpretation must still be given.  It
 cannot correspond to the photon, since the photon couples to both
left-handed and right-handed fermions.  At the same time, the photon is
somehow involved in the weak interactions associated with $W$ exchange.
In particular, the $W^\pm$ themselves are charged, so any theory in which
electromagnetic current is conserved must involve a $\gamma W^+ W^-$
coupling.  Moreover, the charge is sensitive to the third component of
the SU(2) algebra we have just introduced.  We shall refer to this as
\SUL, recognizing that only left-handed fermions $\psi_L$ transform
non-trivially under it.  Then we can define a {\it weak hypercharge} $Y$ in
terms of the difference between the electric charge $Q$ and the third component
$I_{3L}$ of \SUL~({\it weak isospin}):
\beq \label{eqn:wkh}
Q = I_{3L} + \frac{Y}{2}~~~.
\eeq
Values of $Y$ for quarks and leptons are summarized in Table \ref{tab:wkh}.

\begin{table}
\caption{Values of charge, $I_{3L}$, and weak hypercharge $Y$ for quarks
and leptons.
\label{tab:wkh}}
\bigskip
\begin{center}
\begin{tabular}{|c c c c|} \hline
Particle(s) &  $Q$   & $I_{3L}$ &   Y    \\ \hline
$\nu_{eL}$  &   0    &   1/2    &  $-1$  \\
$e^-_L$     &  $-1$  &  $-1/2$  &  $-1$  \\ \hline
$u_L$       &  2/3   &  $1/2$   &   1/3  \\
$d_L$       & $-1/3$ &  $-1/2$  &   1/3  \\ \hline
$e^-_R$     &  $-1$  &     0    &  $-2$  \\
$u_R$       &  2/3   &     0    &  4/3   \\
$d_R$       & $-1/3$ &     0    & $-2/3$ \\ \hline
\end{tabular}
\end{center}
\end{table}

If you find these weak hypercharge assignments mysterious, you are not alone.
They follow naturally in unified theories ({\it grand unified theories}) of the
electroweak and strong interactions.  A ``secret formula'' for $Y$, which may
have deeper significance (Pati and Salam 1973), is $Y = 2 I_{3R} +
(B-L)$, where $I_{3R}$ is the third component of ``right-handed'' isospin, $B$
is baryon number (1/3 for quarks), and $L$ is lepton number (1 for leptons
such as $e^-$ and $\nu_e$).  The orthogonal component of $I_{3R}$ and
$B-L$ may correspond to a higher-mass, as-yet-unseen vector boson, an
example of what is called a $Z'$.  The search for $Z'$ bosons with various
properties is an ongoing topic of interest; current limits are quoted
by the Particle Data Group (2000).

The gauge theory of charged and neutral $W$'s thus must involve the photon in
some way.  It will then be necessary, in order to respect the formula
(\ref{eqn:wkh}), to introduce an additional U(1) symmetry associated with
weak hypercharge.  The combined electroweak gauge group will have the form
\SUL~$\otimes$ \U1Y.

\subsection{Symmetry breaking}

Any unified theory of the weak and electromagnetic interactions must be
broken, since the photon is massless while the $W$ bosons (at least) are
not.  An explicit mass term in a gauge theory of the form
$m^2 A^i_\mu A^{\mu i}$ violates gauge invariance.  It is not invariant
under the replacement (\ref{eqn:gtn}).  Another means must be found to
introduce a mass.  The symmetry must be broken in such a way as to preserve
gauge invariance.

A further manifestation of symmetry breaking is the presence of fermion mass
terms.  Any product $\pb \psi$ may be written as
\beq
\pb \psi = (\pb_L + \pb_R)(\psi_L + \psi_R) = \pb_L \psi_R + \pb_R \psi_L~~~,
\eeq
using the fact that $\pb_L = \pb(1+\gf)/2$, $\pb_R = \pb(1-\gf)/2$.  Since
$\psi_L$ transforms as an \SUL~doublet but $\psi_R$ as an \SUL~ singlet, a mass
term proportional to $\pb \psi$ transforms as an overall
\SUL~doublet.  Moreover, the weak hypercharges of left-handed fermions
and their right-handed counterparts are different.  Hence one cannot even have
explicit {\it fermion} mass terms in the Lagrangian and hope to preserve
local gauge invariance.

One way to generate a fermion mass without explicitly violating gauge
invariance is to assume the existence of a complex scalar \SUL~doublet $\phi$
coupled to fermions via a Yukawa interaction:
\beq \label{eqn:phi}
\cL_Y = - g_Y( \pb_L \phi \psi_R + \hc)~~,~~~
\phi \equiv \left[ \begin{array}{c} \phi^+ \\ \phi^0 \end{array} \right]~~~.
\eeq
Thus, for example, with $\pb_L = (\bar \nu_e, e)_L$ and $\psi_R = e_R$,
we have
\beq
\cL_{Y,e} = - g_{Ye}(\bar \nu_{eL} \phi^+ e_R + \bar e_L \phi^0 e_R +
\hc)~~~.
\eeq
If $\phi^0$ acquires a vacuum expectation value, $\vev{\phi^0} \ne 0$, this
quantity will automatically break \SUL~and \U1Y, and will give rise to
a non-zero electron mass.  A neutrino mass is not generated, simply because
no right-handed neutrino has been assumed to exist.  (We shall see in the last
Section how to generate the tiny neutrino masses that appear to be present in
nature.)  The gauge symmetry is not broken in the Lagrangian, but only in the
solution.  This is similar to the way in which rotational invariance is broken
in a ferromagnet, where the fundamental interactions are rotationally
invariant but the ground-state solution has a preferred direction along
which the spins are aligned.

The $d$ quark masses are generated by similar couplings involving
$\pb_L = (\bar u, \bar d)_L$, $\psi_R = d_R$, so that
\beq
\cL_{Y,d} = - g_{Yd}(\bar u_L \phi^+ d_R + \bar d_L \phi^0 d_R + \hc)~~~.
\eeq
To generate $u$ quark masses one must either use the multiplet
\beq
\tilde{\phi} \equiv \left[ \begin{array}{c} \bar \phi^0 \\ - \phi^- \end{array}
\right] = i \tau^2 \phi^*~~~,
\eeq
which also transforms as an SU(2) doublet, or a separate doublet of scalar
fields
\beq
\phi' = \left[ \begin{array}{c} {\phi'}^0 \\ {\phi'}^- \end{array} \right]~~~.
\eeq
With $\pb_L = (\bar u, \bar d)_L$ and $\psi_R = u_R$, we then find
\beq
\cL_{Y,u} = -g_{Yu}(\bar u_L \bar \phi^0 u_L - \bar d_L \phi^- u_L + \hc)
\eeq
if we make use of $\tilde \phi$, or
\beq
\cL_{Y,u} = -g_{Yu}(\bar u_L {\phi'}^0 u_L + \bar d_L {\phi'}^- u_L + \hc)
\eeq
if we use $\phi'$.  For present purposes we shall assume the existence of a
single complex doublet, though many theories (notably, some grand unified
theories or supersymmetry) require more than one.

\subsection{Scalar fields and the Higgs mechanism}

Suppose a complex scalar field of the form (\ref{eqn:phi}) is described by a
Lagrangian
\beq \label{eqn:Lphi}
\cL_\phi = (\dm \phi)^\dag(\dM \phi) - \frac{\lambda}{4}(\phi^\dag \phi)^2 +
 \frac{\mu^2}{2} \phi^\dag \phi~~~.
\eeq
Note the ``wrong'' sign of the mass term.  This Lagrangian is invariant under
\SUL~$\otimes$ \U1Y.  The field $\phi$ will acquire a constant vacuum
expectation value which we calculate by asking for the stationary
value of $\cL_\phi$:
\beq
\frac{\partial \cL_\phi}{\partial(\phi^\dag \phi)} = 0 \Rightarrow
\vev{\phi^\dag \phi} = \frac{\mu^2}{\lambda}~~~.
\eeq
We still have not specified which component of $\phi$ acquires the
vacuum expectation value.  At this point only $\phi^\dag \phi = |\phi^+|^2 +
|\phi^0|^2$ is fixed, and $(\re~\phi^+,~\im~\phi^+,~\re~\phi^0,~\im~\phi^0)$
can range over the surface of a four-dimensional sphere.  The Lagrangian
(\ref{eqn:Lphi}) is, in fact, invariant under rotations of this
four-dimensional sphere, a group SO(4) isomorphic to \ew.  A lower-dimensional
analogue of this surface would be the bottom of a wine bottle along which a
marble rolls freely in an orbit a fixed distance from the center.

Let us {\it define} the vacuum expectation value of $\phi$ to be a real
parameter in the $\phi^0$ direction:
\beq
\vev{\phi} = \left[ \begin{array}{c} 0 \\ v/\s \end{array} \right]~~~.
\eeq
The factor of $1/\s$ is introduced for later convenience.  We then find,
from the discussion in the previous section, that Yukawa couplings of $\phi$
to fermions $\psi_i$ generate mass terms $m_i = g_{Yi} v/\s$.  We must now
see what such vacuum expectation values do to gauge boson masses.
(For numerous illustrations of this phenomenon in simple field-theoretical
models see Abers and Lee 1973, Quigg 1983, and Peskin and Schroeder 1995.)

In order to introduce gauge interactions with the scalar field $\phi$, one must
replace $\dm$ by $\bDm$ in the kinetic term of the Lagrangian (\ref{eqn:Lphi}).
Here
\beq
\bDm = \dm - i g \frac{\tau^i W^i_\mu}{2} - i g' \frac{Y}{2}B_\mu~~~,
\eeq
where the \U1Y~interaction is characterized by a coupling constant $g'$ and
a gauge field $B_\mu$, and we have written $g$ for the $SU(2)$ coupling
discussed earlier.  It will be convenient to write $\phi$ in terms of four
independent real fields $(\xi^i,~\eta)$ in a slightly different form:
\beq \label{eqn:ung}
\phi = \exp \left( \frac{i \xi^i \tau^i}{2 v} \right) \left[ \begin{array}{c}
0 \\ \frac{v + \eta}{\s} \end{array} \right]~~~.
\eeq
We then perform an \SUL~gauge transformation to remove the $\xi$ dependence
of $\phi$, and rewrite it as
\beq \label{eqn:gau}
\phi = \left[ \begin{array}{c} 0 \\ \frac{v + \eta}{\s}\end{array}\right]~~~.
\eeq
The fermion and gauge fields are transformed accordingly; we rewrite the
Lagrangian for them in the new gauge.  The resulting kinetic term for the
scalar fields, taking account that $Y = 1$ for the Higgs field (\ref{eqn:phi}),
is
$$
\cL_{K,\phi} = (\bDm \phi)^\dag(\bDM \phi)
$$
\beq
= \left| \left\{ \dm - \frac{i g}{2} \left[ \begin{array}{c c}
 W^3_\mu & W^1_\mu - i W^2_\mu \\ W^1_\mu + i W^2_\mu & - W^3_\mu \end{array}
\right] - \frac{i g'}{2}B_\mu \right\} \left[ \begin{array}{c} 0 
\\ \frac{v + \eta}{\s}\end{array}\right] \right|^2~~~.
\eeq
This term contains several contributions.
\begin{enumerate}

\item There is a kinetic term $\frac{1}{2}(\dm \eta)(\dM \eta)$ for the scalar
field $\eta$.

\item A term $v \dm \eta$ is a total divergence and can be neglected.

\item There are $WW \eta$, $BB \eta$, $W W \eta^2$, and $B B \eta^2$
interactions.

\item The $v^2$ term leads to a mass term for the Yang-Mills fields:

\end{enumerate}
\beq \label{eqn:YMm}
\cL_{m,YM} = \frac{v^2}{8} \left\{ g^2[(W^1)^2 + (W^2)^2] 
+ (g W^3 - g' B)^2 \right\} ~~~.
\eeq
The spontaneous breaking of the \ew~symmetry thus has led to the appearance of
a mass term for the gauge fields.  This is an example of the {\it Higgs
mechanism} (Higgs 1964).  An unavoidable consequence is the appearance of the
scalar field $\eta$, the {\it Higgs field}.  We shall discuss it further in
Section 5.

The masses of the charged $W$ bosons may be identified by comparing
Eqs.~(\ref{eqn:YMm}) and (\ref{eqn:Wdef}):
\beq
(gv)^2/8 = M^2_W/2~~,~~~{\rm or}~M_W = gv/2~~~.
\eeq
Since the Fermi constant is related to $g/M_W$, one finds
\beq
\frac{G_F}{\s} = \frac{g^2}{8 M_W^2} = \frac{1}{2v^2}~~,~~~
{\rm or}~~v = 2^{-1/4} G_F^{-1/2} = 246~{\rm GeV}~~~.
\eeq

The combination $gW^3_\mu - g' B_\mu$ also acquires a mass.  We must
normalize this combination suitably so that it contributes properly in
the kinetic term for the Yang-Mills fields:
\beq
\cL_{K,YM} = -\frac{1}{4}W^i_{\mu \nu} W^{\mu \nu i} - \frac{1}{4}
B_{\mu \nu} B^{\mu \nu}~~~,
\eeq
where
\beq
W^i_{\mu \nu} \equiv \dm W^i_\nu - \dn W^i_\mu + g \epsilon_{ijk} W^j_\mu
W^k_\nu~~,~~~
B_{\mu \nu} \equiv \dm B_\nu - \dn B_\mu~~~.
\eeq
Defining
\beq \label{eqn:thd}
\cos \theta \equiv \frac{g}{\gz}~~\left[ {\rm so~that~}~~ \sin \theta =
 \frac{g'}{\gz} \right]~~~,
\eeq
we may write the normalized combination $\sim g W^3_\mu - g' B_\mu$ which
acquires a mass as
\beq \label{eqn:Zdef}
Z_\mu \equiv W^3_\mu \cos \theta - B_\mu \sin \theta~~~.
\eeq
The orthogonal combination does not acquire a mass.  It may then be
identified as the photon:
\beq \label{eqn:Adef}
A_\mu = B_\mu \cos \theta + W^3_\mu \sin \theta~~~.
\eeq

The mass of the $Z$ is given by
\beq
\frac{(g^2 + {g'}^2) v^2}{8} = \frac{M_Z^2}{2}~~,~~~
{\rm or}~~ M_Z = M_W \gz/g = M_W/\cos \theta~~~,
\eeq
using (\ref{eqn:thd}) in the last relation.  The $W$'s and $Z$'s have
acquired masses, but they are not equal unless $g'$ were to
vanish.  We shall see in the next subsection that both $g$ and $g'$ are
nonzero, so one expects the $Z$ to be heavier than the $W$.

It is interesting to stop for a moment to consider what has taken place.  We
started with four scalar fields $\phi^+$, $\phi^-$, $\phi^0$, and
$\bar \phi^0$.  Three of them [$\phi^+$, $\phi^-$, and the combination
$(\phi^0 - \bar \phi^0)/i\s$] could be absorbed in the gauge transformation
in passing from (\ref{eqn:ung}) to (\ref{eqn:gau}), which made sense only as
long as $(\phi^0 + \bar \phi)/\s$ had a vacuum expectation value $v$.  The
net result was the generation of mass for three gauge bosons $W^+$, $W^-$,
and $Z$.

If we had not transformed away the three components $\xi^i$ of $\phi$ in
(\ref{eqn:ung}), the term $\cL_{K,\phi}$ in the presence of gauge fields
would have contained contributions $W_\mu \dM \phi$ which mixed gauge fields
and derivatives of $\phi$.  These can be expressed as
\beq
W_\mu \dM \phi = \dM(W_\mu \phi) - (\dM W_\mu) \phi
\eeq
and the total divergence (the first term) discarded.  One thus sees that such
terms {\it mix longitudinal components of gauge fields} (proportional to
$\dM W_\mu$) {\it with scalar fields}.  It is necessary to redefine the
gauge fields by means of a gauge transformation to get rid of such mixing
terms.  It is just this transformation that was anticipated in passing
from (\ref{eqn:ung}) to (\ref{eqn:gau}).

The three ``unphysical'' scalar fields provide the necessary longitudinal
degrees of freedom in order to convert the massless $W^\pm$ and $Z$ to 
massive fields.  Each massless field possesses only two polarization states
($J_z = \pm J$), while a massive vector field has three ($J_z = 0$ as well).
Such counting rules are extremely useful when more than one Higgs field is
present, to keep track of how many scalar fields survive being ``eaten''
by gauge fields.

\subsection{Interactions in the \ew~theory}

By introducing gauge boson masses via the Higgs mechanism, and letting
the simplest non-trivial representation of scalar fields acquire a vacuum
expectation value $v$, we have related the Fermi coupling constant to $v$,
and the gauge boson masses to $gv$ or $\gz v$.  We still have two arbitrary
couplings $g$ and $g'$ in the theory, however.  We shall show how to relate
the electromagnetic coupling to them, and how to measure them separately.

The interaction of fermions with gauge fields is described by the kinetic
term $\cL_{K,\psi} = \pb \bDs \psi$.  Here, as usual,
\beq
\bDs = \ds - i g \frac{\tau^i \not \! W^i}{2} - i g' \frac{Y}{2} \not \! B~~~.
\eeq
The charged-$W$ interactions have already been discussed.  They are described
by the terms (\ref{eqn:Lintl}) for leptons and (\ref{eqn:Lintq}) for quarks.
The interactions of $W^3$ and $B$ may be re-expressed in terms of $A$ and $Z$
via the inverse of (\ref{eqn:Zdef}) and (\ref{eqn:Adef}):
\beq
W^3_\mu = Z_\mu \cos \theta + A_\mu \sin \theta~~,~~~
B_\mu = - Z_\mu \sin \theta + A_\mu \cos \theta~~~.
\eeq
Then the covariant derivative for neutral gauge bosons is
\beq \label{eqn:covn}
\bDs|_{\rm neutral} = \ds - i g I_{3L} (\Zs \cos \theta + \As \sin \theta)
- i g'(Q - I_{3L})(- \Zs \sin \theta + \As \cos \theta)~~~.
\eeq
Here we have substituted $Y/2 = (Q - I_{3L})$.  We identify the
electromagnetic contribution to the right-hand side of (\ref{eqn:covn})
with the familiar one $-i e Q \As$, so that
\beq \label{eqn:egg}
e = g' \cos \theta = g \sin \theta~~~.
\eeq
The second equality, stemming from the demand that $I_{3L} \As$ terms cancel
one another in (\ref{eqn:covn}), is automatically satisfied as a result of the
definition (\ref{eqn:thd}).  Combining (\ref{eqn:thd}) and (\ref{eqn:egg}),
we find
\beq
e = \frac{gg'}{\sqrt{g^2 + {g'}^2}}~~~{\rm or}~~~
\frac{1}{e^2} = \frac{1}{g^2} + \frac{1}{{g'}^2}~~~,
\eeq
the result advertised in the Introduction.

The interaction of the $Z$ with fermions may be determined from Eq.\
(\ref{eqn:covn}) with the help of (\ref{eqn:thd}), noting that $g \cos \theta
+ g' \sin \theta = \gz$ and $g' \sin \theta = \gz \sin^2 \theta$.  We find
\beq
\bDs|_{\rm neutral} = \ds - i e Q \As - i \gz(I_{3L} - Q \sin^2 \theta) \Zs~~~.
\eeq

Knowledge of the weak mixing angle $\theta$ will allow us to predict the
$W$ and $Z$ masses.  Using $G_F/\s = g^2/8M_W^2$ and $g \sin \theta = e$, we
can write
\beq
M_W = \left[ \frac{\pi \alpha}{\s G_F} \right]^{1/2} \frac{1}{\sin \theta}
= \frac{37.3~\g}{\sin \theta}
\eeq
if we were to use $\alpha^{-1} = 137.036$.  However, we shall see in the
next Section that it is more appropriate to use a value of $\alpha^{-1}
\simeq 129$ at momentum transfers characteristic of the $W$ mass.  With this
and other electroweak radiative corrections, the correct estimate is
raised to $M_W \simeq 38.6 \g/\sin \theta$, leading to the successful
predictions (\ref{eqn:MWZ}).  The $Z$ mass is expressed in terms of the $W$
mass by $M_Z = M_W/\cos \theta$.
  
\subsection{Neutral current processes}

The interactions of $Z$'s with matter,
\beq \label{eqn:Zint}
\cL_{{\rm int},Z} = \gz \pb(I_{3L} - Q \sin^2 \theta) \Zs \psi~~~,
\eeq
may be taken to second order in perturbation theory, leading to an
effective four-fermion theory for momentum transfers much smaller than the $Z$
mass.  In analogy with the relation between the $W$ boson interaction terms
(\ref{eqn:Lintl}) and (\ref{eqn:Lintq}) and the four-fermion charged-current
interaction (\ref{eqn:4f}), we may write
\beq \label{eqn:HNC}
\cH_W^{NC} = 4 G_F \s [\pb_1 (I_{3L} - Q \sin^2 \theta) \gM \psi_2]
[\pb_3(I_{3L} - Q \sin^2 \theta) \gm \psi_4]~~~,
\eeq
where we have used the identity $(g^2 + {g'}^2)/8 M_Z^2 = G_F/\s$ following
from relations in the previous subsection.

Many processes are sensitive to the neutral-current interaction
(\ref{eqn:HNC}), but no evidence for this interaction had been demonstrated
until the discovery in 1973 of neutral-current interactions on hadronic
targets of deeply inelastically scattered neutrinos (Hasert \ite~1973;
Benvenuti \ite~1974).  For many years these processes provided the most
sensitive measurement of neutral-current parameters.  Other crucial experiments
(see, e.g., reviews by Amaldi \ite~1987 and Langacker \ite~1992)
included polarized electron or muon scattering
on nucleons, asymmetries and total cross sections in $e^+ e^- \to \mu^+
\mu^-$ or $\tau^+ \tau^-$, parity violation in atomic transitions,
neutrino-electron scattering, coherent $\pi^0$ production on nuclei by
neutrinos, and detailed measurements of $W$ and $Z$ properties.  Let us
take as an example the scattering of leptons on quarks to see how they
provide a value of $\sin^2 \theta$.  In the next subsection we shall turn to
the properties of the $Z$ bosons, which are now the source of the
most precise information.

\begin{figure}[t]
\centerline{\includegraphics[height=4.5in]{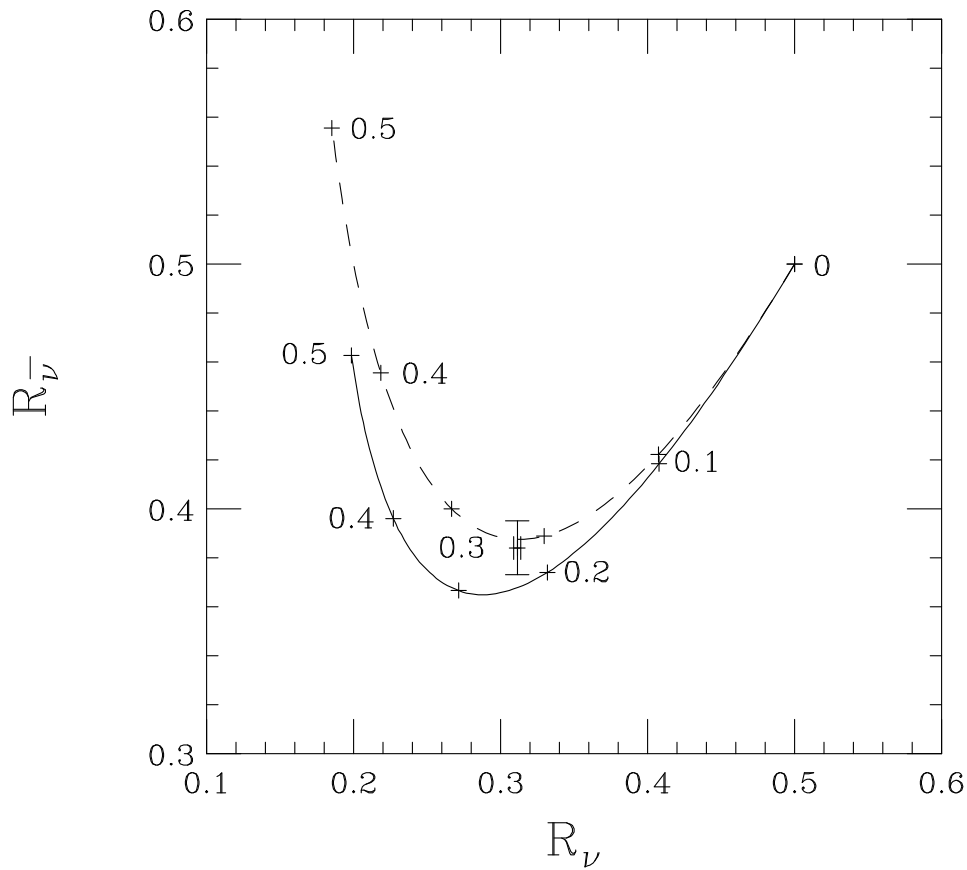}}
\caption{The Weinberg-Salam ``nose'' depicting the relation between $R_\nu$
and $R_{\nb}$.  The solid line corresponds to $r = 0.429$, close to the actual
situation; the dashed line corresponds to the idealized case $r = 1/3$ in which
antiquarks in the nucleon are neglected.  The plotted point with error bars
corresponds to the average of measured values.
\label{fig:nose}}
\end{figure}

One measures quantities
\beq
R_\nu \equiv \frac{\sigma(\nu A \to \nu + \ldots)}
{\sigma(\nu A \to \mu^- + \ldots)}~~~,
~~R_{\nb} \equiv \frac{\sigma(\nb A \to \nb + \ldots)}
{\sigma(\nb A \to \mu^+ + \ldots)}~~~.
\eeq
These ratios may be calculated in terms of the weak Hamiltonians (\ref{eqn:4f})
and (\ref{eqn:HNC}).  It is helpful to note that for states of the same
helicity ($L$ or $R$, standing for left-handed or right-handed) scattering on
one another, the differential cross section is a constant:
\beq
\frac{d \sigma}{d \Omega}(RR \to RR) = \frac{d \sigma}{d \Omega}(LL \to LL)
= \frac{\sigma_0}{4 \pi}~~~,
\eeq
where $\sigma_0$ is some reference cross section, while for states of
opposite helicity,
\beq
\frac{d \sigma}{d \Omega}(RL \to RL) = \frac{d \sigma}{d \Omega}(LR \to LR)
= \frac{\sigma_0}{4 \pi} \left( \frac{1 + \cos \tcm}{2}
\right)^2~~~.
\eeq
Thus
\beq \label{eqn:sigs}
\sigma(RR \to RR) = \sigma(LL \to LL) = 3 \sigma(RL \to RL) = 3 \sigma
(LR \to LR)~~~.
\eeq
We first simplify the calculation by assuming the numbers of protons and
neutrons are equal in the target nucleus, and neglecting the effect of
antiquarks in the nucleon.  (We shall use the shorthand $\nu = \nu_\mu$ and
$\nb = \bar \nu_\mu$.)  Then
\beq
R_\nu = \frac{\sigma(\nu u \to \nu u) + \sigma(\nu d \to \nu d)}
{\sigma(\nu d \to \mu^- u)}~~,~~~
R_{\nb} = \frac{\sigma(\nb u \to \nb u) + \sigma(\nb d \to \nb d)}
{\sigma(\nb u \to \mu^+ d)}~~~.
\eeq
One can write the effective Hamiltonian (\ref{eqn:HNC}) in the form
$$
\cH^{NC}_{\nu q} = \frac{G_F}{\s} [\nb \gm (1 - \gf) \nu ][ \bar u \gM
(1 - \gf) u \epsilon_L(u)
$$
\beq
+ \bar u \gM (1 + \gf) u \epsilon_R(u) + \bar d \gM (1 - \gf) d \epsilon_L(d)
+ \bar d \gM (1 + \gf) d \epsilon_R(d)]~~~,
\eeq
where
\beq
\epsilon_L(u) = \frac{1}{2} - \frac{2}{3} \sin^2 \theta~~,~~~
\epsilon_R(u) = - \frac{2}{3} \sin^2 \theta~~~,
\eeq
\beq
\epsilon_L(d) = - \frac{1}{2} + \frac{1}{3} \sin^2 \theta~~,~~~
\epsilon_R(d) =  \frac{1}{3} \sin^2 \theta~~~.
\eeq
Taking account of the relations (\ref{eqn:sigs}), one finds
\beq
R_\nu = [\epsilon_L(u)]^2 + \frac{1}{3}[\epsilon_R(u)]^2
     + [\epsilon_L(d)]^2 + \frac{1}{3}[\epsilon_R(d)]^2~~~,
\eeq
\beq
R_{\nb} = [\epsilon_L(u)]^2 + 3[\epsilon_R(u)]^2
     + [\epsilon_L(d)]^2 + 3[\epsilon_R(d)]^2~~~,
\eeq
where we have used the fact that $\sigma(\nu d \to \mu^- d)
= 3 \sigma(\nb u \to \mu^+ d)$.  The results are
\beq \label{eqn:noa}
R_\nu = \frac{1}{2} - \sin^2 \theta + \frac{20}{27} \sin^4 \theta~~,~~~
R_{\nb} = \frac{1}{2} - \sin^2 \theta + \frac{20}{9} \sin^4 \theta~~~.
\eeq
If we consider also the antiquark content of nucleons, this result may be
generalized (Llewellyn Smith 1983) by defining
\beq
r \equiv \frac{\sigma(\nb N \to \mu^+ X)}{\sigma(\nu N \to \mu^- X)}~~~.
\eeq
Instead of (\ref{eqn:noa}) one then finds
\beq \label{eqn:aq}
R_\nu = \frac{1}{2} - \sin^2 \theta + \frac{5}{9}(1 + r) \sin^4 \theta~~,~~~
R_{\nb} = \frac{1}{2} - \sin^2 \theta + \frac{5}{9}(1 + \frac{1}{r})
\sin^4 \theta~~~.
\eeq
Some experimental values of $R_\nu$, $R_{\nb}$, and $r$ are shown in Table
\ref{tab:nus} (Conrad \ite~1998). The relation between $R_\nu$
and $R_{\nb}$ as a function of $\sin^2 \theta$ is plotted in Figure
\ref{fig:nose}.  This result has a couple of interesting features.

\begin{table}
\caption{Neutrino neutral-current parameters.
\label{tab:nus}}
\bigskip
\begin{center}
\begin{tabular}{|c c c c|} \hline
Experiment & $R_\nu$ & $R_{\nb}$ & $r$ \\ \hline
CHARM & $0.3091 \pm 0.0031$ & $0.390 \pm 0.014$ & $0.456 \pm 0.011$ \\
CDHS  & $0.3135 \pm 0.0033$ & $0.376 \pm 0.016$ & $0.409 \pm 0.014$ \\ \hline
Average & $0.3113 \pm 0.0023$ & $0.384 \pm 0.011$ & $0.429 \pm 0.011$ \\ \hline
\end{tabular}
\end{center}
\end{table}

The observed $R_{\nb}$ is very close to its minimum possible value of
less than 0.4.  Initially this made the observation of neutral currents quite
challenging.  Note that $R_\nu$ is even smaller.  Its value provides the
greatest sensitivity to $\sin^2 \theta$.  It is
also more precisely measured than $R_{\nb}$ (in part, because
neutrino beams are easier to achieve than antineutrino beams).
The effect of $r$ on the determination of $\sin^2 \theta$ is relatively mild.

A recent determination of $\sin^2 \theta$ (Zeller \ite~1999), based on a
method proposed by Paschos and Wolfenstein (1973), makes use of the ratio

\beq
R^- \equiv \frac{\sigma(\nu N \to \nu X) - \sigma(\nb N \to \nb X)}
{\sigma(\nu N \to \mu^- X) - \sigma(\nb N \to \mu^+ X)}
= \frac{R_\nu - r R_{\nb}}{1 - r} = \frac{1}{2} - \sst~~~.
\eeq
In these differences of neutrino and antineutrino cross sections, effects
of virtual quark-antiquark pairs in the nucleon (``sea quarks,'' as opposed
to ``valence quarks'') cancel one another, and an important systematic error
associated with heavy quark production (as in $\nu s \to \mu^- c$) is greatly
reduced.  The result is
\beq
\sst^{~(\rm on-shell)} = 0.2253 \pm 0.0019({\rm stat.}) \pm
 0.0010({\rm syst.})~~~,
\eeq
which implies a $W$ mass
\beq
M_W \equiv M_Z \cos \theta^{~(\rm on-shell)} = 80.21 \pm 0.11~\g~~~.
\eeq
The ``on-shell'' designation for $\sst$ is necessary when discussing
higher-order electroweak radiative corrections, which we shall do in the next
Section.

[Note added:  a more recent analysis by Zeller \ite~(2001) finds
\beq
\sst^{~(\rm on-shell)} = 0.2277 \pm 0.0014({\rm stat.}) \pm 0.0008({\rm syst.})
~~~,
\eeq
equivalent to $M_W = 80.136 \pm 0.084$ GeV.  Incorporation of this result into
the electroweak fits described in the next Section is likely to somewhat relax
constraints on the Higgs boson mass:  See Rosner (2001).]
\subsection{$Z$ and top quark properties}

We have already noted the prediction and measurement of the $W$ mass and
width.  The $Z$ mass and width are very precisely determined by studying
the shape of the cross section for electron-positron annihilation as
one varies the energy across the $Z$ pole.  The results (LEP Electroweak
Working Group [LEP EWWG] 2001) are
\beq \label{eqn:mgz}
M_Z = 91.1875 \pm 0.0021~\g~~,~~~\Gamma_Z = 2.4952 \pm 0.0023~\g~~~.
\eeq
In much of the subsequent discussion we shall make use of the very precise
value of $M_Z$ as one of our inputs to the electroweak theory; the two
others, which will suffice to specify all parameters at lowest order of
perturbation theory, will be the Fermi coupling constant $G_F = 1.16637(1)
\times 10^{-5}~\g^{-2}$ and the electromagnetic fine-structure constant,
evolved to a scale $M_Z^2$: $\alpha^{-1~(\MSb)}(M_Z^2) = 128.933 \pm 0.021$
(Davier and H\"ocker 1998).  This last quantity depends for its determination
upon a precise evaluation of hadronic contributions to vacuum polarization,
and is very much the subject of current discussion.

The relative branching fractions of the $Z$ to various final states may be
calculated on the basis of Eq.~(\ref{eqn:Zint}).  One may write this expression
as
\beq \label{eqn:Zff}
\cL_{Z f \bar f} = \gz \bar f \Zs [(1 - \gf) a_L + (1 + \gf) a_R ] f~~~.
\eeq
The values of $a_L$ and $a_R$ for each fermion are shown in Table \ref{tab:gz}.

The partial width of $Z$ into $f \bar f$ is
\beq
\Gamma(Z \to f \bar f) = \frac{4 G_F}{3 \pi \s}M_Z^3 (a_L^2 + a_R^2) n_c~~~,
\eeq
where $n_c$ is the number of colors of fermions $f$:  1 for leptons, 3 for
quarks.

The predicted partial width for each $Z \to \nu \nb$ channel is independent
of $\sst$:
\beq
\Gamma(Z \to \nu \nb) = \frac{G_F}{\s} \frac{M_Z^3}{12 \pi} = 165.9~\m
\eeq
using the observed value of $M_Z$.  The partial decay rates to other channels
are expected to be in the ratios
$$
\nu \nb:~e^+ e^-:~u \bar u:~d \bar d = 
$$
\beq
1: 1 - 4 \sst + 8 \sft~:~3 - 8 \sst + \frac{32}{3} \sft~:~
3 - 4 \sst + \frac{8}{3} \sft~~~,
\eeq
or 1: 0.503: 1.721: 2.218 for $\sst = 0.231$.  A small kinematic correction
for the non-zero $b$ quark mass leads to a suppression factor
\beq
\Phi_{b \bar b} = \left( 1 - \frac{4 m_b^2}{M_Z^2} \right)^{1/2}
\left[ f_V \left( 1 + \frac{2 m_b^2}{M_Z^2} \right)
+ f_A \left( 1 - \frac{4 m_b^2}{M_Z^2} \right) \right]~~~,
\eeq
where $f_V$ and $f_A = 1 - f_V$ are the relative fractions of the
partial decay width proceeding via the vector ($\sim a_L + a_R$) and
axial-vector ($\sim a_L -a_R$) couplings.  For $\sst = 0.23$,
$f_V \simeq 1/3$, $f_A \simeq 2/3$, and $\Phi_{b \bar b} \simeq 0.988$.
A further correction to $\Gamma(Z \to b \bar b)$, important for the precise
determinations in the next Section, is associated with loop graphs associated
with top quark exchange (see the review by Chivukula 1995), and is of the
same size, about 0.988.  Taking a correction factor $(1 + \alpha_S/\pi)$ with
$\alpha_S(M_Z^2) = 0.12$ for the hadronic partial widths of the $Z$, we then
predict the contributions to $\Gamma_Z$ listed in Table \ref{tab:gz}.  (The $t
\bar t$ channel is, of course, kinematically forbidden.)

\begin{table}
\caption{Contributions to $\Gamma_Z$ predicted in lowest-order electroweak
theory (including leading-order QCD corrections to hadronic channels).  Here
we have taken $\sst = 0.231$ and $\alpha_S(M_Z^2) = 0.12$.
\label{tab:gz}}
\begin{center}
\begin{tabular}{|c c c c c c|} \hline
Channel    & $a_L$ & $a_R$   & Partial & Number of & Subtotal \\
           &       &         & width (MeV) & channels & (MeV) \\ \hline
$\nu \nb$  &  $\frac{1}{4}$ & 0 & 165.9 & 3 & 498 \\
$l \bar l$ & $\frac{1}{2} \left( - \frac{1}{2} + \sst \right)$ 
  & $\frac{1}{2}\sst$ & 83.4  & 3 & 250 \\
$u \bar u,~c \bar c$ & $\frac{1}{2} \left(\frac{1}{2} - \frac{2}{3} \sst
 \right)$
& $\frac{1}{2} \left( - \frac{2}{3} \sst \right)$ & 296.5 & 2 & 593 \\
$d \bar d,~s \bar s$ & $\frac{1}{2} \left( - \frac{1}{2} + \frac{1}{3} \sst
  \right)$ & $\frac{1}{2} \left( \frac{1}{3} \sst \right)$ & 382.1 & 2 & 764 \\
$b \bar b$           & $\frac{1}{2} \left( - \frac{1}{2} + \frac{1}{3} \sst
  \right)$ & $\frac{1}{2} \left( \frac{1}{3} \sst \right)$ & 372.8 & 1 & 373 \\
\hline
Total      & &               &       &   & 2478 \\ \hline
\end{tabular}
\end{center}
\end{table}

The measured $Z$ width (\ref{eqn:mgz}) is in qualitative agreement with the
prediction, but above it by about 0.7\%.  This effect is a signal of
higher-order
electroweak radiative corrections such as loop diagrams involving the top quark
and the Higgs boson.  Similarly, the observed value of $\Gamma(Z \to e^+ e^-)$,
assuming lepton universality, is $83.984 \pm 0.086$ MeV, again higher by 0.7\%
than the predicted value of 83.4 MeV.  We shall return to these effects in the
next Section.

The width of the $Z$ is sensitive to additional $\nu \nb$ pairs.  Clearly
there is no room for an additional light pair coupling with full strength.
Taking account of all precision data and electroweak corrections, the latest
determination of the ``invisible'' width of the $Z$ (see the compilations by
the LEP EWWG 2001 and by Langacker 2001)
fixes the number of ``light'' neutrino species as $N_\nu = 2.984 \pm 0.008$.

The $Z$ is produced copiously in $e^+ e^-$ annihilations when the
center-of-mass energy $\sqrt{s}$ is tuned to $M_Z$.  The Stanford Linear
Collider (SLC) and the Large Electron-Positron Collider at CERN
(LEP) exploited this feature.
The cross section of production of a final state $f$ near the resonance,
ignoring the effect of the virtual photon in the direct channel, should be
\beq
\sigma(e^+ e^- \to Z^0 \to f) = 12 \pi \frac{\Gamma(Z \to e^+ e^-)
 \Gamma(Z \to f)}{(s - M_Z^2)^2 + M_Z^2 \Gamma_Z^2}~~~.
\eeq
At resonance, the peak total cross section should be $\sigma_{\rm peak} =
12 \pi \Bee/M_Z^2 \simeq 59.4$ nb, corresponding to
\beq
R_{\rm pk} \equiv \frac{\sigma(e^+ e^- \to Z^0 \to {\rm all})}
{\sigma(e^+ e^- \to \mu^+ \mu^-)} = \frac{9 \Bee}{\alpha^2} \simeq 5000~~~.
\eeq
Here $\Bee \equiv \Gamma(Z^0 \to e^+ e^-)/\Gamma_Z \simeq 3.37\%$.
This is a spectacular value of $R$, which is only a few units
in the range of lower-energy $e^+ e^-$ colliders.  Of course, not all
of the cross section at the $Z$ peak is visible:  Nearly 12 nb goes into
neutrinos!  Another 6 nb goes into charged lepton pairs, leaving $\sigma_{\rm
peak,~hadrons} = 41.541 \pm 0.037$ nb.

We close this subsection with a brief discussion of spin-dependent
asymmetries at the $Z$.  These are some of the most powerful sources of
information on $\sst$.  They have been measured both at LEP (through
forward-backward asymmetries) and at SLC (through the use of polarized
electron beams).

The discussion makes use of an elementary feature of vector- and axial-vector
couplings.  Processes involving such couplings to a real or virtual
particle (such as the $Z$) always conserve chirality.  In the direct-channel
reactions $e^- e^+ \to Z \to f \bar f$ this means that a left- (right-)handed
electron only interacts with a right- (left-) handed positron, and if the
final fermion $f$ is left- (right-)handed then the final antifermion $\bar f$
will be right- (left-) handed.  Moreover, such reactions have characteristic
angular distributions, with
\beq
\frac{d \sigma(e^-_{L} \to f_{L})}{d \Omega} = \sigma_0 (a_L^e)^2 (a_L^f)^2
\left( \frac{1 + \cos \tcm}{2} \right)^2~~~;
\eeq
\beq
\frac{d \sigma(e^-_{R} \to f_{R})}{d \Omega} = \sigma_0 (a_R^e)^2 (a_R^f)^2
\left( \frac{1 + \cos \tcm}{2} \right)^2~~~;
\eeq
\beq
\frac{d \sigma(e^-_{L} \to f_{R})}{d \Omega} = \sigma_0 (a_L^e)^2 (a_R^f)^2
\left( \frac{1 - \cos \tcm}{2} \right)^2~~~;
\eeq
\beq
\frac{d \sigma(e^-_{R} \to f_{L})}{d \Omega} = \sigma_0 (a_R^e)^2 (a_L^f)^2
\left( \frac{1 - \cos \tcm}{2} \right)^2~~~;
\eeq
where $\sigma_0$ is some common factor, and the $a_{L,R}$ are given in
Table \ref{tab:gz}.  Several asymmetries can be formed using these results.

The {\it polarized electron left-right asymmetry} compares the cross sections
for producing fermions using right-handed and left-handed polarized electrons,
as can be produced and monitored at the SLC.  The cross section asymmetry is
given by
$$
A^e_{LR}({\rm hadrons}) \equiv
 \frac{\sigma(e^-_L e^+ \to {\rm hadrons})
- \sigma(e^-_R e^+ \to {\rm hadrons})}
{\sigma(e^-_L e^+ \to {\rm hadrons})
+ \sigma(e^-_R e^+ \to {\rm hadrons})}
$$
\beq
= \frac{(a_L^e)^2 - (a_R^e)^2}{(a_L^e)^2 + (a_R^e)^2}
= \frac{1 - 4 \sst}{1 - 4 \sst + 8 \sft}~~~.
\eeq
The measured value (LEP EWWG 2001) $A^e_{LR}({\rm hadrons})=0.1514 \pm 0.0022$
corresponds to $\sst = 0.23105 \pm 0.00028$ using this formula.  (We shall
discuss small corrections in the next Section.) 

The {\it forward-backward asymmetry} in $e^+ e^- \to f \bar f$ uses the
fact that
\beq
\left( \int_0^1 - \int_{-1}^0 \right) d \cos \theta (1 \pm \cos \theta)^2
= \pm 2~~,~~~
\left( \int_0^1 + \int_{-1}^0 \right) d \cos \theta (1 \pm \cos \theta)^2
= \frac{8}{3}~~~,
\eeq
so that
$$
A_{FB}^f \equiv \frac{\sigma(e^- e^+ \to f \bar f)_{\rm fwd}
- \sigma(e^- e^+ \to f \bar f)_{\rm back}}
{\sigma(e^- e^+ \to f \bar f)_{\rm fwd}
+ \sigma(e^- e^+ \to f \bar f)_{\rm back}}
$$
\beq
= \frac{3}{4} \frac{(a_L^e)^2 - (a_R^e)^2}{(a_L^e)^2 + (a_R^e)^2}
\frac{(a_L^f)^2 - (a_R^f)^2}{(a_L^f)^2 + (a_R^f)^2}
= \frac{3}{4}A^e_{LR} A^f_{LR}~~~
\eeq

These quantities can be measured not only for charged leptons, but also
for quarks such as the $b$, whose decays allow for a distinction to
be made (at least on a statistical basis) between $b$ and $\bar b$.

The discovery of the top quark by the CDF (1994) and D0 (1995) Collaborations
culminated nearly two decades of detector and machine work at the Fermilab
Tevatron.  A ring of superconducting magnets was added to the 400 GeV Fermilab
accelerator, more than doubling its energy.  Low-energy rings were added
to accumulate and store antiprotons, which were then injected into the
superconducting ring and made to collide with oppositely-directed protons
at a center-of-mass energy of 1.8 TeV.  The top quarks were produced in
the reaction $p \bar p \to t \bar t + \ldots.$  

The top quark's mass is currently measured to be $m_t = 174.3 \pm 5.1~\g$.
It couples mainly to $b$, as expected in the pattern of couplings discussed
in Section 3.  One determination (see Gilman, Kleinknecht, and Renk 2000
for details) is that
\beq
\frac{|V_{tb}|^2}{|V_{td}|^2 + |V_{ts}|^2 + |V_{tb}|^2} = 0.99 \pm 0.29~~~.
\eeq
This result makes use of the measured fraction of the decays
$t \to b e^+ \nu_e$ in top semileptonic decays.

The top quark is the only quark heavy enough to decay directly to another
quark (mainly $b$) and a real $W$.  Its decay width is
\beq
\Gamma(t \to W^+ b) = \frac{G_F m_t^3}{8 \pi \s} \left[ \left( 1 -
\frac{M_W^2}{m_t^2} \right)^2 \left( 1 + 2 \frac{M_W^2}{m_t^2} \right)
\right] \simeq 1.53~\g~~~.
\eeq
This is larger than the typical spacing between quarkonium levels (see Figures
\ref{fig:chm} and \ref{fig:ups}), and so there is not expected to be a rich
spectroscopy of $t \bar t$ levels, but only a mild enhancement near threshold
of the reaction $e^+ e^- \to t \bar t$, associated with the production of the
1S level (Kwong 1991, Strassler and Peskin 1991).  A good review of present and
anticipated top quark physics is given by Willenbrock (2000).

\section{Higgs boson and beyond}

\subsection{Searches for a standard Higgs boson}

Let us assume that all quark and lepton masses and all $W$ and $Z$ masses
arise from the vacuum expectation value of a single Higgs boson: $\vev
{\phi^0} = v/\s$, where the strength of the Fermi coupling requires
$v = 246$ GeV.  The Yukawa coupling $g_{Yf}$ (\ref{eqn:phi}) for a fermion
$f$ is related to the fermion's mass:  $g_{Yf} = \s m_f/v$.  (It is a
curious feature of the top quark's mass that, within present errors,
$g_{Yt} = 1$.  Since fermion masses ``run'' with scale $\mu$, it is not clear
how fundamental this relation is.)  Those quarks with the greatest mass then
are expected to have the greatest coupling to the physical Higgs boson $H = \s
\phi^0 - v$.  (Here we use $H$ to denote the field represented by $\eta$
in the previous Section.)

The Higgs boson has a well-defined coupling to $W$'s and $Z$'s as a result
of the discussion in the previous Section.  The term $(\bDm \phi)^\dag
(\bDM \phi)$ in the Lagrangian leads to
\beq \label{eqn:HVV}
\cL_{HWW} = g H M_W (W_\mu^- W^{\mu +} )~~,~~~
\cL_{HZZ} = \frac{\gz H M_Z}{2} (Z_\mu Z^\mu)~~~.
\eeq
To lowest order, one find $\cL_{HZ \gamma} = \cL_{H \gamma \gamma} = 0$.

Processes involving the couplings (\ref{eqn:HVV}) include $q \bar q \to
W_{\rm virtual} \to W + H$ and especially
\beq
e^+ e^- \to Z_{\rm real~or~virtual} \to Z_{\rm virtual~or~real} + H~~~,
\eeq
where the final $Z^0$ can be detected (for example) via its decay to
$e^+ e^-$, $\mu^+ \mu^-$, or even its existence inferred from its $\nu
\nb$ decay.  For a virtual intermediate and real final $Z$, the cross
section (Quigg 1983) is
\beq
\sigma(e^+ e^- \to Z H) = \frac{\pi \alpha^2 (p^{*2} + 3 M_Z^2)}{24 \sft \cft
(M_Z^2 - s)^2} \left( 1 - 4 \sst + 8 \sft \right) \frac{2 p^*}{\sqrt{s}}~~~,
\eeq
 where $p^*$ is the final c.m. 3-momentum.  This cross section behaves as
$1/s$ for large $s$ (as does any cross section for production of $q \bar q$,
$\mu^+ \mu^-, \ldots$), so that as $s \to \infty$,
\beq
\frac{\sigma(e^+ e^- \to Z H)}{\sigma(e^+ e^- \to \mu^+ \mu^-)}
\to \frac{1 - 4 \sst + 8 \sft}{128 \sft \cft} \simeq \frac{1}{8}~~~.
\eeq
At very high energies, the Higgs boson can be produced by means of $W^+
W^-$ and $ZZ$ fusion; the (virtual) $W$'s and $Z$'s can be produced
in either hadron-hadron or lepton-lepton collisions.  A further proposal
for producing Higgs bosons is by means of muon-muon collisions.

For Higgs bosons far above $WW$ and $ZZ$ threshold, one expects (Eichten
\ite~1984)
\beq
\Gamma_H = \Gamma(H \to W^+ W^-) + \Gamma(H \to ZZ) = \frac{3 G_F}{16 \pi \s}
M_H^3 \simeq 60~\g~ \left( \frac{M_H}{500~\g} \right)^3~~~,
\eeq
as one can show with the help of (\ref{eqn:HVV}).  The longitudinal degrees
of freedom of the $W$ and $Z$ provide the dominant contribution to the
decay width in this limit.  For $M_H = 1$ TeV, this relation implies that
the Higgs boson's width will be nearly 500 GeV.  Such a broad object will
be difficult to separate from background.  However, mixed signals for a
a much lighter Higgs boson have already been received at LEP.

At the very highest LEP energies attained, $\sqrt{s} \le 209~\g$, the four
LEP collaborations ALEPH, DELPHI, L3, and OPAL have presented combined results
(LEP Higgs Working Group 2001) which may be interpreted either as a
lower limit on the Higgs boson mass of $114.1$ GeV, or as a weak signal for a
Higgs boson of mass $M_H \simeq 115.6$ GeV produced by the above process.  This
latter interpretation is driven in large part by the ALEPH data sample
(Barate \ite~2001).  The main decay mode of a Higgs boson in this mass range is
expected to be $b \bar b$, with $\tau^+ \tau^-$ taking second place.

LEP now has ceased operation in order to make way for the Large Hadron
Collider (LHC), which will collide 7 TeV protons with 7 TeV protons and
should have no problem producing such a boson.  The LHC is scheduled to
begin operation in 2006.  In the meantime, the Fermilab Tevatron has resumed
$p \bar p$ collider operation after a hiatus of 5 years.  Its scheduled
``Run II'' is initially envisioned to provide an integrated luminosity
of 2 fb$^{-1}$, which is thought to be sufficient to rival the sensitivity of
the LEP search (Carena \ite~2000), making use of the subprocess $q \bar q \to
W_{\rm virtual} \to W + H$.  With 10 fb$^{-1}$ per detector, a benchmark
goal for several years of running with luminosity improvements, it should be
possible to exclude a Higgs boson with standard couplings nearly up to the
$Z Z$ threshold of 182 GeV, and to see a $3 \sigma$ signal if $M_H \le 125$
GeV.  Other scenarios, including the potential for discovering the Higgs
boson(s) of the Minimal Supersymmetric Standard Model (MSSM) are given
by Carena \ite~(2000).  Meanwhile, we shall turn to the wealth of precise
measurements
of electroweak properties of the $Z$, $W$, top quark, and lighter fermions as
indirect sources of information about the Higgs boson and other new physics.

\subsection{Precision electroweak tests}

We have calculated processes to lowest electroweak order in the previous
Section, with the exception that we took account of vacuum polarization
in the photon propagator, which leads to a value of $\alpha^{-1}$ closer
to 129 than to 137.037 at the mass scale of the $Z$.  The lowest-order
description was found to be adequate at the percent level, but many electroweak
measurements are now an order of magnitude more precise.  As one example, we
found that the predicted total and leptonic $Z$ widths both fell short of
the corresponding experimental values by about 0.7\%.  Higher-order
electroweak corrections are needed to match the precision of the new data.
These corrections can shed fascinating light on new physics, as well as
validating the original motivation for the electroweak theory (which was
to be able to perform higher-order calculations).

We shall describe a language introduced by Peskin and Takeuchi (1990) for
precise electroweak tests which
allows the constraints associated with nearly every observable to be
displayed on a two-dimensional plot.  The Standard Model implies a particular
locus on this plot for every value of $m_t$ and $M_H$, so one can see how
observables can vary with $m_t$ (not much, now that $m_t$ is so well
measured) and $M_H$.  Moreover, one can spot at a glance if a particular
measurement is at variance with others; this can either signify physics
outside the purview of the two-dimensional plot, or systematic experimental
error.

The corrections which fall naturally into the two-dimensional description are
those known as {\it oblique corrections}.  The name stems from the fact
that they do not directly affect the fermions participating in the
processes of interest, but appear as vacuum polarization corrections in
gauge boson propagators.  In that sense processes which are sensitive to
oblique corrections have a broad reach for discovering new physics, since they
do not rely on a new particle's having to couple directly to the external
fermion in question.

The oblique correction first identified by Veltman (1977), still the most
important, is that due to top quarks in $W$ and $Z$ boson propagators.
The large splitting between the top and bottom quarks' masses violates
a {\it custodial SU(2)} symmetry (Sikivie \ite~1980) responsible for preserving
the tree-level relation $M_W = M_Z \cos \theta$ mentioned in the previous
Section.  As a result, an effect is generated which is equivalent to having a
Higgs {\it triplet} vacuum expectation value.

For the photon, gauge invariance prohibits contributions quadratic in fermion
masses, but for the $W$ and $Z$, no such prohibition applies.
The vacuum polarization diagrams with $W^+ \to t \bar b \to W^+$ and
$Z \to (t \bar t, b \bar b) \to Z$ lead to a modification of the
relation between $G_F$, coupling constants, and $M_Z$ for neutral-current
exchanges:
\beq \label{eqn:GFr}
\frac{G_F}{\s} = \frac{g^2 + {g'}^2}{8 M_Z^2} ~~~\to~~~
\frac{G_F}{\s} \rho = \frac{g^2 + {g'}^2}{8 M_Z^2}~~~,~~
\rho \simeq 1 + \frac{3 G_F m_t^2}{8 \pi^2 \s}~~~.
\eeq
The $Z$ mass is now related to the weak mixing angle by
\beq
M_Z^2 = \frac{\pi \alpha}{\s G_F \rho \sst \cst}~~~,
\eeq
where we have omitted some small terms logarithmic in $m_t$.  A precise
measurement of $M_Z$ now specifies $\theta$ only if $m_t$ is known, so
$\theta = \theta(m_t)$ and hence $M_W^2 = \pi \alpha/(\s G_F \sst)$ is also
a function of $m_t$.

The factor of $\rho$ in (\ref{eqn:GFr}) will multiply every neutral-current
four-fermion interaction in the electroweak theory.  Thus, for, example,
cross sections for charge-preserving interactions of neutrinos with matter
will be proportional to $\rho^2$, while parity-violating neutral-current
amplitudes (to be discussed below) will be proportional to $\rho$.  Partial
decay widths of the $Z$, since they involve the combination $g^2 + {g'}^2$,
will be proportional to $\rho$.  A large part of the 0.7\% correction
mentioned previously is due to $\rho > 1$.  The observed values of $M_W/M_Z
= \rho \cos \theta$ and $\sst$ also are much more compatible with each other
for a value of $\rho$ exceeding 1 by about a percent.

The $W$ and $Z$ propagators are also affected by virtual Higgs-boson
states due to the couplings (\ref{eqn:HVV}).  Small corrections, logarithmic
in $M_H$, affect all the observables, but notably $\rho$.

In order to display dependence of electroweak observables on such quantities as
the top quark and Higgs boson masses $m_t$ and $M_H$, we choose to expand the
observables about ``nominal'' values calculated by Marciano (2000) for specific
$m_t$ and $M_H$.  We thereby bypass a discussion of ``direct'' radiative
corrections which are independent of $m_t,~M_H$, and new particles.  We isolate
the dependence on $m_t,~M_H$, and new physics arising from ``oblique''
corrections associated with loops in the $W$ and $Z$ propagators.

For $m_t = 174.3$ GeV, $M_H = 100$ GeV, the measured value of $M_Z$ leads to a
nominal expected value of $\sin^2 \theta_{\rm eff} = 0.2314$.  In what follows
we shall interpret the effective value of $\sin^2 \theta$ as that measured via
leptonic vector and axial-vector couplings: $\sin^2 \theta_{\rm eff} \equiv
(1/4)(1 - [g_V^{\ell}/g_A^{\ell}])$.

Defining the parameter $T$ by $\Delta \rho \equiv \alpha T$, we find
\begin{equation} \label{eqn:Teq}
T \simeq \frac{3}{16 \pi \sin^2 \theta} \left[ \frac{m_t^2 - (174.3
~{\rm GeV})^2}{M_W^2} \right] - \frac{3}{8 \pi \cos^2 \theta}
\ln \frac{M_H}{100~{\rm GeV}} ~~~.
\end{equation}
The weak mixing angle $\theta$, the $W$ mass, and other electroweak observables
now depend on $m_t$ and $M_H$.

The weak charge-changing and neutral-current interactions are probed under a
number of different conditions, corresponding to different values of momentum
transfer.  For example, muon decay occurs at momentum transfers small with
respect to $M_W$, while the decay of a $Z$ into fermion-antifermion pairs
imparts a momentum of nearly $M_Z/2$ to each member of the pair. Small
``oblique'' corrections, logarithmic in $m_t$ and $M_H$, arise from
contributions of new particles to the photon, $W$, and $Z$ propagators. Other
(smaller) ``direct'' radiative corrections are important in calcuating actual
values of observables.

We may then replace the lowest-order relations between $G_F$, couplings, and
masses by
\beq
\frac{G_F}{\sqrt{2}} = \frac{g^2}{8 M_W^2} \left( 1 + \frac{\alpha S_W}{4
\sin^2 \theta} \right)~~~,~~~
\frac{G_F \rho}{\sqrt{2}} = \frac{g^2 + {g'}^2}{8M_Z^2} \left( 1 + \frac{\alpha
S_Z}{4 \sin^2 \theta \cos^2 \theta} \right)~~~,
\eeq
where $S_W$ and $S_Z$ are coefficients representing variation with momentum
transfer. Together with $T$, they express a wide variety of electroweak
observables in terms of quantities sensitive to new physics.  (The presence of
such corrections was noted quite early by Veltman 1977.)  The Peskin and
Takeuchi (1990) variable $U$ is equal to $S_W - S_Z$, while $S \equiv S_Z$.

Expressing the ``new physics'' effects in terms of deviations from nominal
values of top quark and Higgs boson masses, we have the expression
(\ref{eqn:Teq}) for $T$, while contributions of Higgs bosons and of possible
new fermions $U$ and $D$ with electromagnetic charges $Q_U$ and $Q_D$ to $S_W$
and $S_Z$, in a leading-logarithm approximation, are (Kennedy and Langacker
1990)
\beq \label{eqn:sz}
S_Z = \frac{1}{6 \pi} \left [
\ln \frac{M_H}{100~\g/c^2} + \sum N_C \left ( 1 - 4 \overline Q \ln
\frac{m_U}{m_D} \right ) \right ] ~~~,
\eeq
\beq \label{eqn:sw}
S_W = \frac{1}{6 \pi} \left [
\ln \frac{M_H}{100 ~\g/c^2} + \sum N_C \left ( 1 - 4 Q_D \ln \frac{m_U}{m_D}
\right ) \right ]~~.
\eeq
The expressions for $S_W$ and $S_Z$ are written for doublets of fermions with
$N_C$ colors and $m_U \geq m_D \gg m_Z$, while $\overline Q \equiv (Q_U + Q_D )
/2$. The sums are taken over all doublets of new fermions. In the limit $m_U =
m_D$, one has equal contributions to $S_W$ and $S_Z$. For a single Higgs boson
and a single heavy top quark, Eqs.~(\ref{eqn:sz}) and (\ref{eqn:sw}) become
$$
S_Z = \frac{1}{6 \pi} \left [ \ln \frac{M_H}{100~\g/c^2} - 2 \ln
\frac{m_t}{174.3~\g/c^2} \right ] ~,~~
$$
\beq
S_W = \frac{1}{6 \pi} \left [ \ln \frac{M_H}{100~\g/c^2} + 4 \ln
\frac{m_t}{174.3~\g/c^2} \right ] ~,
\eeq
where the leading-logarithm expressions are of limited validity for $M_H$
and $m_t$ far from their nominal values.  (We shall plot contours of $S$
and $T$ for fixed $m_t$ and $M_H$ values without making these approximations.)
A degenerate heavy fermion doublet
with $N_c$ colors thus contributes $\Delta S_Z = \Delta S_W = N_c/6 \pi$.
For example, in a minimal dynamical symmetry-breaking (``technicolor'')
scheme, with a single
doublet of $N_c = 4$ fermions, one will have $\Delta S = 2/3 \pi \simeq
0.2$.  This will turn out to be marginally acceptable, while many non-minimal
schemes, with large numbers of doublets, will be seen to be ruled out.

Many analyses of present electroweak data within the $S$, $T$ rubric are
available (e.g., Swartz 2001).  We shall present a ``cartoon'' version after
discussing possible extensions of the Higgs system.  Meanwhile we note briefly
a topic which will not enter that discussion.

The anomalous magnetic moment of the electron and muon have been measured
ever more precisely.  The latest measurement of the $\mu^+$ (Brown \ite~2001),
performed in a special storage ring at Brookhaven National Laboratory, gives
\beq
a_{\mu^+,{\rm exp}} \equiv \left( \frac{g-2}{2} \right)_{\mu^+} = 
11~659~202(14)(6) \times 10^{-10}~~ (1.3~{\rm ppm})~~~.
\eeq
The theoretical value (CPT invariance implies $a_{\mu^+} = a_{\mu^-}$) is
\beq
a_{\mu,{\rm th}} \simeq 11~659~177(7) \times 10^{-10}~~(0.6~{\rm ppm})~~~,
\eeq
the sum of $a_{\mu,{\rm QED}} = 11~658~470.56(0.29) \times 10^{-10}$
(0.025 ppm), $a_{\mu,{\rm weak}} = 15.1(0.4) \times 10^{-10}$ (0.03 ppm), and
$a_{\mu, {\rm had}} \simeq 691(7) \times 10^{-10}$ (0.6 ppm), where we
have incorporated a recently-discovered sign change in the hadronic
light-by-light scattering contribution (Knecht and Nyffeler 2001; Hayakawa
and Kinoshita 2001).  The difference,
\beq
a_{\mu^+,{\rm exp}} - a_{\mu^+,{\rm th}} = 25(17) \times 10^{-10}~~~,
\eeq
is not yet known precisely enough to test the expected weak contribution.
Results of analyzing a larger data sample are expected shortly.

\subsection{Multiple Higgs doublets and Higgs triplets}

There are several reasons for introducing a more complicated Higgs boson
spectrum.  Reasons for introducing separate Higgs doublets for
$u$-type and $d$-type quarks include higher symmetries following from
attempts to unify the strong and electroweak interactions, and supersymmetry.
We examine the simplest model with more than one Higgs doublet,
in which a single doublet couples to $d$-type quarks and charged leptons, and
a different doublet couples to $u$-type quarks.  This model turns out to
naturally avoid flavor-changing neutral currents associated with Higgs
exchange (Glashow and Weinberg 1977).

Let us denote by $\phi_u$ the Higgs boson coupling to $u$-type quarks and by
$\phi_d$ the boson coupling to $d$-type quarks and charged leptons.  We let
\beq
\vev{\phi_u} = v_u/\s~~,~~~\vev{\phi_d} = v_d/\s~~~.
\eeq
The contribution of $\phi_u$ and $\phi_d$ to $W$ and $Z$ masses comes from
\beq
\cL_K + (\bDm \phi_u)^\dag (\bDM \phi_u) +
(\bDm \phi_d)^\dag (\bDM \phi_d)~~~.
\eeq
We find the same $W_\mu^3 - B_\mu$ mixing pattern as before, and in fact
this pattern would remain the same no matter how many Higgs doublets were
introduced.  The parameters $v_u$ and $v_d$ may be related to the
quantity $v = 246$ GeV introduced earlier by $v_u^2 + v_d^2 = v^2$, whereupon
all previous expressions for $M_W$ and $M_Z$ remain valid.  One would have $v^2
 = \sum_i v_i^2$ for any number of doublets.

The quark and lepton couplings to Higgs doublets are enhanced if there are
multiple doublets.  Since $m_q = g_Y v_q/\s$ ($q = u$ or $d$) and $v_q < v$,
one has larger Yukawa couplings than in the standard single-Higgs model.
A more radical consequence, however, of multiple doublets in the \SUL~gauge
theory is that there are not enough gauge bosons to ``eat'' all the
scalar fields.  In a two-doublet model, five ``uneaten'' scalars remain:
two charged and three neutral.  The phenomenology of these is well-described
by Gunion \ite~(1990).

The prediction $M_Z = M_W/\cos \theta$ is specific to the assumption that only
Higgs doublets of \SUL~ exist.  [SU(2)$_L$ singlets which are neutral also
have $Y=0$, and do not affect $W$ and $Z$ masses.]  If triplets or higher
representations of SU(2) exist, the situation is changed.  We shall examine
two cases of triplets:  a complex triplet with charges (++,+,0) and one with
charges (+,0,--).

Consider first a complex triplet of the form
\beq
\Phi \equiv \left[ \begin{array}{c} \Phi^{++} \\ \Phi^+ \\ \Phi^0
\end{array} \right]~~,~~~  I_{3L} = \left\{ \begin{array}{c} +1 \\ 0 \\ -1
\end{array} \right.~~~.
\eeq
Since $Q = I_{3l} + \frac{Y}{2}$, one must have $Y = 2$ for this triplet.
In calculating $|\bDm \Phi|^2$ we will need the triplet representation for
weak isospin:
\beq
I_3=\left[ \begin{array}{c c c}1 & & \\ & 0 & \\ & & -1 \end{array} \right]~~,
~~~I_1 = \frac{1}{\s} \left[ \begin{array}{c c c} 0 & 1 & 0 \\ 1 & 0 & 1 \\
0 & 1 & 0 \end{array} \right]~~,~~~
I_2= \frac{i}{\s} \left[ \begin{array}{c c c} 0 & -1 & 0 \\ 1 & 0 & -1 \\
o & 1 & 0 \end{array} \right]~~~.
\eeq
The result, if $\vev{\Phi^0} = V_{1,-1}/\s$, is that
\beq
\vev{|\bDm \Phi|^2} = \frac{V^2_{1,-1}}{2} \left\{ \frac{g^2}{2} 
[(W^1)^2 + (W^2)^2] + (-g W^3 + g' B)^2 \right\}~~~.
\eeq
The same combination of $W^3$ and $B$ gets a mass as in the case of one or
more Higgs doublets, simply because we assumed that it was a neutral
Higgs field which acquired a vacuum expectation value.  Electromagnetic
gauge invariance remains valid; the photon does not acquire a mass.
However, the ratio of $W$ and $Z$ masses is altered.  In the presence
of doublets and this type of triplet, we find
\beq
M_W^2 = \frac{g^2}{4}(v^2 + 2 V^2_{1,-1})~~,~~~
M_Z^2 = \left( \frac{g^2 + {g'}^2}{4} \right) (v^2 + 4 V^2_{1,-1})~~~,
\eeq
so the ratio $\rho = (M_W/M_Z \cos \theta)^2$ is no longer 1, but becomes
\beq
\rho = \frac{v^2 + 2 V^2_{1,-1}}{v^2 + 4 V^2_{1,-1}}~~~.
\eeq
This type of Higgs boson thus leads to $\rho < 1$.

A complex triplet
\beq
\Phi \equiv \left[ \begin{array}{c} \Phi^+ \\ \Phi^0 \\
\Phi^- \end{array} \right]
\eeq
is characterized by $Y = 0$.  If we let $\vev{\Phi^0} = V_{1,0}/\s$, we find
by an entirely similar calculation, that
\beq
M_W^2 = \frac{g^2}{4}(v^2 + 4 V^2_{1,0})~~,~~~
M_Z^2 = \left( \frac{g^2 + {g'}^2}{4} \right) v^2~~~.
\eeq
Here we predict
\beq
\rho = 1 + \frac{4 V_{1,0}^2}{v^2}~~~,
\eeq
so this type of Higgs boson leads to $\rho > 1$.

We now examine a simple set of electroweak data (Rosner 2001), updating an
earlier analysis (Rosner 1999) which may be consulted for further references.
(See also Peskin and Wells 2001.)  We omit some data which provide similar
information but are less constraining.  Thus, we take only the observed values
of $M_W$ as measured at the Fermilab Tevatron and LEP-II, the leptonic width of
the $Z$, and the value of $\sin^2 \theta_{\rm eff}$ as measured in various
asymmetry experiments at the $Z$ pole in $e^+ e^-$ collisions.  We also include
parity violation in atoms, stemming from the interference of $Z$ and photon
exchanges between the electrons and the nucleus.  The most precise constraint
at present arises from the measurement of the {\it weak charge} (the coherent
vector coupling of the $Z$ to the nucleus), $Q_W = \rho(Z - N - 4 Z \sin^2
\theta)$, in atomic cesium.  The prediction $Q_W({\rm Cs}) = -73.19 \pm 0.13$
is insensitive to standard-model parameters once $M_Z$ is specified;
discrepancies are good indications of new physics. 

\begin{figure}
\centerline{\includegraphics[height=4.4in]{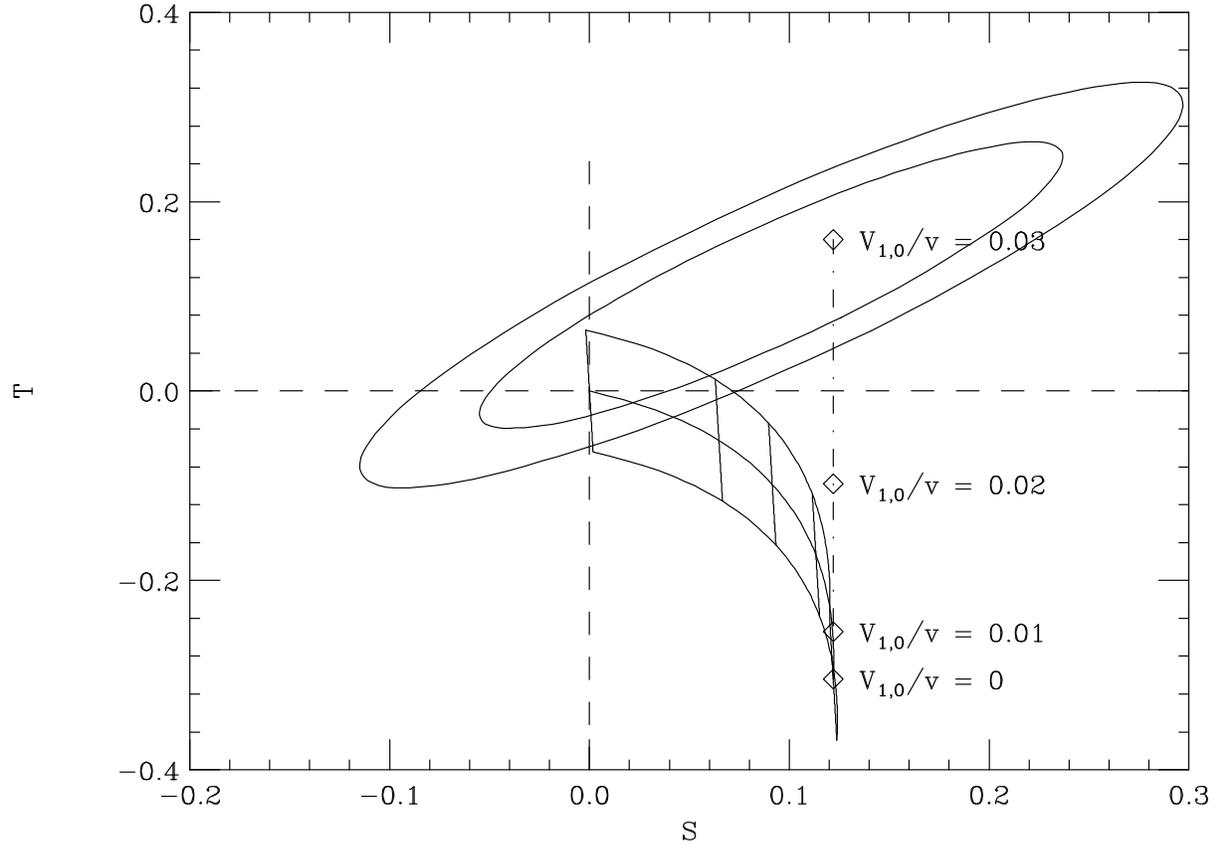}}
\caption{Regions of 68\% (inner ellipse) and 90\% (outer ellipse) confidence
level values of $S$ and $T$ based on the comparison of the theoretical and
experimental electroweak observables shown in Table \ref{tab:ST}.  Details
are given in the text.
\label{fig:ST}}
\end{figure}

The inputs, their nominal values for $m_t = 174.3$ GeV and $M_H = 100$ GeV, and
their dependences on $S$ and $T$ are shown in Table \ref{tab:ST}.
We do not constrain the top quark mass; we display its effect on
$S$ and $T$ explicitly.  Each observable specifies a pair of parallel lines
in the $S-T$ plane.  The leptonic width mainly constrains $T$; $\sef$
provides a good constraint on $S$ with some $T$-dependence; and $M_W$ lies
in between.  Atomic parity violation experiments constrain $S$ with almost no
$T$ dependence.  Although the errors on $S$ they entail are too large to have
much impact, we include them for illustrative purposes.  Since the slopes 
associated with constraints are very different, the resulting allowed region is
an ellipse, shown in Figure \ref{fig:ST}.  [Note added: Milstein and Sushkov
(2001) have noted that a correction due to the strong nuclear
field changes the central value of $Q_W({\rm Cs})$ in Table \ref{tab:ST} to
$\simeq -72.2$, while Dzuba {\it et al.} (2001) include this and further
corrections to obtain $Q_W = -72.39 \pm 0.58$.]

\begin{table}
\begin{center}
\caption{Electroweak observables described in fit.  References for atomic
physics experiment and theory are given by Rosner (2001). \label{tab:ST}}
\medskip
\begin{tabular}{c c c} \hline
Quantity      &   Experimental   &   Theoretical \\
              &      value       &    value      \\ \hline
$M_W~(\g/c^2)$ & $80.451 \pm 0.033^{~a)}$  & $80.385^{~b)} -0.29S + 0.45T$ \\
$\Gamma_{\ell\ell}(Z)$ (MeV) & $83.984 \pm 0.086^{~c)}$ & $84.011^{~b)} -0.18S
+ 0.78T$ \\
$\sef$ & $0.23152 \pm 0.00017^{~c)}$ & $0.23140^{~b)}
 + 0.00362 S - 0.00258T$ \\
$Q_W({\rm Cs})$ & $-72.5 \pm 0.8$ & $-73.19 - 0.800 S - 0.007 T$ \\
$Q_W({\rm Tl})$ & $-115.0 \pm 4.5$ & $-116.8 - 1.17S - 0.06T$ \\
 \hline
\end{tabular}
\leftline{\qquad $^{a)}$ Charlton (2001).
$^{b)}$ Marciano (2000). $^{c)}$ LEP EWWG (2001).}
\end{center}
\end{table}

Figure \ref{fig:ST} also shows predictions by Peskin and Wells (2001) of the
standard
electroweak theory.  Nearly vertical lines correspond, from left to right,
to Higgs boson masses $M_H = 100,$ 200, 300, 500, 1000 GeV; drooping curves
correspond, from top to bottom, to $+1 \sigma$, central, and $- 1 \sigma$
values of $m_t = 174.3 \pm 5.1$ GeV.

In the standard model, the combined constraints of electroweak observables
such as those in Table \ref{tab:ST} and the top quark mass favor a very light
Higgs boson, with most analyses favoring a value of $M_H$ so low that the
Higgs boson should already have been discovered.  The efficacy of a small
amount of triplet symmetry breaking has recently been stressed in a nice paper
by Forshaw \ite~(2001). It is also implied in the discussions of Dobrescu and
Hill (1998), Collins \ite~(2000), He \ite~(2001), and Peskin (2001).

The standard model prediction for $S$ and $T$ curves down quite sharply in $T$
as $M_H$ is increased, quickly departing from the region allowed by the fit to
electroweak data.  (Useful analytic expressions for the contribution of a Higgs
boson to $S$ and $T$ are given by Forshaw \ite~2001.) However, if a small
amount of triplet symmetry breaking is permitted, the agreement with the
electroweak
fit can be restored.  As an example, a value of $V_{1,0}/v = 0.03$ permits
satisfactory agreement even for $M_H = 1$ TeV, as shown by the vertical
line in the Figure.

\subsection{Supersymmetry, technicolor, and alternatives}

What could lie beyond the standard model?  The odds-on favorite among most
theorists is {\it supersymmetry}, an extremely beautiful idea which may or may
not be realized at the electroweak scale, but which almost certainly plays a
role at the Planck scale at which space and time first acquire their meaning.

The simplest illustration of supersymmetry (in one time and no space
dimensions!) does back to Darboux in 1882, who factored second-order
differential operators into the product of two first-order operators.
Dirac's famous treatment of the harmonic oscillator, writing its
Hamiltonian as $H = \hbar \omega (a^\dag a + \frac{1}{2})$, is an example of
this procedure, which was generalized by Schr\"odinger in 1941 and
Infeld and Hull in 1951.  Some of this literature is reviewed by Kwong and
Rosner (1986).  The Hamiltonian is the generator of time translations, so
this form of supersymmetry essentially amounts to saying that a time
translation can be expressed as a composite of more fundamental operations.

Modern supersymmetry envisions both spatial and time translations as
belonging to a super-algebra.  The Lorentz group is isomorphic to
SU(2) $\otimes$ SU(2) (with factors of $i$ thrown in to account for the
Minkowski metric); under this group space and time translations transform
as (1/2,1/2).  The supercharges transform as (1/2,0) and (0,1/2), clearly
more fundamental objects.

Electroweak-scale supersymmetry is motivated by several main points.
You will hear further details in this lecture series from Abel (2001).

\begin{enumerate}

\item In any gauge theory beyond the standard SU(3)$_{\rm color}
\otimes$ \SUL, if the scale $\Lambda$ of new physics is very high, this scale
tends to make its way into the Higgs sector through loop diagrams,
leading to quadratic contributions $\sim g^2 \Lambda^2$ to the Higgs boson
mass.  Unless something cancels these contributions, one has to fine-tune
counterterms in the Lagrangian to exquisite accuracy, at each order of
perturbation theory.  This is known as the ``hierarchy problem.''

\item The very nature of a $\lambda (\phi^\dag \phi)^2$ term in the
Lagrangian is problematic when considered from the standpoint of scale
changes.  This is known as the ``triviality problem.''

\item In the simplest theory by Georgi and Glashow (1974) unifying
SU(3)$_{\rm color} \otimes$
\SUL, based on the gauge group SU(5), the coupling constants approach one
another at high scale, but there is some ``astigmatism.'' In a
non-supersymmetric model, they do not all come together at the same scale.
This is known as the ``unification problem.'' It is cured in the simplest
supersymmetric model, as a result of the different particle content in loop
diagrams contributing to the running of the coupling constants.  The model
has a problem, however, in predicting too large a rate for $p \to K^+
\bar \nu$ (Murayama and Pierce 2001, Peskin 2001).

\end{enumerate}

An alternative scheme for solving these problems, which has had a much
poorer time constructing any sort of self-consistent theory, is
{\it technicolor}, the notion that the Higgs boson is a bound state of
more fundamental constituents in the same way that the pion is really a
bound state of quarks.  This mechanism works beautifully when applied
to the generation of gauge boson masses, but fails spectacularly (and
requires epicyclic patches!) when one attempts to describe fermion masses.
The basic idea of technicolor is that there is no hierarchy problem because
there is no hierarchy; a wealth of TeV-scale new physics awaits to be
discovered in the simplest version (applied to gauge bosons) of the theory.

A further, even more radical notion, is that both Higgs bosons and fermions
are composite.  This scheme so far has run aground on the difficulty of
constructing quarks and leptons, keeping their masses light by nearly
preserving a chiral symmetry ('t Hooft 1980).  One can make guesses as to
quantum numbers of constituents (Rosner and Soper 1992), but a sensible
dynamics remains completely elusive.

\subsection{Fermion masses}

We finessed the question of the origin of the Cabibbo-Kobayashi-Maskawa
(CKM) matrix.  It comes about in the following way.

The electroweak Lagrangian, before electroweak symmetry breaking, may be
written in flavor-diagonal form as
\beq
\cL_{\rm int} = - \frac{g}{\sqrt{2} }[ \overline{U '}_L
\gamma^\mu W_\mu^{(+)} {D'}_L + \hc]~~~,
\eeq
where $U' \equiv (u',c',t')$ and $D' \equiv (d',s',b')$ are column vectors
decribing {\em weak eigenstates}. Here $g$ is the weak $SU(2)_L$ coupling
constant, and $\psi_L \equiv (1 - \gf) \psi /2$ is the left-handed
projection of the fermion field $\psi = U$ or $D$.

Quark mixings arise because mass terms in the Lagrangian are permitted to
connect weak eigenstates with one another. Thus, the matrices ${\cal M}_{U,~D}$
in
\beq
\cL_m = - [\overline{U '}_R {\cal M}_U {U'}_L + \overline {D '}_R {\cal
M}_D {D'}_L + \hc]
\eeq
may contain off-diagonal terms. One may diagonalize these matrices by separate
unitary transformations on left-handed and right-handed quark fields:
\beq
R_{Q}^+ {\cal M}_{Q} L_{Q} = L_{Q}^+ {\cal M}_{Q}^+ R_Q = \Lambda_Q ~~~.
\eeq
where
\beq
{Q'}_L = L_Q Q_L ; ~~ {Q'}_R = R_Q Q_R ~~~ (Q = U, D)~~~.
\eeq
Using the relation between weak eigenstates and mass eigenstates:
${U'}_L = L_U U_L , ~ {D'}_L = L_D D_L$, we find
\beq
\cL_{\rm int} = - \frac{g}{\sqrt{2}} [ \overline{U}_L \gamma^\mu W_\mu
V D_L + \hc] ~~~,
\eeq
where $U \equiv (u,c,t)$ and $D \equiv (d,s,b)$ are the mass eigenstates, and
$V \equiv L_U^+ L_D$. The matrix $V$ is just the Cabibbo-Kobayashi-Maskawa
matrix. By construction, it is unitary: $V^+V = VV^+ = 1$. It carries no
information about $R_U$ or $R_D$. More information would be forthcoming from
interactions sensitive to right-handed quarks or from a genuine theory of
quark masses.

Quark mass matrices can yield the
observed hierarchy in CKM matrix elements.  As an example (Rosenfeld and
Rosner 2001), the regularities of quark masses evolved to a common high mass
scale can be reproduced by the choice
\beq \label{eqn:mhier}
\cM_Q \; = \; m_3 \left ( \matrix{
0       & \epsilon^3 e^{i \phi} & 0 \cr
\epsilon^3 e^{-i \phi}  & \epsilon^2    & \epsilon^2 \cr
0       & \epsilon^2    & 1 } \right ) \; ,
\eeq
where $m_3$ denotes the mass eigenvalue of the third-family quark ($t$ or $b$),
and $\epsilon \simeq 0.07$ for $u$ quarks, $\simeq 0.21$ for $d$ quarks.
Hierarchical descriptions of this type were first introduced by Froggatt and
Nielsen (1979).  The present ansatz is closely related to one described by
Fritzsch and Xing (1995).  This
type of mass matrix leads to acceptable values and phases of CKM elements.

The question of neutrino masses and mixings has entered a whole new phase
with spectacular results from neutrino observatories such as super-Kamiokande
(``Super-K'') in Japan and the Sudbury Neutrino Observatory (SNO) in
Canada.  These indicate that:

\begin{enumerate}

\item Atmospheric muon neutrinos oscillate in vacuum, probably
to $\tau$ neutrinos, with near-maximal mixing and a difference in
squared mass $\Delta m^2 \simeq 3 \times 10^{-3}$ eV$^2$.

\item Solar electron neutrinos oscillate, most likely in matter,
to some combination of muon and $\tau$ neutrinos.  All possible $\Delta m^2$
values are at most about $10^{-4}$ eV$^2$; several ranges of parameters are
permitted, with large mixing favored by some analyses.

\end{enumerate}

In addition one experiment, the Liquid Scintillator Neutrino Detector (LSND) at
Los Alamos National Laboratory, suggests $\bar \nu_\mu
\to \bar \nu_e$ oscillations with $\Delta m^2 \simeq 0.1$ to 1 eV, with small
mixing.  This possibility is difficult to reconcile with the previous two,
and a forthcoming experiment at Fermilab (Mini-BooNE) is scheduled to check
the result.  For late news on neutrinos see the Web page maintained
by Goodman (2001).

A possible explanation of small neutrino masses (Gell-Mann, Ramond, and Slansky
1979, Yanagida 1979) is that they are Majorana masses
of order $m_M = m_D^2/M_M$, where $m_D$ is a typical Dirac mass and $M_M$ is a
large Majorana mass acquired by right-handed neutrinos.  Such a mass term
is invariant under \SUL, and hence is completely acceptable in the
electroweak theory.  The pattern of neutrino Majorana and Dirac masses,
and the mixing pattern, is likely to provide us with fascinating clues over the
coming years as to the fundamental origin and nature of mass.

\section{Summary}

The Standard Model of electroweak and strong interactions has been in place for
nearly thirty years, but precise tests have entered a phase that permits
glimpses of physics beyond this impressive structure, most likely associated
with the yet-to-be discovered Higgs boson.  Studies of mixing between neutral
kaons or neutral $B$ mesons, covered by Stone (2001) in these lectures, are
attaining impressive accuracy as well, and could yield cracks in the Standard
Model at any time.  It is time to ask what lies behind the pattern of fermion
masses and mixings.  This is an {\it input} to the Standard Model,
characterized by many free parameters all of which await explanation.

\section*{Acknowledgments}

I wish to thank Fred Harris for an up-to-date copy of Figure 2, and Christine
Davies, Zumin Luo, and Denis Suprun for careful reading of the manuscript.
This work was supported in part by the United States Department of Energy
through Grant No.\ DE FG02 90ER40560.

\def \ajp#1#2{{\it American Journal of Physics} {\bf#1} #2}
\def \apny#1#2{{\it Annals of Physics (N.Y.)} {\bf#1}, #2}
\def \app#1#2{{\it Acta Physica Polonica} {\bf#1} #2}
\def \arnps#1#2{{\it Annual Review of Nuclear and Particle Science} {\bf#1} #2}
\def \art{and references therein}
\def \cmts#1#2{{\it Comments on Nuclear and Particle Physics} {\bf#1} #2}
\def \cn{Collaboration}
\def \cp89{{\it CP Violation,} edited by C. Jarlskog (World Scientific,
Singapore, 1989)}
\def \econf#1#2#3{Electronic Conference Proceedings {\bf#1}, #2 (#3)}
\def \efi{Enrico Fermi Institute Report No.\ }
\def \epjc#1#2{{\it European Journal of Physics} C {\bf#1}, #2}
\def \f79{{\it Proceedings of the 1979 International Symposium on Lepton and
Photon Interactions at High Energies,} Fermilab, August 23-29, 1979, ed. by
T. B. W. Kirk and H. D. I. Abarbanel (Fermi National Accelerator Laboratory,
Batavia, IL, 1979}
\def \hb87{{\it Proceeding of the 1987 International Symposium on Lepton and
Photon Interactions at High Energies,} Hamburg, 1987, ed. by W. Bartel
and R. R\"uckl (Nucl.\ Phys.\ B, Proc.\ Suppl., vol. 3) (North-Holland,
Amsterdam, 1988)}
\def \ib{{\it ibid.}~}
\def \ibj#1#2#3{~{\bf#1}, #2 (#3)}
\def \ichep72{{\it Proceedings of the XVI International Conference on High
Energy Physics}, Chicago and Batavia, Illinois, Sept. 6 -- 13, 1972,
edited by J. D. Jackson, A. Roberts, and R. Donaldson (Fermilab, Batavia,
IL, 1972)}
\def \ichepb{{\it Proceedings of the XIII International Conference on High
Energy Physics} (Berkeley, CA: Univ.~of Calif.~Press)}
\def \ijmpa#1#2#3{Int.\ J.\ Mod.\ Phys.\ A {\bf#1}, #2 (#3)}
\def \jhep#1#2#3{JHEP {\bf#1}, #2 (#3)}
\def \jpb#1#2#3{J.\ Phys.\ B {\bf#1}, #2 (#3)}
\def \jpg#1#2{{\it Journal of Physics} G {\bf#1}, #2}
\def \lg{{\it Proceedings of the XIXth International Symposium on
Lepton and Photon Interactions,} Stanford, California, August 9--14, 1999,
edited by J. Jaros and M. Peskin (World Scientific, Singapore, 2000)}
\def \lkl87{{\it Selected Topics in Electroweak Interactions} (Proceedings of
the Second Lake Louise Institute on New Frontiers in Particle Physics, 15 --
21 February, 1987), edited by J. M. Cameron \ite~(World Scientific, Singapore,
1987)}
\def \kaon{{\it Kaon Physics}, edited by J. L. Rosner and B. Winstein,
University of Chicago Press, 2001}
\def \kdvs#1#2#3{{Kong.\ Danske Vid.\ Selsk., Matt-fys.\ Medd.} {\bf #1}, No.\
#2 (#3)}
\def \ky{{\it Proceedings of the International Symposium on Lepton and
Photon Interactions at High Energy,} Kyoto, Aug.~19-24, 1985, edited by M.
Konuma and K. Takahashi (Kyoto Univ., Kyoto, 1985)}
\def \lpRoma{XX International Symposium on Lepton and Photon Interactions
at High Energies, Rome, Italy, July 23--27, 2001}
\def \mpla#1#2#3{Mod.\ Phys.\ Lett.\ A {\bf#1}, #2 (#3)}
\def \nat#1#2#3{Nature {\bf#1}, #2 (#3)}
\def \nc#1#2{{\it Nuovo Cimento} {\bf#1} #2}
\def \nima#1#2{{\it Nuclear Instruments and Methods} A {\bf#1} #2}
\def \np#1#2#{{\it Nuclear Physics} {\bf#1} #2}
\def \npps#1#2#3{Nucl.\ Phys.\ Proc.\ Suppl.\ {\bf#1}, #2 (#3)}
\def \os{XXX International Conference on High Energy Physics, Osaka, Japan,
July 27 -- August 2, 2000}
\def \PDG{Particle Data Group (Groom D E \ite), 2000, \epjc{15}{1}}
\def \pisma#1#2#3#4{Pis'ma Zh.\ Eksp.\ Teor.\ Fiz.\ {\bf#1}, #2 (#3) [JETP
Lett.\ {\bf#1}, #4 (#3)]}
\def \pl#1#2{{\it Physics Letters} {\bf#1} #2}
\def \pla#1#2#3{Phys.\ Lett.\ A {\bf#1}, #2 (#3)}
\def \plb#1#2{{\it Physics Letters} B {\bf#1} #2}
\def \ppmsj#1#2{{\it Proceedings of the Physical and Mathematical Society of
Japan} #1:#2} 
\def \pr#1#2{{\it Physical Review} {\bf#1} #2}
\def \prd#1#2{{\it Physical Review} D {\bf#1} #2}
\def \prl#1#2{{\it Physical Review Letters} {\bf#1} #2}
\def \prp#1#2{{\it Physics Reports} {\bf#1} #2}
\def \ptp#1#2{{\it Progress of Theoretical Physics} (Kyoto) {\bf#1} #2}
\def \ptps#1#2{{\it Progress of Theoretical Physics (Suppl.)} {\bf#1} #2}
\def \rmp#1#2{{\it Reviews of Modern Physics} {\bf#1} #2}
\def \si90{25th International Conference on High Energy Physics, Singapore,
Aug. 2-8, 1990}
\def \slc87{{\it Proceedings of the Salt Lake City Meeting} (Division of
Particles and Fields, American Physical Society, Salt Lake City, Utah, 1987),
ed. by C. DeTar and J. S. Ball (World Scientific, Singapore, 1987)}
\def \slac89{{\it Proceedings of the XIVth International Symposium on
Lepton and Photon Interactions,} Stanford, California, 1989, edited by M.
Riordan (World Scientific, Singapore, 1990)}
\def \smass82{{\it Proceedings of the 1982 DPF Summer Study on Elementary
Particle Physics and Future Facilities}, Snowmass, Colorado, edited by R.
Donaldson, R. Gustafson, and F. Paige (World Scientific, Singapore, 1982)}
\def \smass90{{\it Research Directions for the Decade} (Proceedings of the
1990 Summer Study on High Energy Physics, June 25--July 13, Snowmass, Colorado),
edited by E. L. Berger (World Scientific, Singapore, 1992)}
\def \tasi{{\it Testing the Standard Model} (Proceedings of the 1990
Theoretical Advanced Study Institute in Elementary Particle Physics, Boulder,
Colorado, 3--27 June, 1990), edited by M. Cveti\v{c} and P. Langacker
(World Scientific, Singapore, 1991)}
\def \yaf#1#2#3{{\it Yadernaya Fizika} {\bf#1} #2 [{\it Soviet Journal of
Nuclear Physics} {\bf#1} #3]}
\def \yafo#1#2#3#4{{\it Yadernaya Fizika} {\bf#1} #2 [#4 {\it Soviet Journal of
Nuclear Physics} {\bf#1} #3]}
\def \zhetf#1#2#3#4#5#6{Zh.\ Eksp.\ Teor.\ Fiz.\ {\bf #1}, #2 (#3) [Sov.\
Phys.\ - JETP {\bf #4}, #5 (#6)]}
\def \zpc#1#2#3{Zeit.\ Phys.\ C {\bf#1}, #2 (#3)}
\def \zpd#1#2#3{Zeit.\ Phys.\ D {\bf#1}, #2 (#3)}

\section*{References}
\frenchspacing
\begin{small}


\reference{Abachi S \ite~(D0 \cn), 1995, \prl{74}{2632}.}

\reference{Abe F (CDF \cn), 1994, \prl{73}{225};
1995, \prd{50}{2966}.}

\reference{Abe K \ite~(Belle \cn), 2001, \prl{87}{091802}.}

\reference{Abel S, 2001, lectures at Scottish Universities'
Summer School in Physics (these Proceedings).}

\reference{Abers E S and Lee B W, 1973, \prp{9C}{1}.}

\reference{Amaldi U \ite, 1987, \prd{36}{1385}.}

\reference{Appelquist T and Politzer H D, 1975, \prl{34}{43}.}

\reference{Aubert B \ite~(BaBar \cn), 2001a, \prl{87}{091801};
SLAC Report No.\ SLAC-PUB-9060, hep-ex/0201020, submitted to Phys.\ Rev.\ D.}

\reference{Aubert B \ite~(BaBar \cn), 2001b, SLAC Report No.\ SLAC-PUB-9012,
hep-ex/0110062, submitted to Phys.\ Rev.\ D.}

\reference{Aubert J J \ite, 1974, \prl{33}{1404}.}

\reference{Augustin J J \ite, 1974, \prl{33}{1406}.}

\reference{Bai J Z \ite~(BES \cn), preprint hep-ex/0102003 v2, 31 May 2001.}

\reference{Barate R \ite~(ALEPH \cn), 2001, \plb{495}{1}.}

\reference{Bardeen W A, Buras A J, Duke D W, and Muta, T, 1978,
\prd{18}{3998}.}

\reference{Benvenuti A \ite, 1974, \prl{32}{800}.}

\reference{Bjorken J D and Glashow S L, 1964, \pl{11}{255}.}

\reference{Bouchiat C, Iliopoulos, J, and Meyer Ph, 1972, \pl{38B}{519}.}

\reference{Brown H B \ite~(Brookhaven E821 \cn), 2001, \prl{86}{2227}.}

\reference{Buchalla G, 2001, lectures at Scottish Universities' Summer School
in Physics (these Proceedings).}

\reference{Cabibbo N, 1963, \prl{10}{531}.}

\reference{Carena M, Conway J S, Haber H E, Hobbs J D \ite, 2000, ``Report of
the Higgs Working Group of the Tevatron Run 2
SUSY/Higgs Workshop,'' hep-ph/0010338.}

\reference{Caswell W E, 1974, \prl{33}{244}.}

\reference{Cazzoli E G \ite, 1975, \prl{34}{1125}.}

\reference{Charlton D, 2001, plenary talk at International Europhysics
Conference on High Energy Physics, Budapest, Hungary, July 12--18, Univ.\ of
Birmingham report BHAM-HEP/01-02,\\ hep-ex/0110086, to be published by JHEP.}

\reference{Chivukula S in {\it The Albuquerque Meeting} (Proceedings of the 8th
Meeting, Division of Particles and Fields, American Physical Society,
Albuquerque, NM, August 2--6, 1994) editor Seidel S, 1995
(World Scientific), p 273.}

\reference{Christenson J H, Cronin J W, Fitch V L, and Turlay R E, 1964, 
\prl{13}{138}.}

\reference{Coleman S, 1971, in {\it Properties of the
Fundamental Interactions} (1971 Erice Lectures), editor Zichichi, A (Editrice
Compositori, Bologna), p 358.}

\reference{Collins H, Grant A K, and Georgi H, 2000, \prd{61}{055002}.}

\reference{Conrad J C, Shaevitz M H, and Bolton T, 1998 \rmp{70}{1341}.}

\reference{Dalitz R H, 1967, \ichepb,~p 215.}

\reference{Davier M and H\"ocker A, 1998, \plb{419}{419}; \plb{435}{427}.}

\reference{De R\'ujula A, Georgi H, and Glashow S L, 1975, \prd{12}{147}.}

\reference{Dobrescu B and Hill C T, 1998, \prl{81}{2634}.}

\reference{Drees J, \lpRoma, hep-ex/0110077.}

\reference{Dzuba V A, Flambaum V V, and Ginges J S M, hep-ph/0111019, submitted
to Phys.\ Rev.\ A.}

\reference{Eichten E \ite, 1984, \rmp{56}{579}; 1986, \rmp{58}{1065(E)}.}

\reference{Ellis J, 1977, in {\it Weak and Electromagnetic
Interactions at High Energies} (1976 Les Houches Lectures), editors
Balian R and Llewellyn Smith C H (North-Holland), p 1.}

\reference{Forshaw J R, Ross D A, and White B E,
University of Manchester report MC-TH-01/07, hep-ph/0107232.}

\reference{Fritzsch H and Xing Z Z, 1995,
\plb{353}{114}; Xing Z Z, 1997, \jpg{23}{1563}.}
 
\reference{Froggatt C D and Nielsen H B, 1979, \np{B147}{277}.}
 
\reference{Gaillard M and Lee B W, 1974, \prd{10}{894}.}

\reference{Gell-Mann M, 1964, \pl{8}{214}.}

\reference{Gell-Mann M and L\'{e}vy M, 1960, \nc{19}{705}.}

\reference{Gell-Mann M and Ne'eman Y, 1964, {\it The Eightfold Way} (Benjamin).}

\reference{Gell-Mann M, Ramond P, and Slansky R, in {\it Supergravity}, editors
van Nieuwenhuizen P and Freedman D Z, 1979 (North-Holland), p.~315.
See also Gell-Mann M, Slansky R, and Stephenson G, 1979 (unpublished).}

\reference{Georgi H and Glashow S L, 1972a, \prl{28}{1494}.}

\reference{Georgi H and Glashow S L, 1972b, \prd{6}{429}.}

\reference{Georgi H and Glashow S L, 1974, \prl{32}{438}.}

\reference{Gilman F, Kleinknecht K, and Renk Z, 2000, in \PDG, pp 110-114.}

\reference{Glashow S, 1961, \np{22}{579}.}

\reference{Glashow S L, Ilipoulos J, and Maiani L, 1970, \prd{2}{1285}.}

\reference{Glashow S L and Weinberg S, 1977, \prd{15}{1958}.}

\reference{Goldhaber G \ite, 1976, \prl{37}{255}.}

\reference{Goodman M, 2001,
http://www.hep.anl.gov/ndk/hypertext/nu\_industry.html.}

\reference{Gronau M and Rosner J L, 2002, \prd{65}{013004}.}

\reference{Gross D J and Jackiw R, 1972, \prd{6}{477}.}

\reference{Gross D J and Wilczek F, 1973, \prd{8} {3633}; 1974 \prd{9}{980}.}

\reference{Gunion J, Haber H E, Kane G, and Dawson S, 1990, {\it The Higgs
Hunter's Guide} (Addison-Wesley).}

\reference{Hara Y, 1964, \pr{134}{B701}.}

\reference{Hasert F J \ite, 1973, \pl{46B}{121,138}; 1974, \np{B73}{1}.}

\reference{Hayashi M and Kinoshita M, KEK Report No.\ KEK-TH-793,
hep-ph/0112102.}

\reference{He H J, Polonsky N, and Su S, \prd{64}{053004}
{2001}; He H-J, Hill C T, and Tait T M P, Univ.\ of Texas
Report No.\ UTEXAS-HEP-01-013, hep-ph/0108041 (unpublished).}
 
\reference{Higgs P W, 1964, \pl{12}{132}; \pl{13}{508};  see also
Englert F and Brout R, 1964, \prl{13}{321}; Guralnik G S, Hagen C R, and Kibble
T W B, 1964, \prl{13}{585}.}

\reference{H\"ocker A, Lacker H, Laplace S, and Le Diberder F, 2001,
\epjc{21}{225}; see
http://www.slac.stanford.edu/~laplace/ckmfitter.html for periodic updates.}

\reference{'t Hooft G, 1973, \np{B61}{455}.}

\reference{'t Hooft G, 1980, in {\it Recent Developments in
Gauge Theories} (Carg\`ese Summer Institute, Aug. 26 - Sept. 8, 1979) 't Hooft
G \ite~editors (Plenum, New York), p 135.}

\reference{Hughes R J, 1980, \pl{97B}{246}: 1981, \np{B186}{376}.}

\reference{Kennedy D C and Langacker P G,
1990, \prl{65}{2967}; 1991, \prl{66}{395(E)}.}

\reference{Khriplovich I B, 1969, \yaf{10}{409}{235}.}

\reference{Kim Y K, 2001, \lpRoma.}

\reference{Klein O, 1938, in {\it Les Nouvelles Th\'eories de
la Physique}, Paris, Inst.~de Co\"operation Intellectuelle (1939), p 81,
reprinted in {\it Oskar Klein Memorial Lectures} vol 1, editor Ekspong G, 1991
(World Scientific, Singapore).}

\reference{Knecht M and Nyffeler A, 2001, Centre de Physique
Th\'eorique (Marseille) Report No.\
CPT-2001/P.4253, hep-ph/0111058.}

\reference{Kobayashi M and Maskawa T, 1973, \ptp{49}{652}.}

\reference{Kosteleck\'y V A and Roberts A, 2001, \prd{63}{096002}.}

\reference{Kwong W, 1991, \prd{43}{1488}.}

\reference{Kwong W and Rosner J L, 1986, \ptps{86}{366}.}

\reference{Langacker P, ``Precision Electroweak Data:
Phenomenological Analysis,'' talk at Snowmass 2001 Workshop, Univ.\ of
Pennsylvania report UPR-0959T, hep-ph/0110129.}

\reference{Langacker P, Luo M, and Mann A K, 1992, \rmp{64}{87}.}

\reference{LEP Electroweak Working Group, 2001: see
http://lepewwg.web.cern.ch/LEPEWWG for periodic updates.}

\reference{LEP Higgs Working Group, LHWG Note/2001-03, hep-ex/0107029, paper
contributed to EPS '01 Conference, Budapest, and \lpRoma.}

\reference{Lipkin H J, 1973, \prp{8}{173}, and references therein.}

\reference{Llewellyn Smith C H, 1983, \np{B228}{205}.}

\reference{Luo Z and Rosner J L, 2001, \efi 01-28, hep-ph/0108024, to be
published in {\it Physical Review D}.}

\reference{Maki Z and Ohnuki Y, 1964, \ptp{32}{144}.}

\reference{Marciano W, Brookhaven National Laboratory
report BNL-HET-00/04, hep-ph/0003181, published in {\it Proceedings of
MuMu 99} (5th International Conference on Physics Potential and Development
of $\mu^+ \mu^-$ Colliders, San Francisco, CA, December 1999), editor
Cline D, AIP Conference Proceedings v 542 (American Institute of Physics).}

\reference{Milstein A I and Sushkov A P, 2001, Novosibirsk and Univ.\ of
New South Wales preprint, hep-ph/0109257.}

\reference{Murayama H and Pierce A, 2001, University of California at Berkeley
preprint, hep-ph/0108104.}

\reference{Nambu Y, 1966, in De-Shalit A, Feshbach H and Van Hove L eds
{\it Preludes in Theoretical Physics in Honor of V. F. Weisskopf} (Amsterdam:
North-Holland and New York: Wiley) pp 133-42.}

\reference{Nambu Y, 1974, \prd{10}{4262}.}

\reference{Nir Y, 2001, lectures at Scottish Universities'
Summer School in Physics (these Proceedings).}

\reference{Niu K, Mikumo E, and Maeda Y, 1971, \ptp{46}{1644}.}

\reference{Paschos E A and Wolfenstein L, 1973, \prd{7}{91}.}

\reference{\PDG.  All experimental values are quoted from this source unless
otherwise noted.}

\reference{Pati J C and Salam A, 1973, \prl{31}{661};
1974, \prd{10}{275}; see also
Mohapatra R N and Pati J C, 1975, \prd{11}{566, 2558}.}

\reference{Perl M L \ite~(SLAC-LBL \cn),
1975, \prl{35}{1489}; 1976, \pl{63B}{466}.}

\reference{Peruzzi I \ite, 1976, \prl{37}{569}.}

\reference{Peskin M E, 2001, ``Interpretation of Precision
Electroweak Data, or Should We Really Believe There is a Light Higgs
Boson?'', seminar at Snowmass 2001 Workshop, July, 2001.}

\reference{Peskin M E and Schroeder D V, 1995,
{\it An Introduction to Quantum Field Theory} (Addison-Wesley).}

\reference{Peskin M E and Takeuchi T, 1990, \prl{65}{964}; 1992,
\prd{46}{381}.}

\reference{Peskin M E and Wells J, 2001, \prd{64}{093003}.}

\reference{Politzer H D, 1974, \prp{14C}{129}.}

\reference{Quigg C, 1983, {\it Gauge Theories of the Strong,
Weak, and Electromagnetic Interactions} (Benjamin--Cummings).}

\reference{Rosenfeld R and Rosner J L, 2001, \plb{516}{408}.}

\reference{Rosner J, 1988, An Introduction to Standard Model
Physics, in {\it The Santa Fe TASI-87} (Proceedings of the 1987 Theoretical
Advanced Study Institute, St. John's College, Santa Fe, NM, July 6--10, 1987),
editors Slansky R and West G (World Scientific), p 3.}

\reference{Rosner J, 1997, Top Quark Mass, in {\it Masses of
Fundamental Particles -- Carg\'ese 1996}, editors L\'evy M \ite~(Plenum),
p 43.}

\reference{Rosner J L, 1999, \prd{61}{016006}.}

\reference{Rosner J L, 2001, \efi 01-43, hep-ph/0109239, to be published in
{\it Physical Review D}.}

\reference{Rosner J L and Slezak S, 2001, \ajp{69}{44}.}

\reference{Rosner J and Soper D E, 1992, \prd{45}{3206}.}

\reference{Ross D A, 1978, \app{B10}{189}.}

\reference{Sakurai J J, 1967, {\it Advanced Quantum
Mechanics} (Addison-Wesley), pp 282 ff.}

\reference{Salam A, 1968, {\it Proceedings of the Eighth Nobel
Symposium}, editor Svartholm N (Almqvist and Wiksell, Stockholm; Wiley, New
York, 1978), p 367.}

\reference{Schubert K, 2001, lectures at Scottish
Universities' Summer School in Physics (these Proceedings).}

\reference{Sikivie P, Susskind L, Voloshin M B, and Zakharov V, 1980,
\np{B173}{189}.}

\reference{Slansky R, 1981, \prp{79}{1}.}

\reference{Stone S, 2001, lectures at Scottish Universities'
Summer School in Physics (these Proceedings).}

\reference{Strassler M J and Peskin M E, 1991, \prd{43}{1500}.}

\reference{Swartz M, lecture at Snowmass 2001 Workshop,
see http://pha.jhu.edu/\~{}morris/higgs.pdf for transparencies.}

\reference{Veltman M, 1977, \np{B123}{89}; \app{B8}{475}.}

\reference{Weinberg S, 1967, \prl{19}{1264}.}

\reference{Willenbrock S, 2000, \rmp{72}{1141}.}

\reference{Wolfenstein L, 1983, \prl{51}{1945}.}

\reference{Yanagida T, 1979, in {\it Proceedings of the Workshop on
Unified Theory and Baryon Number in the Universe}, editors Sawada O and
Sugamoto A, 1979, (Tsukuba, Japan, National Laboratory for High Energy
Physics), p 95.}

\reference{Yang C N, 1974, \prl{33}{445}.}

\reference{Yang C N and Mills R L, 1954, \pr{96}{191}.}

\reference{Yukawa H, 1935, \ppmsj{17}{48}.}

\reference{Zel'dovich Ya B and Sakharov A D, 1966, \yafo{4}{395}{283}{1967}.}

\reference{Zeller G P \ite~(NuTeV \cn), presented
at 1999 Division of Particles and Fields Meeting, UCLA, 5--9 January 1999,
hep-ex/9906024.}

\reference{Zweig G, 1964, CERN Reports 8182/TH 401 and 8419/TH
412 (unpublished), second paper reprinted in {\it Developments in the Quark
Theory of Hadrons}, 1980, editors Lichtenberg D B and Rosen S P (Hadronic
Press), v 1, pp 22--101.}

\end{small}
\end{document}